\documentclass{article}

\usepackage{xr}
\makeatletter

\newcommand*{\addFileDependency}[1]{
\typeout{(#1)}
\@addtofilelist{#1}
\IfFileExists{#1}{}{\typeout{No file #1.}}
}\makeatother

\usepackage{t1enc}
\usepackage[flushleft]{threeparttable} 
\usepackage{moreverb,url}
\usepackage{pdflscape,lipsum}
\usepackage{longtable}
\usepackage{float}
\usepackage[colorlinks,bookmarksopen,bookmarksnumbered,citecolor=red,urlcolor=red]{hyperref}
\usepackage{multirow} 
\usepackage{makecell} 
\usepackage{caption}
\DeclareCaptionLabelFormat{AppendixTables}{Table A.#2}

\usepackage{subcaption}
\usepackage{graphicx}

\newcommand\BibTeX{{\rmfamily B\kern-.05em \textsc{i\kern-.025em b}\kern-.08em
T\kern-.1667em\lower.7ex\hbox{E}\kern-.125emX}}

\def\volumeyear{2016}

\usepackage[utf8]{inputenc} 
\usepackage[english]{babel} 
\usepackage{amsmath}
\usepackage{amssymb}
\usepackage{txfonts}
\usepackage{mathdots}
\usepackage[classicReIm]{kpfonts}
\usepackage{graphicx}
\usepackage{tabulary}

\usepackage[verbose=true,letterpaper]{geometry}
\AtBeginDocument{
  \newgeometry{
    textheight=9in,
    textwidth=5.75in,
    top=1in,
    headheight=14pt,
    headsep=25pt,
    footskip=30pt
  }
}

\usepackage{authblk}
\RequirePackage{caption}
\title{\bf \Large \rule{\textwidth}{2pt}\\
Methods for non-proportional hazards in clinical trials: A systematic review\\
\rule{\textwidth}{2pt}}

\author[1,*]{Maximilian Bardo}
\author[1,*]{Cynthia Huber}
\author[1,2]{Norbert Benda}
\author[3]{Jonas Brugger}
\author[4]{Tobias Fellinger}
\author[5]{Vaidotas Galaune} 
\author[1]{Judith Heinz} 
\author[6]{Harald Heinzl} 
\author[5]{Andrew C. Hooker}
\author[4]{Florian Klinglm\"{u}ller} 
\author[3]{Franz K\"{o}nig} 
\author[1]{Tim Mathes}
\author[6]{Martina Mittlb\"{o}ck} 
\author[3]{Martin Posch}
\author[3]{Robin Ristl}
\author[1]{Tim Friede}

\affil[1]{University Medical Center G\"{o}ttingen, Department of Medical Statistics, G\"{o}ttingen, Germany}
\affil[2]{Federal Institute for Drugs and Medical Devices, Bonn, Germany}
\affil[3]{Medical University of Vienna, Center for Medical Data Science, Section of Medical Statistics, Vienna, Austria}
\affil[4]{Agentur f\"{u}r Gesundheit und Ern\"{a}hrungssicherheit (AGES), Vienna, Austria}

\affil[5]{Uppsala University, Department of Pharmacy, Uppsala, Sweden}
\affil[6]{Medical University of Vienna, Center for Medical Data Science, Section of Clinical Biometrics, Vienna, Austria}
\affil[*]{Maximilian Bardo and Cynthia Huber contributed equally to this study.\newline Corresponding author: cynthia.huber@med.uni-goettingen.de }

\newcommand{\NEW}[2][black]{\textcolor{#1}{#2}}
\begin{document}

\maketitle
\normalsize

\begin{abstract}

\noindent For the analysis of time-to-event data, frequently used methods such as the log-rank test or the Cox proportional hazards model are based on the proportional hazards assumption, which is often debatable. Although a wide range of parametric and non-parametric methods for non-proportional hazards (NPH) has been proposed, there is no consensus on the best approaches. To close this gap, we conducted a systematic literature search to identify statistical methods and software appropriate under NPH. 
Our literature search identified 907 abstracts, out of which we included 211 articles, mostly methodological ones.
Review articles and applications were less frequently identified.
The articles discuss effect measures, effect estimation and regression approaches, hypothesis tests, and sample size calculation approaches, which are often tailored to specific NPH situations. 
Using a unified notation, we provide an overview of methods available. 
Furthermore, we derive some guidance from the identified articles.\\
\textbf{Keywords:} Cox model, log-rank test, survival analysis, right-censored observations, non-proportional hazards
\end{abstract}
\newpage

 \section{Introduction}\label{sec.intro}
\begin{figure}[!ht]
\centering
\begin{subfigure}{0.3\textwidth}
    \includegraphics[width=\textwidth]{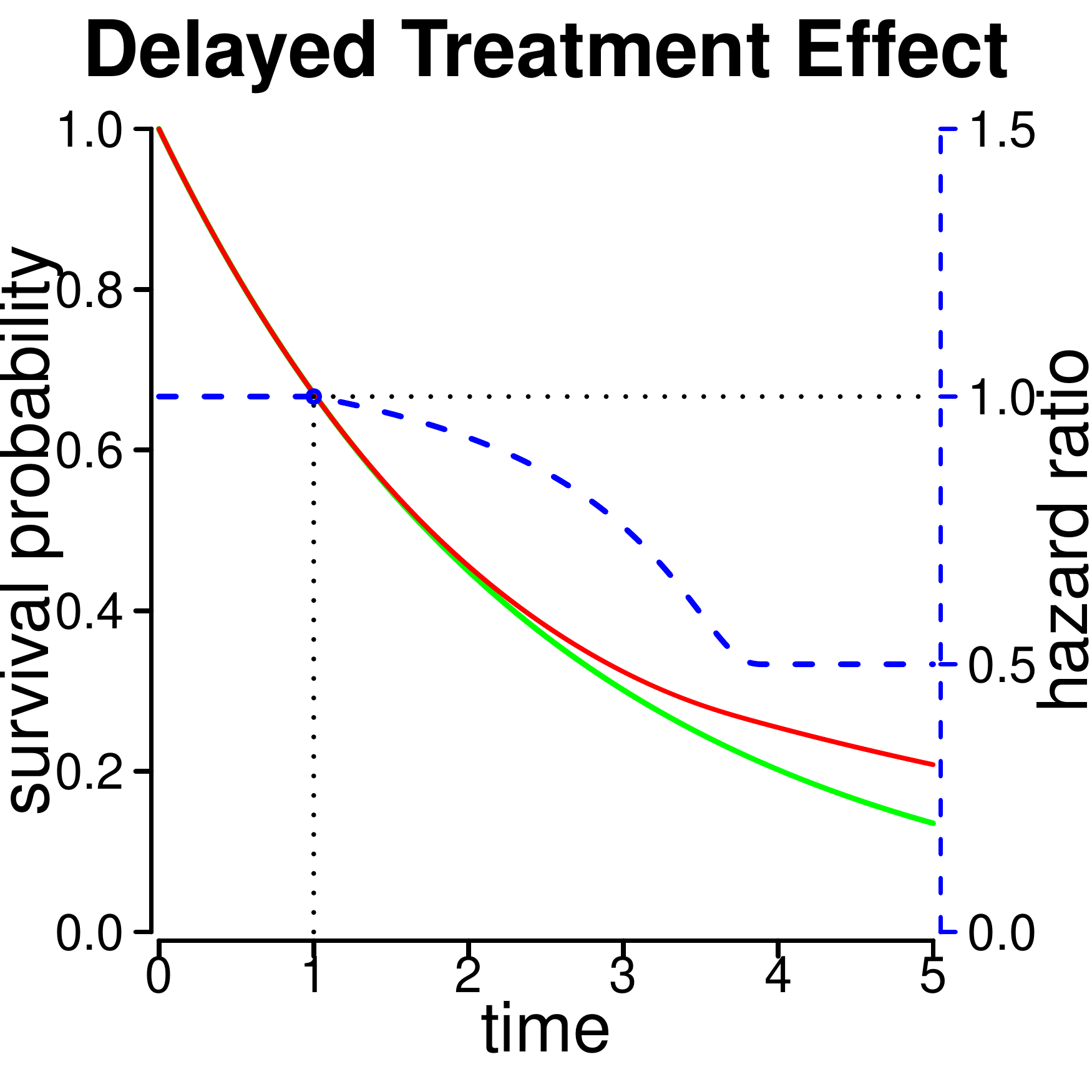}
    \caption{Delayed effect.
    }
    \label{fig:delay}
\end{subfigure}
\hfill
\begin{subfigure}{0.3\textwidth}
    \includegraphics[width=\textwidth]{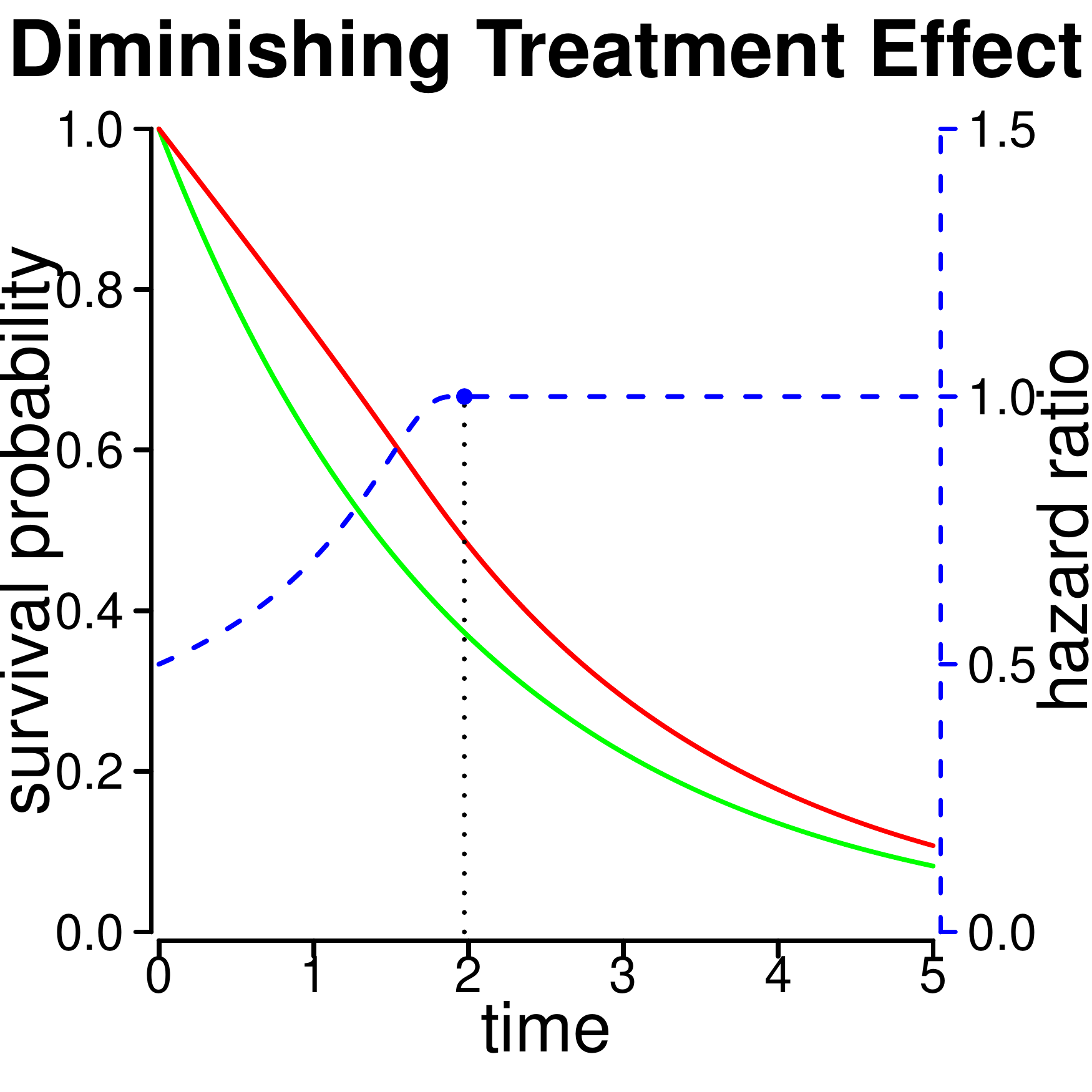}
    \caption{Diminishing effect.}
    \label{fig:dim}
\end{subfigure}
\hfill
\begin{subfigure}{0.3\textwidth}
    \includegraphics[width=\textwidth]{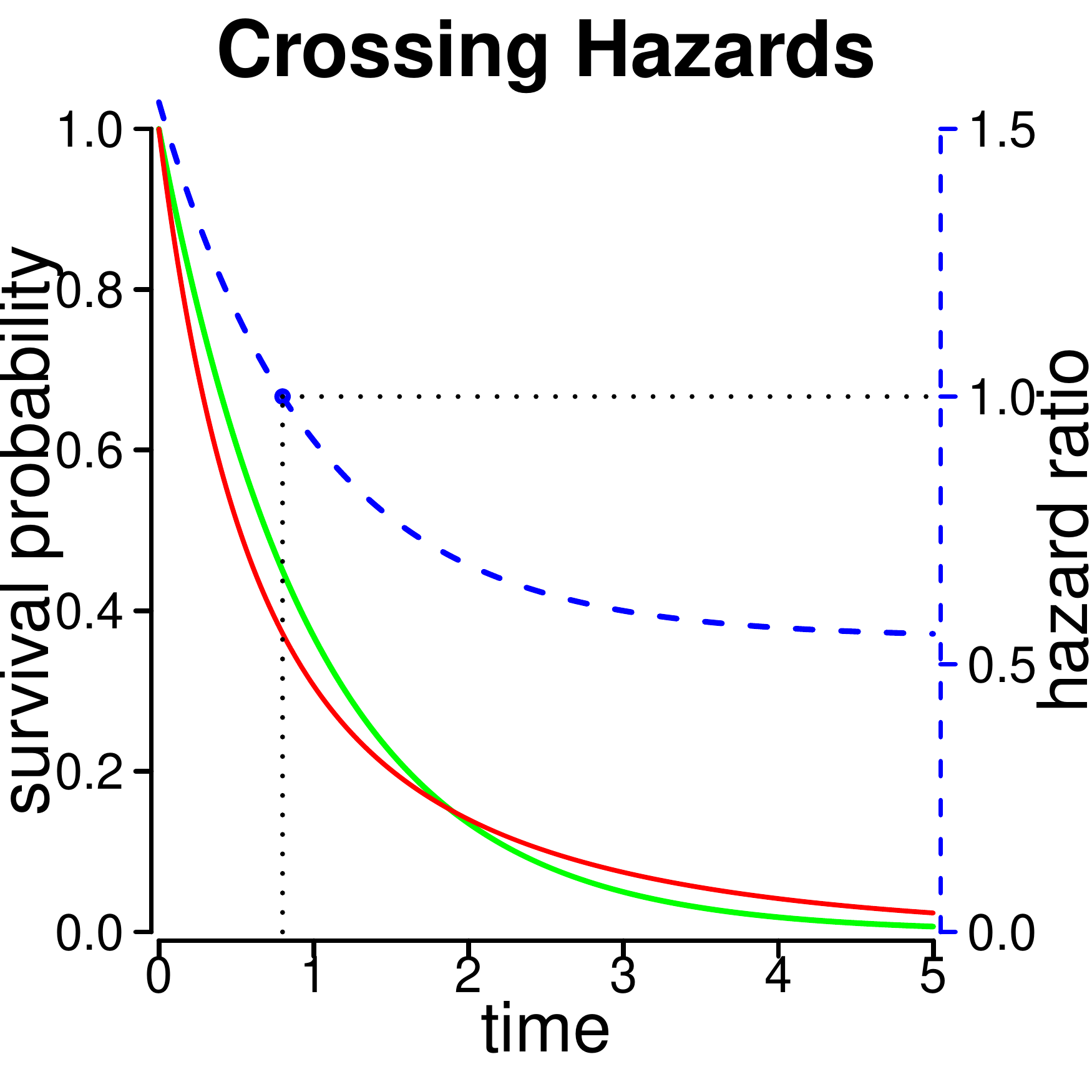}
    \caption{Crossing hazards.}
   \label{fig:cross}
\end{subfigure}
\caption{Stylized NPH treatment effect scenarios with hazard ratio (blue dashed line) and corresponding survival curves (red and green solid lines).
The green line represents the reference group.
The dotted line with black solid points refers to the time when the hazard ratio is equal to $1$.}
\label{fig:HRscenario}
\end{figure}

\noindent
In clinical studies with time-to-event outcomes, it is commonly assumed that the hazard functions of the treatment groups are proportional. 
However, several scenarios can lead to non-proportional hazards (NPH).
Figure \ref{fig:delay} and \ref{fig:dim} illustrate the hazard ratio of a delayed and a diminishing treatment effect, respectively.
A delayed treatment effect for the experimental arm can also lead to crossing hazards (see Figure \ref{fig:cross}) if the comparator is an active treatment with an immediate response as is often the case in trials concerning immuno-oncology drugs.
Other scenarios of crossing hazards are experiments where the treatment effect is non-homogeneous across subgroups, i.e. if the treatment is harmful in a subgroup but beneficial in its complement \cite{Ananthakrishnan.2021}.
NPH can also occur in settings with long-term survivors in one treatment arm or if there is treatment switching to another arm after disease progression on the original arm.

Under proportional hazards (PH), comparisons of hazard ratios or cumulative hazard ratios result in equivalent conclusions, whereas under NPH these results may vary substantially. Standard statistical tests for the comparison of time-to-event outcomes between groups such as the log-rank test or tests based on Cox regression models are not optimal for detecting relevant differences under NPH. Additionally, the hazard ratio estimate of the standard Cox regression model, a commonly used effect measure, is neither robust nor meaningful under NPH \cite{Xu.2002}. 
In contrast to PH, the interpretation of estimates of a specific effect measure, such as the hazard ratio or the cumulative hazard ratio, depend on the follow-up considered for evaluation in the presence of NPH.\\
Well-established methods for time-to-event data are available when the PH assumption holds. However, there is no consensus on best practices under NPH.  
Moreover, approaches to deal with NPH are not globally optimal but depend on the specific NPH scenario.
A variety of parametric and non-parametric methods for treatment effect estimation and hypothesis testing in NPH settings have been proposed. 
We aim to identify statistical methods and, if available, the corresponding software that is suitable for NPH.
In contrast to other overview articles that focus on specific disease areas (e.g., oncology \cite{Ananthakrishnan.2021}), NPH patterns (e.g., switching treatment \cite{Latminer.2018}), or specific methods (e.g., statistical testing \cite{Dormuth.2022, Royston.2020}), the scope of this literature review is broader and based on a systematic approach to identifying relevant literature.

\noindent The remainder of this paper is organized as follows.
In Section \ref{motiv}, we show the relevance of scenarios with NPH by investigating reconstructed data from \NEW{a} clinical trial.
In Section \ref{systematic}, we describe the literature search, data extraction and summarize the quantitative results of the review.
The identified approaches are presented in a common notation, which can be found in Section \ref{classification}, where we focus on NPH for the treatment indicator. 
We categorize and discuss approaches to estimate and model treatment or covariate effects under NPH in Section \ref{estimation}.
Testing and sample size calculation approaches under NPH are discussed in Section \ref{test}. 
We compare the flexibility of the proposed methods presented in Sections \ref{estimation} and \ref{test} on theoretical grounds and highlight results of conducted comparison studies if available.
Finally, we summarize and discuss the findings in Section \ref{discussion}.
The Appendix \ref{Appendix} provides more detailed information on the literature search and data extraction.
The Online Supplement \ref{Supplement} provides more detailed information on the estimation and testing approaches identified as appropriate for NPH.


\section{{Motivation}} \label{motiv} 
\noindent \NEW{Borghaei et al \cite{Borghaei.2015} report a phase 3 trial comparing the effect of nivolumab versus docetaxel in nonsquamous non-small lung cancer concerning overall survival. 
For illustration, we consider the study's secondary endpoint, progression-free survival (PFS).
Using the webplotdigitizer \cite{Rohatgi.2022} and the method described in \cite{Guyot.2012}, we reconstructed the individual patient data by digitizing the Kaplan-Meier (KM) estimates of the survival functions. }

\begin{figure}[!ht]
\centering
\begin{subfigure}{0.495\textwidth}
    \includegraphics[width=\textwidth]{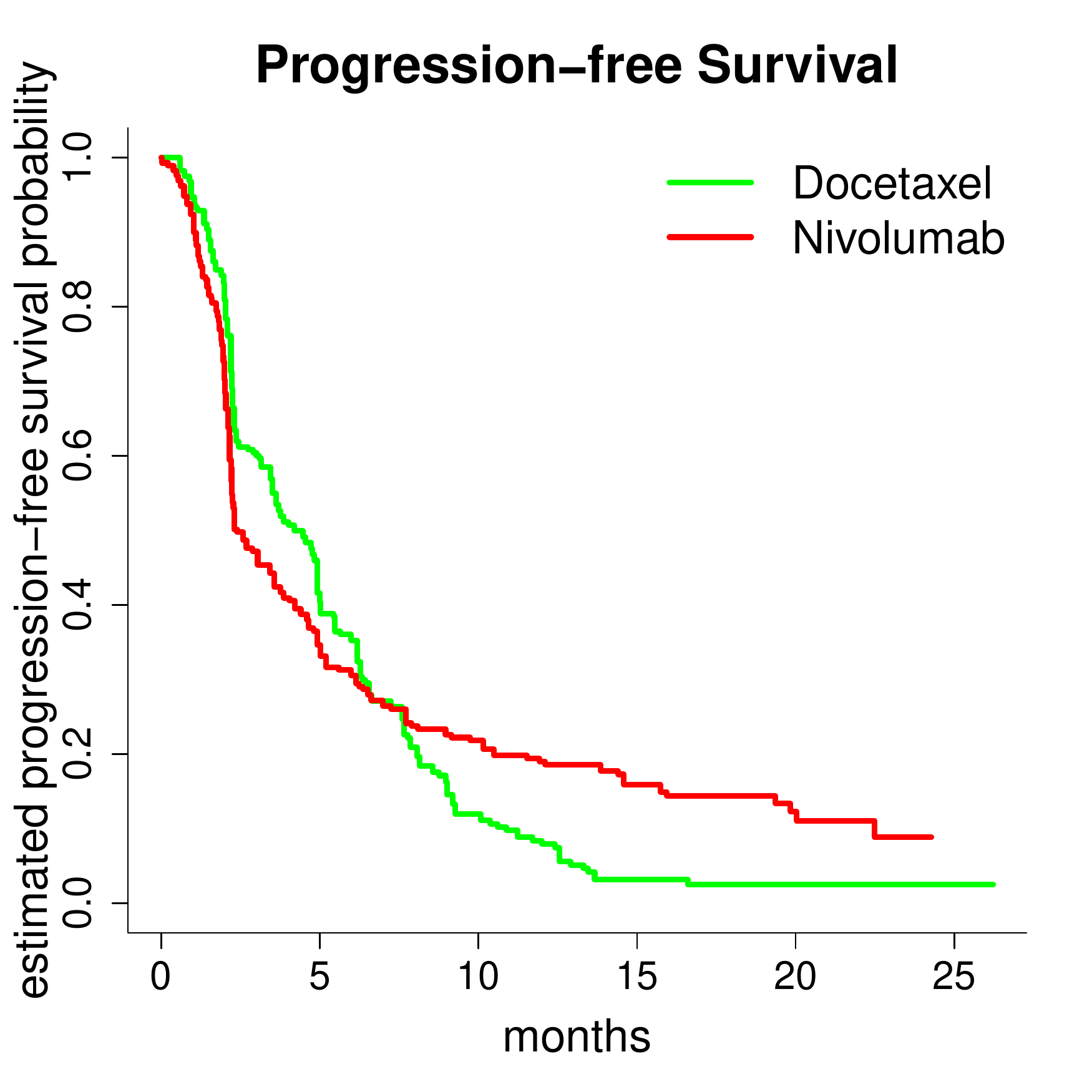}
    \caption{Data is reconstructed from Borghaei et al \cite{Borghaei.2015}.}
    \label{fig:BorgS}
\end{subfigure}
\hfill
\begin{subfigure}{0.495\textwidth}
    \includegraphics[width=\textwidth]{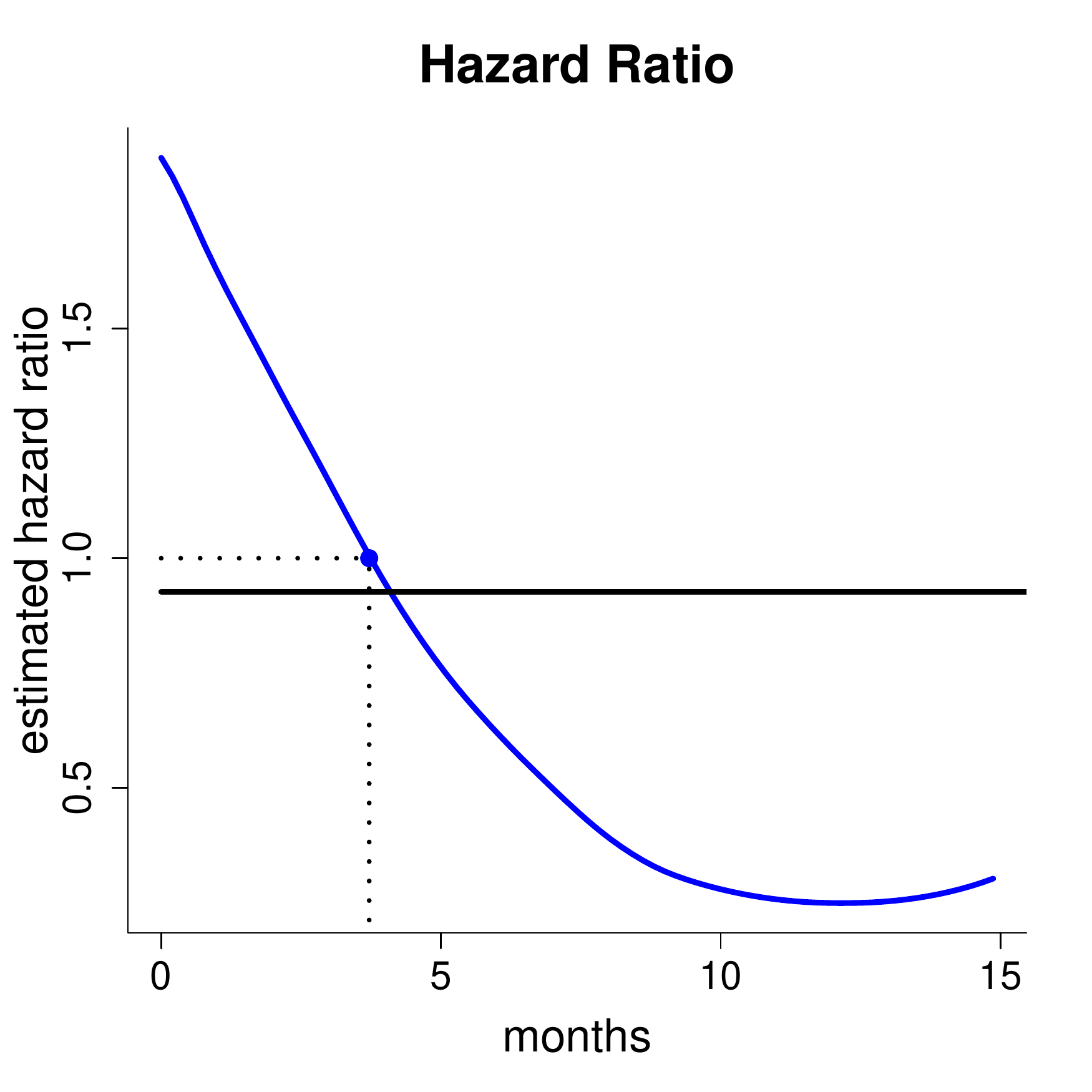}
    \caption{Nivolumab vs docetaxel}
    \label{fig:BorgHR}
\end{subfigure}

\caption{Left-hand side figure shows Kaplan-Meier estimates of the survival function, and right-hand side figure shows estimated hazard ratio. The solid blue line is an estimate of the time-varying hazard ratio obtained through smoothing the increments of the Nelson-Aalen estimate of the cumulative hazard function, the solid black line is an estimate from the Cox PH model, the dotted line with a blue solid point indicates a hazard ratio of $1$, i.e. the time point, where the estimated hazard rates are equal.}
\end{figure}

\noindent \NEW{During follow-up, $238$ of $292$ patients in the nivolumab arm and $248$ of $290$ patients in the docetaxel arm either died or had lung cancer progression (reconstructed data). 
Figure \ref{fig:BorgS} shows the re-estimated KM estimates of the PFS curve for the nivolumab and docetaxel group. 
The estimated KM curves are crossing, indicating a non-constant, crossing-hazards treatment effect.
This is further investigated in Figure \ref{fig:BorgHR}. 
The blue line shows the estimated (time-dependent) hazard ratio which is obtained by smoothing the increments of the cumulative hazard rates which are computed via the Nelson-Aalen estimator.
Smoothening was done via kernel-based methods and global bandwidth as implemented in the R package \texttt{muhaz}.
The estimated curve of the hazard ratio indicates an inferior treatment effect of nivolumab as compared to docetaxel early on. After approximately $4$ months, however, the hazard ratio falls below one, favouring nivolumab.
The crossing hazards result in crossing PFS curves, approximately two months after the hazard ratio crosses the threshold one, suggesting better performance of nivolumab.
Borghaei et al \cite{Borghaei.2015} suspect that a delayed effect of the nivolumab treatment causes this effect which is typical for immunotherapy.
The time-invariant estimate of the hazard ratio under the PH assumption is indicated by the solid black line in Figure \ref{fig:BorgHR}. 
While the estimate of the constant hazard ratio is close to 1, indicating no treatment effect, the estimates of the time-varying hazard ratio and the KM curves suggest otherwise and provide additional insights.
A log-rank test (on the reconstructed data) yields a test statistic of $0.6$, resulting in a p-value of $0.4$.}

\noindent \NEW{This type of statistical analysis is typical for clinical studies.
Jachno et al \cite{Jachno.2019} review 66 trials with time-to-event outcomes. 
For analysis, the majority of papers reported KM curves ($98$\%) and the Cox PH model ($97$\%) and inference was based on the log-rank test in $88$\% of the papers.
Only $11$\% of the reviewed papers in \cite{Jachno.2019} reported either testing for or visual inspection of NPH.
Moreover, at the stage of trial planning, only $11$\% considered non-constant hazard rates or NPH,
i.e. trial analysis is often restricted to PH methods.
}



\noindent \NEW{
This could be problematic as the Cox PH model is misspecified under NPH.
Consequently, its parameters are inappropriate to be interpreted as parameters of the time-to-event distribution,
as it may not capture the nature of the treatment effect, as illustrated in the example above.
Alternatively, the HR could be interpreted as a summary measure that quantifies the treatment effect in a single number, while the time-to-event distributions can be investigated using the KM estimates.
However, the HR estimate under NPH depends on the censoring distribution and therefore lacks a clear interpretation \cite{Xu.2000}.
In addition, the log-rank test loses power under NPH, which could lead to medical advances not being detected as such.}

\noindent \NEW{Ignoring the methodology for NPH and not testing for NPH makes it challenging to understand the impact of the PH assumption on the analysis of a specific trial. 
Dormuth et al \cite{Dormuth.2022} re-examined $18$ clinical trials characterized by crossing survival curves and inconclusive log-rank tests. They discovered significant differences in survival outcomes in $9$ of these trials when using testing procedures appropriate for NPH.}




\section{Systematic literature search and study selection} \label{systematic}
\begin{figure}
    \includegraphics[scale=0.75]{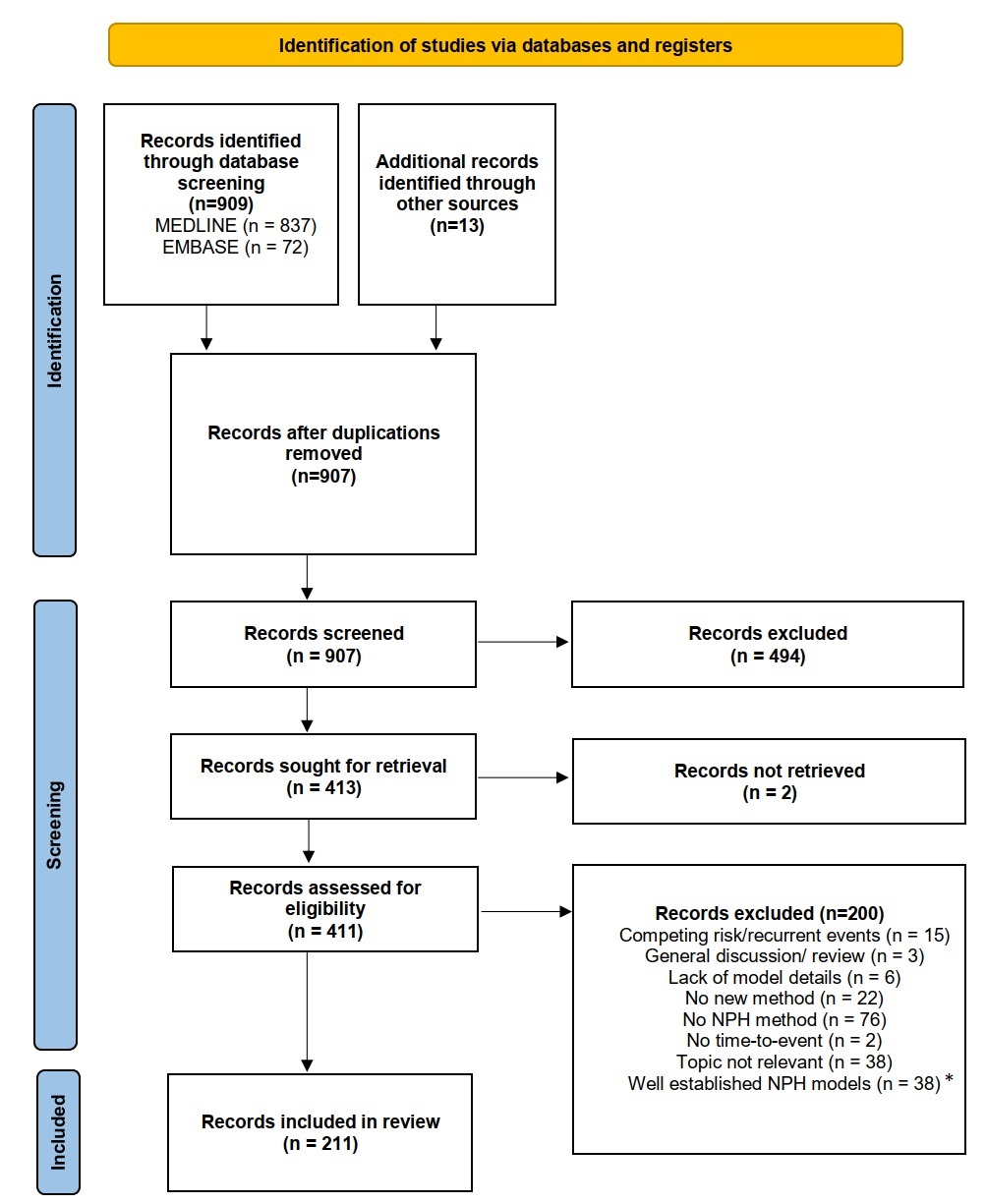}
         \caption{PRISMA 2020 flow diagram \cite{prisma} – identified and included studies from the database searches (MEDLINE and EMBASE). \\
         \NEW{*e.g. Stratified Cox PH model or use of time-dependent covariates in PH models as described in Klein and Moeschberger \cite[Chapter 9]{Klein.2003}.}}
    \label{fig:prisma}
\end{figure}

We performed a comprehensive literature search using two electronic databases, MEDLINE and EMBASE, on March 15th, 2022. Details on the literature search and the data extraction are provided in Appendix \ref{detailedMethodSection}.
In total 907 articles were identified, which were screened for eligibility. After the abstract screening and retrieval of full texts, a total of 411 articles were assessed for eligibility. In total, 200 articles (49\%) were excluded. The most frequent reason for exclusion was that the articles neither developed nor applied any NPH method. The final analysis included 211 publications, see PRISMA flow chart in Figure \ref{fig:prisma}.  

\begin{figure}
    \centering
    \includegraphics*[width=2.53in, height=2.53in]{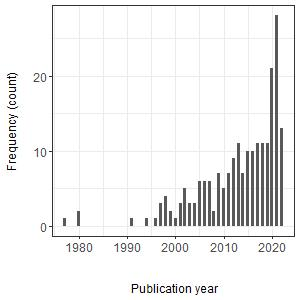}
    \caption{Publication year of the 211 included articles. Note that the number for 2022 is based on the articles published until 15th March 2022 and is therefore incomplete.}
    \label{fig:pubyear}
\end{figure}

\noindent The complete list of included articles is available in Table \ref{tab:S3} of the Online Supplement.\\
Figure \ref{fig:pubyear} shows the publication years of the articles included. In our review more than 70\% of the articles were published in 2010 or later and only a few before 2000. However, it has to be considered that the total number of published articles grew over the last years \cite{publications}. \\
The vast majority of articles ($\mathrm{>}$80\%) introduce statistical methods for NPH; reviews and applications were less frequent ($10\%$).


\begin{table*}[h]

    \caption{Absolute and relative frequencies of publications discussing a method class. Note that publications may discuss methods belonging to multiple classes of method categories. Therefore the classes are not mutually exclusive.}
    \label{tab:methodclass}
    \centering
    \begin{tabular}{|p{3.4in}|p{1in}|} \hline 
\multicolumn{2}{|c|}{\textbf{Estimation approaches (n=139)}}\\
\hline
 Kaplan-Meier based estimation approaches & 18 (12.9\%) \\ \hline
 Stratified Cox model & 3 (2.2\%) \\ \hline 
 Time-varying coefficients for the hazard rates & 47 (33.8\%) \\ \hline
 Transformation models with time-covariate interaction &  9 (6.5\%) \\ \hline
 Short- and long-term HR & 9 (6.5\%) \\ \hline 
 Joint Models & 3  (2.2\%) \\ \hline
 Frailty models & \NEW{17}  (\NEW{12.2}\%) \\ \hline 
Parametric models & 38 (27.3\%) \\ \hline
 Machine learning approaches & 11 (7.9\%) \\ \hline 
 Other & 8 (5.8\%) \\ \hline \hline
\multicolumn{2}{|c|}{\textbf{Hypothesis testing approaches (n=98)}}\\
\hline
Log-rank tests & 63 (64.3\%) \\ \hline 
Kaplan-Meier-based tests & 26 (26.5\%) \\ \hline 
Combination tests & 20 (20.4\%) \\ \hline 
Other tests & 12 (12.2\%) \\ \hline 
\end{tabular}

\end{table*}

\noindent \NEW{Concerning} the methods introduced in the identified articles we distinguished whether articles include methods for estimation of time-varying covariate/treatment effects and/or hypothesis testing. 
To further characterize the articles, we additionally introduce categories for articles focusing on estimation and/or testing methods. These categories are displayed in Table \ref{tab:methodclass}. Note that these categories are non-exclusive.
The categories summarize the core contribution of the methods discussed in the corresponding articles.
\NEW{The category "Kaplan-Meier based estimation approaches" includes articles discussing for example approaches based on Kaplan-Meier or Nelson-Aalen estimates, pseudo values and quantile regression. Articles grouped into "Time-varying coefficients for the hazard rates" discuss for example change-point approaches, splines and fractional polynomials. Further} explanation of the categories is given below in Section \ref{estimation} and Section \ref{test}.

\noindent The allocation of each paper to at least one of the categories according to Table \ref{tab:methodclass} can be found in the Online Supplement, see Table \ref{tab:S1}. 
A categorization of estimation approaches into more detailed categories discussed in Section \ref{estimation} can also be found in the supplement, see Table \ref{tab:S2}.
\noindent In Tables \ref{tab:S1} and \ref{tab:S2}, we attempted to identify the main NPH contribution for each paper, sometimes ignoring possible extensions that might have been mentioned. However, papers often cover multiple topics resulting in papers being classified into more than one category. Table \ref{tab:S4} gives details regarding the null hypothesis considered in the proposed testing approaches.

\noindent In total, 113 out of 211 (53\%) articles identified in the review include estimation methods, 72 out of 211 (24\%) involve hypothesis testing methods, and 26 out of 211 (13\%) involve both hypothesis and estimation methods. 

\noindent 
Log-rank test approaches are the most frequent hypothesis test methods that we identified in our literature review (Table \ref{tab:methodclass}). Methods for trials including an interim analysis are considered in $12\%$ of the articles.\\
The literature review identified articles covering different aspects of survival analysis in NPH settings. We identified articles proposing new test statistics for testing whether the survival is different in two treatment groups, as well as articles proposing new effect measures or regression models for quantifying the treatment effect in settings violating the common PH assumption.\\
In 72 out of 211 (34\%) articles freely accessible software is provided. Another 13\% of the articles provide the code for the methods upon request.
Software was considered to be freely available code in form of e.g. R packages, code snippets given in the text, or freely accessible code (e.g. supplement of article or online repository). Additionally, publicly available code or code snippets for commercial software are also categorized as freely available although the software needed to run the code is not freely available. Code snippets published in the articles sometimes implement only specific features or are used to deepen the understanding of the methods. Moreover, the code snippets are usually intended to enable users to apply the methods proposed. 

\noindent Simulation studies are reported in 158 (75\%) articles.
This is more pronounced in articles considering testing procedures, where 86 out of 98 (88\%) papers provide simulation studies.
For 91 out of 139  (65\%) papers that focus on estimation procedures simulation studies are provided.


\section{Notation and summary effect measures} \label{classification}

Before we proceed with describing the method categories according to Table \ref{tab:methodclass}, we introduce the notation that is used throughout this paper and the supplement.
We also define the identified treatment effect measures.


\subsection{Notation} \label{notation}
\noindent The number of subjects included in a trial is denoted by $N$. $Z$ is a treatment indicator with $Z=0$ indicating the control (placebo or comparator) and $Z=1$ the experimental treatment arm. The outcome of interest is the time to event $T$, whereas $C$ denotes the censoring time. The event indicator is denoted by $\delta =I(T\le C)$. The maximum follow-up time of the trial is set to $\tilde{t}$ and a specific time point during the follow-up time is denoted by $t^*$. The distinct ordered event-times are indicated by $t_{\left(i\right)}$, i.e. $0<t_{\left(1\right)}<t_{\left(2\right)}<\dots $ where at each time $t_{\left(i\right)}$ at least one event occurred. Note that we define $t_{\left(0\right)}=0$. Covariates or factors other than the treatment indicator are denoted by $x$. The regression coefficients for the treatment indicator and covariates are denoted $\gamma$ and $\beta$, respectively. Note that we use $\gamma (t)$ to denote a time-dependent treatment effect.

\noindent Additionally, we will use indexing of $\gamma$ if more than one parameter is required to specify the treatment effect. With ${\lambda }^{\left(Z\right)}(t)$ we denote the hazard rate of the treatment group $Z$ with covariates $x$ at time $t$, i.e. ${\mathop{\mathrm{lim}}_{\epsilon \mathrm{\to }\mathrm{0}} \frac{P\mathrm{(}t\mathrm{\le }T\mathrm{<}t\mathrm{+}\epsilon \mathrm{|}T\mathrm{\ge }t,Z,x\mathrm{)}}{\epsilon }\ }$, and with ${\lambda }_0(t)$ the baseline hazard rate respectively. The cumulative hazard rate ${\mathrm{\Lambda }}^{\left(Z\right)}(t)$ equals $\int^t_0{{\lambda }^{\left(Z\right)}(u)du}$. The survival function of treatment group $Z$ with covariates $x$, $P\mathrm{(}T\mathrm{>}t\mathrm{|}Z\mathrm{,x)}$ is indicated by $S^{\left(Z\right)}\left(t\right).$ Note that potential dependence of ${\lambda }^{\left(Z\right)},{\mathrm{\Lambda }}^{\left(Z\right)}$, ${\mathrm{S}}^{\left(Z\right)}$ on covariates $x$ is suppressed in the notation. The at-risk indicator $Y_i(t)$ denotes whether patient $i$ is uncensored and event-free at time $t$, $Y_i(t)=1$, or not, $Y_i(t)=0$.
The number of patients at risk at time $t$ is denoted by $Y\left(t\right)=\sum_i{Y_i\left(t\right)}$. In general, we use `$(Z)$' in the superscript to denote group-specific quantities. \\
Table \ref{tab:notation} of the Supplement gives an overview of the used notation and the quantities defined in Section \ref{effect}.

\subsection{Effect measures} \label{effect}
\noindent The treatment effect can be quantified, e.g.\@, by the difference or ratio of the survival function at a chosen landmark time $t^*\mathrm{\le }\tilde{t}$, i.e. $S^{\left(1\right)}\left(t^*\right)-S^{\left(0\right)}(t^*)$ or ${S^{\left(1\right)}(t^*)}/{S^{\left(0\right)}(t^*)}$ \cite{Uno.2014}.
Equivalent conclusions can be obtained by the cumulative $\text{HR}$ ($\text{cHR}$) at time $t^{\mathrm{*}}$, $\text{cHR}\left(t^{\mathrm{*}}\right)\mathrm{=\ }{\mathrm{\Lambda }}^{\left(\mathrm{1}\right)}\mathrm{(}t^{\mathrm{*}}\mathrm{)/}{\mathrm{\Lambda }}^{\mathrm{(0)}}\mathrm{(}t^{\mathrm{*}}\mathrm{)}$, with ${\mathrm{\Lambda }}^{\mathrm{(}Z\mathrm{)}}\left(t^{\mathrm{*}}\right)\mathrm{=}\int^{t^{\mathrm{*}}}_0{{\lambda }^{\mathrm{(}Z\mathrm{)}}\mathrm{(}s\mathrm{)}ds}$.
Alternatively, the ${\tau }^{\mathrm{th}}$ quantile of $T^{\left(0\right)}$ and $T^{\left(1\right)}$ may be compared, i.e. taking differences or ratios of $t^{\left(1\right)}_{\tau }={S^{\left(\mathrm{1}\right)}}^{-1}\left(1-\tau \right)$ and $t^{\left(0\right)}_{\tau }={S^{\left(0\right)}}^{-1}\left(1-\tau \right)$, where ${S^{\left(Z\right)}}^{-1}$ denotes the inverse of the survival function of treatment group $Z$ \cite{Uno.2014}.

\noindent The above treatment effect measures are cumulative in that sense that they compare survival functions or cumulative hazard rates.
Instantaneous differences between the treatment and the placebo group can be investigated by the instantaneous hazard ratio at $t^*$, $\text{HR}\left(t^*\right)={\lambda }^{(1)}(t^*)/{\lambda }^{\left(0\right)}(t^*)$.
However, the $\text{HR}(t^*)$ cannot necessarily be interpreted as the current (causal) effect of the treatment, as the population of survivors in the two treatment groups may differ for unmeasured characteristics or unadjusted covariates. 

\noindent 
Furthermore, conclusions based on a single time point $t^*$ (or quantile $\tau$) may not be meaningful\NEW{, because analysis at a single time point (quantile) is not informative for the time points (quantiles) before or after the chosen $t^*$ ($\tau$), where the treatment effect may be substantially different or even in the opposite direction.}
\noindent The dynamic of differences between survival time distributions over the course of time may be investigated by calculating the above effect measures over a suitable grid of time. 
Given that the hazard rate is often the pivot of modeling approaches, the $\text{HR}(t)$ is a common choice. 
However, it may be difficult to assess the overall effectiveness of the treatment from an examination of the effect measure over time.
This is particularly relevant for the $\text{HR}(t)$.
For illustration assume a crossing hazards scenario as depicted by the blue dashed line in Figure \ref{fig:cross}. The trajectory of the $\text{HR}(t)$ alone does not provide relevant information if and when the survival curves cross. For that, the trajectory of the baseline hazard rate in the same time period is also required. 
This illustrates that "extreme" values of the $\text{HR}(t)$ in a certain time period do not tell whether these extreme differences in treated and untreated individuals 
result in relevant differences between survival time distributions.
\NEW{Hence, dynamic effect measures such as the $ \text{HR}(t) $ may be less useful to clinicians who need to make a treatment decision at $t=0$ and therefore need to know which treatment is superior with respect to some feature of the time-to-event distributions.
For such binary decision-making, summary effect measures that summarize the treatment effect in a single number can be helpful.} 

\noindent 
A summary effect measure considers the entire survival curve (within the interval $[0,t^*], t^*\leq \tilde{t}$) and is thus \NEW{usually} of average nature.
An attempt to summarize the treatment effect in a single number is the average $\mathrm{\ }\text{HR}$ $\text{aHR}(t^*)\mathrm{=\ }{\int^{{\mathrm{t}}^{\mathrm{*}}}_0{\frac{{\lambda }^{\left(\mathrm{1}\right)}\mathrm{(}t\mathrm{)}}{\lambda \mathrm{(}t\mathrm{)}}}dG\mathrm{(}t\mathrm{)}}/{\int^{{\mathrm{t}}^{\mathrm{*}}}_0{\frac{{\lambda }^{\left(0\right)}\mathrm{(}t\mathrm{)}}{\lambda \mathrm{(}t\mathrm{)}}}dG\mathrm{(}t\mathrm{)}}$, with $\lambda \left(t\right)\mathrm{=}{\lambda }^{\left(\mathrm{1}\right)}\left(t\right)\mathrm{+\ }{\lambda }^{\left(0\right)}\mathrm{(}t\mathrm{)}$ and $G\left(t\right)$ is a chosen \NEW{weighting function, that often consists of the time-to-event distribution, survival function and inverse probability of censoring weights (see, for example, \cite{Chen.2015,Schemper.2009,Rauch.2018}).}
Note, that competing definitions of the $\text{aHR}$, different weighting functions and various estimation techniques exist. 
\NEW{A related summary effect measure utilized in survival analysis is the concordance probability $P(T^{(0)}<T^{(1)})$ or odds of concordance $\frac{P(T^{(0)}<T^{(1)})}{1-P(T^{(0)}<T^{(1)})}$ \cite{Schemper.2009,Dunkler.2010,Martinussen.2013}}.

\noindent The restricted mean survival time is the {\NEW{mean}} survival time within the time period [0,$\mathrm{\ }t^{\mathrm{*}}$], $t^{\mathrm{*}}\mathrm{\le }\mathrm{\ }\tilde{t}$, i.e. $\text{RMST}\mathrm{(}t^{\mathrm{*}}\mathrm{)=\ }\int^{t^{\mathrm{*}}}_0{S\mathrm{(}t\mathrm{)}dt}$. 
An effect measure can again be constructed \NEW{from} the difference $\text{RMST}^{\left(\mathrm{1}\right)}\mathrm{(}t^{\mathrm{*}}\mathrm{)-\ }\text{RMST}^{\left(0\right)}\mathrm{(}t^{\mathrm{*}}\mathrm{)}$ or the ratio, ${\text{RMST}^{\left(\mathrm{1}\right)}\mathrm{(}t^{\mathrm{*}}\mathrm{)}}/{\text{RMST}^{\left(0\right)}\mathrm{(}t^{\mathrm{*}}\mathrm{)}}$ between the $\text{RMST}$ of the treatment and the control group \cite{Uno.2014}. 

\noindent In a NPH setting summary effect measures do not cover the dynamics of treatment efficacy and hence, do not necessarily deliver an adequate picture of the nature of the treatment effect over time. See \cite{Dehbi.2017} for a discussion and potential remedy that relies on calculating more than one summary effect measure over varying time ranges. 

\noindent For a comprehensive description of the survival distribution, the group-specific quantities ${\lambda }^{\left(Z\right)},{\mathrm{\Lambda }}^{\left(Z\right)}$ or ${\mathrm{S}}^{\left(Z\right)}$ are required. These can be estimated \NEW{from} statistical models or stratification of non-parametric estimation approaches by $Z$. 
Stratification reduces the need for assumptions, such as the proportional hazards (PH) assumption, across different subgroups, as the estimation within each stratum relies solely on the information from that specific subgroup.

\noindent Effect measures unconditional of covariates $x$ might require additional steps in estimation. As the covariates typically enter the survival function in a non-linear fashion, it will, in general, not be sufficient to plug $E[X|Z]$ into $S^{\left(Z\right)}\left(t\right)$ to obtain $E[S^{\left(Z\right)}\left(t\right)|Z]$, where the expectation is taken with respect to the covariates, due to Jensens's inequality. The interested reader is referred to \cite{Keiding.2005} and Chapter 10 of \cite{Therneau.2013}. This also holds for the hazard rate ${\lambda }^{\left(Z\right)}(t)$. Even if covariates would enter the hazard rate linearly, the expected covariate values given survival up to $t$ of each treatment group would be necessary to obtain the \NEW{hazard rate unconditional of covariates other than the treatment indicator.
Further, note that the above-mentioned approaches would result in quantities that depend on the covariate distribution of the treatment group. To isolate the treatment effect, G-computation approaches could be used.}\\ 

\noindent

\section{Estimation approaches for NPH treatment effects}\label{estimation}

\noindent
\noindent This section describes the categories of identified estimation approaches for NPH treatment/ covariate effects.
The first column of Table \ref{tab:est_overview} shows the main categories as introduced in Table \ref{tab:methodclass}. 
Some categories are divided into sub-categories given in the second column of Table \ref{tab:est_overview}.
The third column of Table \ref{tab:est_overview} provides a brief description of the methods.
References to the Supplement are given in the first two columns of Table \ref{tab:est_overview},
where a more detailed overview can be found.
In the corresponding section of the Supplement, references to Table \ref{tab:S2} are given.
In Table \ref{tab:S2} of the Online Supplement, each paper that was considered in this literature review is allocated into one or more sub-categories according to Table \ref{tab:est_overview}, indicating the paper's main contribution to address NPHs.
Table \ref{tab:S2} (column K) also provides information on whether the corresponding paper took a Bayesian estimation approach or not.
The degree of detail and information given is hierarchical: Table \ref{tab:est_overview} gives an overview, the referenced sections in the Online Supplement provide more detailed explanations including model formulas and a discussion of the literature, whereas Table \ref{tab:S2} in combination with the respective papers (and the references therein) provide full information on the specific approaches to cope with NPH that we detected in the literature.

\noindent The $4^{\text{th}}$ and $5^{\text{th}}$ column of Table \ref{tab:est_overview} give a simplified impression of how flexible the corresponding approach is with respect to patterns of the hazard rate $\lambda^{(0)}(t)$ and time-varying NPH-patterns, respectively.
Note that these statements focus on the treatment groups or, more generally, on the covariates where NPH are modeled. 
The assumptions on remaining covariate effects might be strict without mentioning this in Table \ref{tab:est_overview}.
Hence, for many approaches, the flexibility of $\lambda^{(0)}(t)$ is determined by the flexibility of the baseline hazard rate $\lambda_0(t)$.
The statements on the flexibility of $\text{HR}(t)$ can be understood as the flexibility of $\lambda^{(1)}(t)$ in contrast to $\lambda^{(0)}(t)$ and this determines the applicability of the method to arbitrary complex NPH scenarios.
If there are no assumptions on the trajectory of $\lambda^{(0)}(t)$ and the $\text{HR}(t)$ this is specified as "none" in 
Table \ref{tab:est_overview}. 
We define the assumptions to be "strict" if the corresponding methods tend to be restricted to monotonically increasing or decreasing trajectories for the $\text{HR}(t)$ and to a single change in slope for $\lambda^{(0)}(t)$.
The statements "few" and "medium" indicate something in between.
This categorization is closely related to a non- ("none"), semi- ("few"), and -parametric ("medium" to  "strict") handling of $\lambda^{(0)}(t)$ and $\text{HR}(t)$, where non-parametric approaches do not impose assumptions on the corresponding trajectories but parametric approaches tend to be limited in flexibility.

\noindent Note that the $4^{\text{th}}$ and $5^{\text{th}}$ column of Table \ref{tab:est_overview} is a statement about the flexibility of each method in accommodating varying scenarios of hazard rate functions (column 4) and time-varying HRs (column 5). Especially column 5 describes the capability of the methods to cope with varying NPH scenarios.
However, this is not a statement as to whether estimates of the hazard rates and $\text{HR}(t)$ are easy to obtain within the framework of the corresponding approach.
The Kaplan-Meier approach is an example where it is not possible to compute hazard rates and hence the time-varying $\text{HR}(t)$ across two strata without further smoothing approaches.\\
\NEW{The last column of Table \ref{tab:est_overview} gives examples of software, including R-packages and SAS procedures, that are available for the groups of methodological approaches described.
}

\begin{center}

\begin{landscape}
\begin{table*}[!htbpt!]
\caption{Overview of methods suitable to estimate NPH effects. More details on the categories can be found in the Online Supplement as indicated by the references given in brackets.
\NEW{$^*$ The assignment of papers to categories is not mutually exclusive. The symbol \# refers to the number of papers in the corresponding category, ref. is a reference to the corresponding sections in the Supplement for the method description.}}
\label{tab:est_overview}
    \centering
    \small
    \addtolength{\leftskip} {-1cm}
    \addtolength{\rightskip}{-1cm}
      \begin{tabulary}{\linewidth}
    {|p{0.9in}|p{0.95in}|p{4.3in}|p{0.6in}|p{0.6in}|p{0.8in}|}
    
 \hline
 Category & Sub-category & \multirow{2}{*}{  Description} & \multicolumn{2}{c|}{Assumptions on}  & \multirow{2}{*}{Software} \\
 \NEW{(\#, ref.)$^*$} & \NEW{(\#, ref.)$^*$} & & $\lambda^{(0)}(t)$ & $\text{HR}(t)$ & \\
 \hline \hline
\multirow{3}{*}{\makecell[l]{{Kaplan-Meier}\\ and\\ {Nelson-Aalen}\\ based\\ approaches\\ (\NEW{18}, \ref{KM_est})}} & {Stratified Kaplan-Meier, Nelson-Aalen estimates  (\NEW{11}, \ref{KM_est})}  & 
The Kaplan-Meier (KM) estimate of the survival function and the Nelson-Aalen (NA) estimate of the cumulative hazard function are non-parametric estimators. Both do not impose any modeling assumption on the hazard rate. 
Hence, a stratified approach across treatment groups is suitable for any trajectory of $\text{HR}(t)$. 
The estimates of $S^{(Z)}(t)$ and $\Lambda^{(Z)}{(t)}$ might be processed to estimates of (summary) effect measures such as $\text{aHR}(t^*)$, $\text{cHR}(t^*)$ or differences/ratios of the $\text{RMST}^{(Z)}(t^*)$. 
&  none & none & \texttt{survival}, \texttt{km.ci}, \texttt{survRM2}, \texttt{AHR} (R), LIFETEST (SAS)\\
\cline{2-6}
& \makecell[l]{Pseudo values\\ (\NEW{5}, \ref{pseudo})} & 
Pseudo values are usually based on KM estimates of the survival function for the pooled data or summary measures that are computed thereof such as the $\text{RMST}(t^*)$. In an iterated leave-one-out fashion (jackknife) the difference for each individual between $N$ times the whole sample estimate and $N-1$ times the estimate without the individuals’ contribution are calculated for a chosen landmark time $t^*$. 
This difference represents a newly created metric variable for every observation and can be further investigated by linear or generalized linear regression models which include the treatment indicator as well as other covariates.
& none & none & \texttt{pseudo}, \texttt{prodlim} (R), RMSTREG (SAS) \\
\cline{2-6} 
& \makecell[l]{Quantile\\ regression\\ (\NEW{2}, \ref{quantile})} & 
In quantile regression, either the $\tau^{\text{th}}$ quantile or its logarithm are assumed to have a linear relationship with the covariates and the treatment indicator, e.g. $\ln\{t_{\tau}\} = \boldsymbol{x}^T\boldsymbol{\beta}(\tau) + Z\gamma(\tau)$. 
Estimation procedures are either based on a generalized KM estimator or martingale-based estimation equations. 
Treatment efficacy at the $\tau^{\text{th}}$ survival quantile can be evaluated by $\gamma(\tau)$.
This process can be iterated over a grid of quantiles $\tau \in (0,1)$, to get a complete picture of the potentially time-varying treatment effect over the course of time through $\gamma(\tau)$.  & none & none & \texttt{quantreg} (R), QUANT-LIFE (SAS) \\
\hline 
\hline
\end{tabulary}
\end{table*}

\newpage

\begin{table*}[!htbp]
    \centering
    \small
    \addtolength{\leftskip} {-1cm}
    \addtolength{\rightskip}{-1cm}
 \begin{tabulary}{\linewidth}
    {|p{0.9in}|p{0.95in}|p{4.3in}|p{0.6in}|p{0.6in}|p{0.8in}|}
 \hline
 Category & Sub-category & \multirow{2}{*}{  Description} & \multicolumn{2}{c|}{Assumptions on}  & \multirow{2}{*}{Software} \\
 \NEW{(\#, ref.)$^*$} & \NEW{(\#, ref.)$^*$} & & $\lambda^{(0)}(t)$ & $\text{HR}(t)$ & \\
 \hline \hline 
\multicolumn{2}{|c|}{\makecell[l]{Stratified Cox model \\ (\NEW{3}, \ref{StratCox})}} & 
The stratified Cox model relaxes the PH assumption by stratifying the baseline hazard rate along the treatment indicator, i.e. $\lambda^{(Z)}(t) = \exp\{\boldsymbol{x}^T\boldsymbol{\beta}\}\lambda_0^{(Z)}(t)$. 
In case of the non-parametric Breslow estimate of the baseline hazard rates, no structure is placed on the baseline hazards.
The model is suitable for any trajectory of $\text{HR}(t)$. 
However, the PH assumption still holds for the remaining covariates.
An estimate of the $\text{cHR}(t)$ may be utilized to evaluate the treatment effect. 
& none & none   & \texttt{survival} (R), PHREG (SAS) \\
\hline
\multicolumn{2}{|c|} {\makecell[l]{Short- and long-term HR\\ (\NEW{9}, \ref{shortHR})}}  & The models in this category differ from the models of the time-varying coefficient category in that sense that they do not formulate a non-constant $\text{HR}(t)$ via a non-constant coefficient for the treatment effect $\gamma(t)$.
Instead, a specific function is imposed on the hazard rate such that NPH arise.
The approaches in this category differ from fully parametric approaches in that they have semi- or non-parametric components and do not fully specify the distribution of the survival time by assumption.
This category mainly encompasses the Yang and Prentice model. 
The hazard function in the Yang and Prentice model is set to ${\lambda }^{\left(Z\right)}\left(t\right)=\ \frac{{\mathrm{exp} \left(x^T\beta \right)\ }}{{\mathrm{exp} \left(-{\gamma }_1Z\right)\ }S^{\left(0\right)}\left(t\right)+{\mathrm{exp} \left(-{\gamma }_2Z\right){(1-S}^{\left(0\right)}\left(t\right)\ })}{\lambda }_0(t)$. A time-dependent weight function puts all the weight to the first parameter $\gamma_1$ at the beginning. This weight is reduced over time and the weight for the second parameter $\gamma_2$ is analogously increased. Consequently, the hazard ratio is a time-weighted average of the short-term and the long-term hazard-ratio, $\exp\{\gamma_1\}$ and $\exp\{\gamma_2\}$, respectively. 
&  none to few & strict & \texttt{YPPE}, \texttt{YPBP} (R) \\
\hline 
\hline
\end{tabulary}
\end{table*}
\newpage

\begin{table*}[!htbp]
    \centering
    \small
    \addtolength{\leftskip} {-1cm}
    \addtolength{\rightskip}{-1cm}
\begin{tabulary}{\linewidth}
    {|p{0.9in}|p{0.95in}|p{4.3in}|p{0.6in}|p{0.6in}|p{0.8in}|}
 \hline
 Category & Sub-category & \multirow{2}{*}{  Description} & \multicolumn{2}{c|}{Assumptions on}  & \multirow{2}{*}{Software} \\
 \NEW{(\#, ref.)$^*$} & \NEW{(\#, ref.)$^*$} & & $\lambda^{(0)}(t)$ & $\text{HR}(t)$ & \\
 \hline \hline 
\multirow{4}{*}{\makecell[l]{Time-varying\\ coefficients\\ for the\\ hazard rates\\ (\NEW{47}, \ref{time varying})}} & {Change point models\newline (\NEW{11}, \ref{change point})} & 
The hazard rate equals $\lambda^{(Z)}(t) = \exp\{x^T \beta + Z\gamma(t)\}\lambda_0(t)$. 
The time-dependent treatment coefficient is piecewise constant, $\gamma(t)= \gamma_1 + \gamma_2I\{t\geq t^*\}$,
where $I$ is the indicator function which is equal to one if the statement in the brackets is correct and 0 else.
Note that the treatment coefficient is based on a time-covariate interaction and is not restricted to factors.
The change point $t^*$ is pre-specified and more than one change point can be included, but black-box approaches also exist.
Along with $\gamma(t)$, the $\text{HR}(t)$ is also piecewise constant and hence rather restricted.
The model might be estimated via the partial likelihood, which places no modeling assumption on the baseline hazard. Parametric choices for the baseline hazard are also possible.
& depends & medium to strict & \texttt{survival} (R), PHREG (SAS) \\
\cline{2-6}
 & {Smooth time-varying coefficients (\NEW{26}, \ref{timeXcovariate}), e.g. fractional polynomials (\NEW{6}, \ref{fractional}) or splines\newline (\NEW{12}, \ref{Splines})} & 
 The hazard rate equals $\lambda^{(Z)}(t) = \exp\{x^T \beta + Z\gamma(t)\}\lambda_0(t)$. 
The time-dependent treatment coefficient differs along with the chosen complexity and so do the trajectories of $\text{HR}(t)$ that can be represented by the model.
A general depiction is $\gamma(t)=\sum_{d}B_d(t)\gamma_d$, where the basis functions $B_d(t)$ are smooth in $t$.
Examples of basis function are, for example, fractional polynomials or B-Splines.
The baseline hazard rate may be left unspecified by utilizing the partial likelihood or piecewise constant through the use of the Poisson GLM routine.
 & none to few & few to strict &  \texttt{dynsurv}, \texttt{polspline}, \NEW{\texttt{PenCoxFrail}} (R), ICPHREG (SAS)\\
\cline{2-6}
 & Weighted partial likelihood (\NEW{12}, \ref{aHR}) & 
The weighted partial likelihood can be utilized to estimate a representative value for a time-varying treatment effect. This is usually the average hazard ratio $\text{aHR}(t^*)$.
The weights have to be chosen and are not necessarily suitable for any trajectory of the $\text{HR}(t)$.
 Note that other approaches to estimate the $\text{aHR}(t^*)$ exist, e.g. via KM.
 & none & \NEW{none} & \texttt{coxphw} (R), macro WCM (SAS) \\
\cline{2-6}
 & Aalen's additive hazard model\newline (\NEW{4}, \ref{additive}) & The hazard rate is equal to $\lambda^{(Z)}(t) = \lambda_0(t) + x^T \beta(t) + Z\gamma(t)$. 
 Estimation is based on the increments of the martingale process and is estimated via least squares at the distinct event times.
 In Aalen's additive hazard model, there is no smoothness assumption on $\lambda_0(t)$ and $\gamma(t)$ e.g. unlike when $\gamma(t)$ is modeled via fractional polynomials or B-Splines.
 & none & none & \texttt{timereg} (R) \\
\hline 
\hline
\end{tabulary}
\end{table*}
\newpage
\begin{table*}[!htbp]
    \centering
    \small
    \addtolength{\leftskip} {-1cm}
    \addtolength{\rightskip}{-1cm}
  \begin{tabulary}{\linewidth}
    {|p{0.9in}|p{0.95in}|p{4.3in}|p{0.6in}|p{0.6in}|p{0.8in}|}
 \hline
 Category & Sub-category & \multirow{2}{*}{  Description} & \multicolumn{2}{c|}{Assumptions on}  & \multirow{2}{*}{Software} \\
 \NEW{(\#, ref.)$^*$} & \NEW{(\#, ref.)$^*$} & & $\lambda^{(0)}(t)$ & $\text{HR}(t)$ & \\
 \hline \hline 
 \multicolumn{2}{|c|}{\makecell[l]{Frailty models \\ (\NEW{17}, \ref{frailty})}}  & Frailty or random effects models assume a heterogeneous population. Even if PH are assumed on the individual level, i.e. given the unobservable characteristics that make the population heterogeneous, the selection effects across the treatment and the placebo cause NPH on the population level in general, i.e. with unobservable characteristics marginalized out.
In case of a diminishing treatment effect even crossing hazards are possible which might be caused by a catch-up process of the treatment group but not by a toxic treatment effect. \NEW{This category also includes cure rate models that can be motivated by a discrete mixture distribution of a susceptible and non-susceptible population}.
 & model dependent & strict &  \texttt{survival}, \texttt{coxme}, \texttt{frailtyEM}, \texttt{frailty-pack}, \NEW{\texttt{PenCoxFrail}} (R), PHREG, NLMIXED (SAS) \\ \hline
\multirow{4}{*}{\parbox{.7in}{Fully parametric approaches (\NEW{38}, \ref{parametric})}} & Piecewise exponential model\newline (\NEW{9}, \ref{piecewise}) & The hazard rate is piecewise constant for that kind of model, i.e. there is a distinct parameter for each of non-overlapping time intervals. If the hazard rate is stratified by treatment the $\text{HR}(t)$ is piecewise constant. Also termed piecewise constant hazard model. & few to strict & few to strict & \texttt{pch}, \texttt{eha} (R), PHREG (SAS)\\
\cline{2-6}
&  AFT \& GAMLSS models \newline (\NEW{18}, \ref{AFT}) & AFT’s assume a specific distribution for $T$ which typically leads to NPH, the Weibull distribution being a prominent exception. AFT’s formulate a treatment and covariate-specific location parameter, GAMLSS extend this to shape and scale parameters. & medium to strict & medium to strict & \texttt{flexsurv}, \texttt{brms}, \texttt{spBayesSurv}, \texttt{mpr}, \texttt{gamlss.cen} (R), LIFEREG (SAS) \\
\cline{2-6}
& First hitting time models (\NEW{4}, \ref{FHT}) & First hitting time models formulate a health process.
The event occurs once the health process reaches a certain threshold (typically $0$).
Distributional assumptions on the health process determine the distribution of the survival time which typically has NPH.
If the health process is assumed to be a Wiener process the distribution of the survival time is inverse Gaussian. 
& medium to strict & medium to strict & \texttt{thregI} (R) \\
\cline{2-6}
& Other fully parametric approaches (\NEW{10}, \ref{GLM}) & Other parametric approaches that do not fit in the former sub-categories are gathered in this category. This for example encompasses GLMs. & medium to strict & medium to strict & \NEW{E.g. standard software for GLMs.}  \\
\hline 
\hline
\end{tabulary}
\end{table*}
\newpage
\begin{table*}[!htbp]
    \centering
    \small
    \addtolength{\leftskip} {-1cm}
    \addtolength{\rightskip}{-1cm}
 \begin{tabulary}{\linewidth}
    {|p{0.9in}|p{0.95in}|p{4.3in}|p{0.6in}|p{0.6in}|p{0.8in}|}
 \hline
 Category & Sub-category & \multirow{2}{*}{  Description} & \multicolumn{2}{c|}{Assumptions on}  & \multirow{2}{*}{Software} \\
 \NEW{(\#, ref.)$^*$}& \NEW{(\#, ref.)$^*$} & & $\lambda^{(0)}(t)$ & $\text{HR}(t)$ & \\
 \hline \hline
\multicolumn{2}{|l|}{\parbox{1.3in}{Transformation models with time-covariate interaction (\NEW{9}, \ref{trans})}}  &
This category considers the Royston-Parmar and the conditional transformation model. Both approaches do not directly impose a model on the hazard rate, but formulate transformation functions as a spline function in (log) time. Time-varying treatment/covariates effects might be incorporated by spline by covariate interaction. The transformation functions are then brought into a parametric framework that determines their interpretation, e.g. as log cumulative hazard rates.
&  few &  few & \texttt{flexsurv} (R), macro sas\textunderscore stpm2 (SAS) \\
\cline{1-6}
\multicolumn{2}{|c|}{\makecell[l]{Joint models\\ (\NEW{3}, \ref{joint})}}  &
Joint models are typically fully parametric models, where measurements of, for example, drug concentrations over time per individual are modeled simultaneously with the time-to-event endpoint.  The predictions of the other variable (concentration) from the model, or summary measures (exposure, area under the concentration-time curve) are used as a covariate in the time-to-event model. 
& depends  & depends & \texttt{JM} (R and SAS), \NEW{\texttt{INLAjoint}} \\
\cline{1-6}
\multicolumn{2}{|c|}{
\makecell[l]{Machine learning\\ approaches\\ (\NEW{11}, \ref{ML}) }}&
We encountered a couple of machine learning approaches in the context of NPH scenarios such as trees (for example for finding the number and time points of a change point model) and forests, model averaging, k-nearest neighbor (in order to determine weights for weighted KM estimates), kernel smoothing based approaches and neural networks.
& none to few & none to few & \texttt{trtf}, \texttt{BART}, \texttt{mboost}, \NEW{\texttt{ipred}}, \NEW{\texttt{randomForestSRC}}, \NEW{\texttt{Ranger}} (R) \\
\cline{1-6}
\multicolumn{2}{|c|}{\makecell[l]{Other approaches\\ (\NEW{8}, \ref{other_estimation})}}  & This is a collective category of methods that did not fit properly into one of the other categories. Among these methods are the rank preserving structural failure time model, concordance regression, the accelerated hazard model, the semi-parametric proportional likelihood ratio model, a Bayesian dependent Dirichlet process to model the time-to-event distribution, and others. & depends & depends & \texttt{rpsftm} (R) \\
\hline 
\hline
\end{tabulary}
\end{table*}

\end{landscape}

\end{center}

\noindent Approaches with no or few assumptions on $\lambda^{(0)}(t)$ and $\text{HR}(t)$ are suited for NPH scenarios that are more complex than those depicted in Figure \ref{fig:delay} (delayed treatment effect), \ref{fig:dim} (diminishing treatment effect), and \ref{fig:cross} (crossing hazards).
Such methods might also be utilized in the absence of knowledge of the trajectory of the HR and hazard rates or as a validity check of assumptions on its trajectory.

\noindent Stratification along the treatment indicator is a general tool to relax assumptions on the underlying estimation procedure. 
A stratified KM estimation approach along the treatment and placebo (comparator) group is suitable for any two possible trajectories of $S^{(0)}(t)$ and $S^{(1)}(t)$ or $\lambda^{(0)}(t)$ and $\text{HR}(t)$, respectively.
If further (continuous) covariates are present the stratified Cox model might be utilized, where stratification along the treatment indicator is suitable for any trajectory of the $\text{HR}(t)$. 
Time-varying coefficients offer arbitrary flexibility, depending on the complexity one allows for $\gamma(t)$ in $\lambda^{(Z)}(t)= \exp\{x^T\beta +Z\gamma(t)\}$. 
High flexibility of the treatment effect can in particular be achieved if $\gamma(t)$ is modeled via (penalized) splines or Aalen's additive model.
For the Royston-Parmar and the conditional transformation model, the basis functions in time can interact with the treatment indicator (or other covariates) in case of unknown or highly variable NPH scenarios.
Machine learning procedures such as trees and forests as well as k-nearest neighbours or kernel approaches also offer high flexibility on $\lambda^{(0)}(t)$ and $\text{HR}(t)$.

\noindent Procedures with limited flexibility on $\text{HR}(t)$ could be used if the trajectory is known/assumed, or if more flexible procedures suggest the appropriateness of more restrictive models.
Less flexible models will typically be easier to analyze, as relatively few model parameters suffice to explain the trajectory of the treatment effect.
Additionally, processing of model quantities, for example, hazard rates, to effect measures, such as the difference in $\text{RMST}$s or the $\text{aHR}$, might be more frequently analytically tractable than for the more flexible methods.
Hence, results obtained from less flexible procedures might be easier to communicate.
Further on, procedures with limited flexibility might avoid over-fitting the data which might result in more efficient estimates if the chosen procedure is well-suited for the data at hand.

\noindent The short- and long-term HR model introduced by \cite{Yang.2005} is a suitable choice among the less flexible methods.
The model moves the HR from an initial value $\text{HR}(0)$ to $\text{HR}(\infty)$ in a monotone fashion \cite{Yang.2010}.
Hence, the model is suitable if PH or monotonically increasing/decreasing HRs are assumed.
The Yang- and Prentice-model is in particular suited for delayed (Figure \ref{fig:delay}), and diminishing treatment effects (Figure \ref{fig:dim}) as well as crossing hazard (Figure \ref{fig:cross}) \cite{Yang.2010}.

\noindent For the change point model, a delayed or diminishing treatment effect as well as crossing hazards can be modeled by a single change point. 
More complex NPH scenarios might be accommodated by multiple change points, where \cite{Xu.2002} provide a tree-based method to determine the number and position of change points.

\noindent Furthermore, the accelerated failure time (AFT) model and its generalizations that also include covariates in the scale and shape parameters are a suitable choice. 
Note that the restrictions imposed on the $\text{HR}(t)$ differ widely within the mentioned methods; for example, an AFT is way more restrictive than a regression model for location, scale, and shape parameters in general. 
Typically, effect parameters in a fully parametric model cannot be directly interpreted but most summary effect measures can be computed from the model.

\noindent The assumption of a homogeneous population or a homogeneous treatment effect can be dropped by utilizing frailty models.
Both scenarios will typically lead to NPH on the population level, i.e. irrespective of individual, unobservable characteristics \cite{Hernan.2010,Martinussen.2020}.
Individual heterogeneity might even lead to crossing hazards on the population level if the treatment effect is beneficial but diminishing on the conditional level.
This is caused by a catch-up process at later times of high-frail individuals from the treatment group who tend to survive longer due to the beneficial treatment \cite[p.252]{Aalen.2008}.
It also highlights that a population $\text{HR}(t)$ above one not necessarily means that the treatment has a detrimental effect.
\\
\noindent Empirical comparisons of NPH regression and estimation methods with simulated or real data without introducing new methodology have been rare in our literature review.
Indeed, most papers provided simulation studies. 
However, giving recommendations on NPH methods based on the simulation studies is difficult for two reasons.
Firstly, the simulation scenarios and procedures subject to investigation differ across the papers.
Hence, an aggregated result is hard if not impossible to obtain from the existing simulation studies.
Secondly, the simulation scenarios could have been chosen to demonstrate superiority of the new method \cite{Boulesteix.2020}.
\\
\noindent
Based on the frequency of the methods in our literature review,
time-varying coefficients for the hazard rates are the most typical choice for incorporating NPH covariate/treatment effects.
Within that category, the $\text{aHR}(t^*)$ via the weighted partial likelihood and time-varying coefficients via (cubic and B-~) splines, followed by change point models, are the most frequently studied methods.
The second largest group is made up by parametric approaches. AFTs and generalized additive models for location scale and shape (GAMLSS) comprise the largest sub-group within the parametric approaches, followed by NPH approaches via the piecewise exponential model.
\NEW{See Table \ref{tab:est_overview} for the frequencies of the sub-categories.}

\noindent A review and simulation study on the $\text{aHR}(t^*)$ can be found in \cite{Rauch.2018}.
Rauch et al \cite{Rauch.2018} investigated the performance of estimates of the $\text{aHR}(t^*)$ either based on KM curves as discussed in Section \ref{KM_est} or partial likelihood-based fitting procedures as discussed in Section \ref{aHR}. 
Five different simulation scenarios, either with PH, strictly increasing or with strictly decreasing $\text{HR}(t)$, and administrative censoring as well as an administrative- combined with a random-censoring scheme were subject to investigation. 
The authors find few differences across the two estimation procedures when the shape parameter of the weighting function is $1$ for the KM-based estimate.
Inappropriately chosen weights might, however, inflate standard errors and introduce substantial bias. Different choices of weights might even result in opposite inference.
Consequently, Rauch et al \cite{Rauch.2018} suggest to carefully check the estimated survival curves to judge the plausibility of the estimates. 

\NEW{An investigation of weights in the context of weighted partial likelihood estimation of the $\text{aHR}(t^*)$ can also be found in \cite{Schemper.2009}.
The authors note that under PH, the $\text{aHR}(t^*)$ leads to a loss of efficiency.
However, for higher censoring rates and small deviations from PH, the loss in efficiency is reduced.
Furthermore, the weighted partial likelihood has higher power than the Cox PH estimate under diminishing effects (if the weights emphasize early effects).
In addition, the authors also highlight that the $\text{aHR}(t^*)$ simplifies the analysis compared to models with time-varying coefficients for hazard rates.}

\noindent From a theoretical point of view Rauch et al \cite{Rauch.2018} also note that the partial likelihood estimate of the $\text{aHR}(t^*)$ might consider further covariates what is not possible for KM-based estimates of the $\text{aHR}(t^*)$ apart from a stratified analysis. 
The KM-based estimate of the $\text{aHR}(t^*)$, however, fulfills the independent increment property and so group-sequential and adaptive designs for tests relating to KM-based estimates of $\text{aHR}(t^*)$ might be formulated.

\noindent A comparison of a (reduced rank) time-varying coefficient, gamma frailty, relaxed Burr, and a cure-rate model to real-world breast cancer data was conducted by \cite{Perperoglou.2007}. 
The authors emphasize interpretational differences across those models that might highlight different features of the data. 
In this sense, the time-varying coefficient model reveals the nature of the covariate effect, but it is not able to shed light on individual heterogeneity as the frailty model does. 
They conclude, that the specific research question should guide the model choice.
\NEW{Furthermore, the authors observe small differences in survival curves in their application and argue that the choice of how to tackle NPH is less important as long as the models are flexible enough for the data at hand.}

\section{Hypothesis tests for equality of survival curves} \label{test}

\noindent \textbf{}

\noindent For the design and analysis of randomized controlled trials with time-to-event outcomes hypothesis tests for equality of survival curves from experimental and control treatment ($H_0:S^{\left(0\right)}\left(t\right)=S^{\left(1\right)}\left(t\right)\ \forall \ t\ge 0$ ) are routinely applied. The equality of survival functions is also implied by the null hypothesis formulated in terms of the hazard rates $H_0:{\lambda }^{\left(0\right)}\left(t\right)={\lambda }^{\left(1\right)}\left(t\right)\ \forall \ t\ge 0$ or in terms of the hazard ratio $H_0:\ \text{HR}(t)=1$ $\forall \ t\ge 0$. Note, that conclusions for times beyond the maximum follow-up time $\tilde{t}$ should be avoided.

\noindent Moreover, null hypotheses based on summary effect measures, e.g.,  $H_0:\ \mathrm{\Delta }\left(t^*\right)=\text{RMST}^{\left(1\right)}\mathrm{(}{\mathrm{t}}^{\mathrm{*}}\mathrm{)}\mathrm{-}\text{RMST}^{\left(0\right)}\mathrm{(}{\mathrm{t}}^{\mathrm{*}}\mathrm{)=}0$ or $H_0:\ \text{aHR}(t^*)=1$ can also be considered as valid tests of the equality hypothesis, as rejection of a more specific null hypothesis implies rejection of equality. A ``stronger'' null hypothesis that survival in the experimental treatment is less or equal to the survival in the control arm, $H_0:S^{\left(1\right)}\left(t\right)\le S^{\left(0\right)}\left(t\right)\ \forall \ t\ge 0$,  is also of interest. However, the implication of $S^{\left(1\right)}\left(t\right)\le S^{\left(0\right)}\left(t\right)\ \forall \ t\ge 0\Rightarrow {\lambda }^{\left(1\right)}\left(t\right)\geq {\lambda }^{\left(0\right)}\left(t\right)\ $only holds under the PH assumption but is not generally true under NPH \cite{Ristl.2021}.  

\noindent 
Table \ref{tab:S4} gives an overview of the used null hypothesis in the articles focusing on hypothesis tests.
We classified whether the null hypothesis was defined as equality of survival, less or equal survival in the experimental arm, or whether it was an average-based hypothesis.

\noindent 

\noindent Under the assumption of PH the log-rank test is the standard procedure. However, if the PH assumption does not hold, power is reduced and the alternative hypothesis cannot necessarily be interpreted as treatment benefit. 
Moreover, rejecting the null hypothesis $H_0:S^{\left(0\right)}\left(t\right)=S^{\left(1\right)}\left(t\right)\,\forall t\ge 0$ in settings with NPH means that there is a non-zero treatment effect at least in some time interval.  

\noindent For situations in which the PH assumption may not hold, alternative hypothesis tests and sample size calculation approaches have been proposed, which we identified in the literature review.

\noindent In our literature review, we identified three categories of hypothesis tests for the above-mentioned null hypotheses in NPH scenarios: Log-rank tests, Kaplan-Meier-based tests, and combination tests. 
Table \ref{tab:test_overview} gives an overview of these different types of tests \NEW{ and provides examples for software, e.g. R-packages or SAS procedures. }\\
Additionally, Table \ref{tab:S1} shows in which categories the identified articles fall. Table \ref{tab:S4} provides an overview of whether the identified articles consider approaches for sample size calculation. 

\noindent \textit{}

\begin{center}
\begin{landscape}
   
\begin{table*}[!htbpt!]
    \caption{Overview of hypothesis tests for NPH. More details on the categories can be found in the Online Supplement as indicated by the references given in brackets.}
    \label{tab:test_overview}
    \centering
    \small
    \addtolength{\leftskip} {-1cm}
    \addtolength{\rightskip}{-1cm}
    \begin{tabulary}{\linewidth} {|L|L|L|L|L|}
 \hline
Category (Reference$^1$) & Sub-category & Description  & Examples of software \\
 \hline \hline
Log-rank tests (\ref{logrank}) & Log-rank test & Most widely used statistical
test to compare the overall survival of two groups. The null hypothesis is $H_0: \lambda^{(0)} (t) = \lambda^{(1)} (t), t \geq 0$ in case of a two-sided alternative or  $H_0: \lambda^{(0)} (t) \leq \lambda^{(1)} (t), t \geq 0$ in case of a one-sided test. &   \texttt{survival}, \texttt{nph} (R), LIFETEST (SAS) \\
\cline{2-4}
& Weighted log-rank test   & Augments the log-rank test with weights $w(t_{(i)})$ to emphasize observations based on their point in time. Null- and alternative hypotheses are identical to the standard log-rank test if the weights are positive.  & \texttt{survival}, \texttt{nph} (R), LIFETEST (SAS) \\
\cline{2-4}
& Modestly weighted log-rank test  & Robust variation of the weighted Log-rank test. The weights are chosen such that in case of locally favorable hazards alone the test will not wrongly infer superiority of the treatment group. Null- and alternative hypotheses are identical to the standard log-rank test.  & \texttt{nphRCT} (R) \\
\hline
Kaplan-Meier-based tests (\ref{KM_test}) & Weighted Kaplan-Meier tests & Tests based on the weighted sum of the differences of the KM estimates of the survival curves.  &  \\
\cline{2-4}
& RMST & Test for differences in restricted mean survival up to $t^*$ based on the empirical survival curves. $H_0: \Delta (t^*) = \text{RMST}^{(1)}(t^*) - \text{RMST}^{(0)}(t^*) = 0$ in case of a two-sided test and $H_0: \Delta (t^*) = \text{RMST}^{(1)}(t^*) - \text{RMST}^{(0)}(t^*) \leq 0$ in case of a one-sided alternative. Effect sizes are computed using Wald statistics. A maximum test statistic can be obtained from a set of potential time points $t^* \in \{t_1,\ldots, t_K\}$. &  \texttt{nph}, \texttt{survRM2}, \texttt{survRM2adap} (R), LIFETEST, RMSTREG (SAS) \\
\cline{2-4}
& Average hazard ratio test statistic & Test for differences in average hazard ratios of two groups. Null hypotheses for two- and one-sided alternatives are defined as $H_0: \text{aHR} = 1$ and $H_0: \text{aHR} \geq 1$, respectively. & \texttt{nph} (R) \\
\hline
\end{tabulary}
\end{table*}
\newpage
\begin{table*}[!htbpt!]
   \centering
    \small
    \addtolength{\leftskip} {-1cm}
    \addtolength{\rightskip}{-1cm}
     \begin{tabulary}{\linewidth} {|L|L|L|L|L|}
 \hline
 Category (Reference$^1$) & Sub-category & Description & Examples of software \\
 \hline \hline
& Window mean survival time & Test for differences in mean survival time of two groups between two time points $t_1^*$ and $t_2^*$. Null- and alternative hypotheses are analogous to the test for difference in RMST.  &  \\
\hline
Combination tests (\ref{combination}) & Max combo test & Maximum test of K differently weighted log-rank test statistics. The p-value of the largest test statistic is obtained from the joint multivariate normal distribution of individual test statistics. Null- and alternative hypotheses are identical to the standard log-rank test.  & \texttt{nph}, \texttt{maxcombo} (R) \\
\cline{2-4}
& Cox test and RMST difference & Combination of Cox likelihood ratio test in a Cox regression model and test for difference in RMST. The global null hypothesis is the equivalence of survival curves in a two-group setting. &  \\
\cline{2-4}
& \NEW{Multiple direction test} & \NEW{Combination of weighted log-rank statistics targeting a comprehensive range of alternatives. Critical values are calculated using permutation approaches. The null hypothesis is the equivalence of two survival distributions. }  & \NEW{\texttt{mdir.logrank}} \\
\hline
Other tests (\ref{other_test}) & Test by Gorfine et al & Test for equivalence of survival functions of K groups based on sample size partitions. Under the null hypothesis, survival curves of all K groups are identical. & \texttt{KONPsurv} (R) \\
\cline{2-4}
& Modification of the Kolmogorov Smirnov test  & Generalization of the Kolmogorov Smirnov test for use with right-censored data. Null hypothesis is the equivalency of two survival curves. &   \\
\cline{2-4}
& Test by Sooriyarachchi and Whitehead & Test for differences in survival curves based on the log odds ratio of the probability of surviving past a given time point $t^*$. The null hypothesis is the equivalency of two survival curves. &  \\
\hline
\end{tabulary}
\begin{tablenotes}
\item [1] $^1$ See corresponding sections in Appendix for the method description.
\end{tablenotes}
\end{table*}
\end{landscape}
\end{center}
\newpage


\noindent 
With prior knowledge of the NPH pattern, weighted log-rank tests can consider certain time periods to be more relevant than others. Kaplan-Meier-based tests are especially appealing to practitioners due to \NEW{their} intuitive interpretation. Combination tests select a test statistic from a small set of prespecified test statistics based on the data and are therefore useful without any prior knowledge regarding the NPH pattern.
Our literature review also identified articles reviewing and comparing hypothesis testing methods under different NPH settings. For instance, Yang \cite{Yang.2019} applies different tests including weighted log-rank tests, combination tests, and Wald tests based on estimators of the average hazard ratio or RMST to different randomized controlled trials to illustrate the virtually ignorable loss of power for reasonably PH situations and emphasizes the substantial gain of power using these approaches in contrast to the standard log-rank test in NPH situations.
Many new tests are tailored to specific NPH situations\NEW{, see Section \ref{suppl.test}}. Therefore,  Yang \cite{Yang.2019} favors the adaptively weighted log-rank test due to its overall trade-off. 
\\
In the comparison study of Dormuth et al \cite{Dormuth.2022}, in which data sets of oncology trials were reconstructed, the proposed log-rank permutation test of Ditzhaus and Friedrich \cite{Ditzhaus.2020} detected most differences between treatment groups. These results align with those of other articles investigating omnibus tests, e.g. Gorfine et al \cite{Gorfine.2020} and Royston and Parmar \cite{Royston.2020}.
If there is uncertainty regarding the underlying survival time distributions, a more recent article by Dormuth et al \cite{Dormuth.2023} recommends the use of omnibus tests for comparisons between groups.
\\
Li et al \cite{Li.2015}, Callegaro and Spiessens \cite{Callegaro.2017}, Royston and Parmar \cite{Royston.2020} and Lin et al \cite{Lin.2020} perform simulation studies for comparing different test statistics for settings with NPH. 
Li et al \cite{Li.2015} applied amongst others tests of the log-rank test family, Kaplan-Meier-based tests, and combination tests to situations of crossing survival curves at early, middle, and late times. They concluded that the adaptive Neyman's smooth test \cite{Kraus.2009} and the two-stage procedure of Qiu and Sheng \cite{Qiu.2007} have higher power in the considered NPH settings, provide an acceptable power under PH, and their type I error rate is close to the nominal level. Therefore, Li et al \cite{Li.2015} recommend the use of these tests as they are ``the most stable and feasible approaches for a variety of situations and censoring rates''.
\\The comparison study of Callegaro and Spiessens \cite{Callegaro.2017} involves, among others, the weighted log-rank test with weights of  the Fleming-Harrington weight family, max combo tests, and the likelihood ratio test for testing the treatment effect in a Cox model with time-varying coefficients. Callegaro and Spiessens  \cite{Callegaro.2017}  consider the latter to be often more powerful than the weighted log-rank tests.
\\
Lin et al \cite{Lin.2020} compare tests of the class of weighted log-rank, Kaplan-Meier, and combination tests. The comparison study did not identify a single test outperforming the others in all considered scenarios; e.g. delayed treatment onset, diminishing effects, crossing hazards, proportional hazards, and delayed effects with converging tails. The comparison study suggests the max combo test as a robust test across different NPH patterns without prior knowledge of the pattern. The review of Mukhopadhyay et al \cite{Mukhopadhyay.2022} compared the log-rank test to the MaxCombo test in immo-oncology trials identified through a systematic literature review. The authors concluded that the MaxCombo test is a "pragmatic alternative" to the log-rank test assuming NPH.
The simulations of Royston and Parmar \cite{Royston.2020} suggest that the modified versatile weighted log-rank test, an unpublished modification of the versatile weighted log-rank test \cite{Lee.2007} with Stata code available on request from Royston, performs best in terms of power under NPH (early, late or near PH treatment effect) without the preconceived type of treatment effect.
\\

\noindent In the last 20 years, there have been constant publications on log-rank tests. Research on combination tests, Kaplan-Meier-based tests, or other approaches has been comparatively rare. In the last 3 years, however, more research on these testing categories including permutation approaches, e.g. \cite{Ditzhaus.2020, Ditzhaus.2021}, was conducted.

\section{ Conclusions and discussion} \label{discussion}

\noindent 

We conducted a systematic literature review of effect estimation and testing methods that are able to cope with NPH in time-to-event analysis.
Review articles focusing on different methods for NPH have been published previously. These reviews mostly focus either on a quantitative comparison for specific NPH scenarios \cite{Li.2015}, or a specific method class \cite{Rauch.2018}, or on a qualitative overview of available methods for specific NPH situations or disease areas, e.g. oncology \cite{Ananthakrishnan.2021}. We conducted a systematic literature search for methodological approaches for any NPH scenario, any model class, and not restricted to a specific disease area. Therefore, our review gives a comprehensive overview of the methods proposed and applicable to NPH settings.
\\
\noindent
In total, our literature review includes 211 articles for final analysis.
Of those articles, 113 focus on effect estimation, e.g. regression methods, 72 on testing, and 26 articles on both.
In the effects estimation and testing literature, we identified categories to group articles according to their approach to the NPH situation.
With respect to effect estimation, the categories are Kaplan-Meier based estimation approaches, stratified Cox model, time-varying coefficients for the hazard rates, transformation models with time-covariate interaction, short- and long-term HR, joint models, frailty models, parametric models, machine learning approaches and others. 
With respect to testing, the categories are log-rank tests, Kaplan-Meier tests, combination tests, and other tests.
\NEW{We have also broken down some of the categories into smaller sub-categories and assigned each paper to at least one of them.
An overview of the categories and subcategories is given in Table \ref{tab:est_overview} and \ref{tab:test_overview}, for estimation and testing approaches respectively. The tables and Section \ref{estimation} and \ref{test} provide brief explanations of the categories. For a more detailed discussion including references to the original articles proposing specific methods, we refer to the Supplement \ref{Supplement}.}
The most common approaches to tackle NPH for effect estimation are time-varying coefficients for the hazard rates (47 papers), and parametric approaches that assume a distribution for the survival time (38 papers), such as GAMLSS models.
The most common testing approach for NPH are variations of the log-rank test (63 papers).
We extracted and documented the software (R and SAS) utilized in the papers under review.
In addition, well-known software for the individual testing and estimation categories was added by the group of authors.
For a more complete overview of available R packages for time-to-event analysis see \NEW{the CRAN Task View homepage for Survival Analysis} \cite{Allignol.2022}. 
\\
\noindent
For the literature review, we excluded standard methods such as the stratified Cox model, 
unless the baseline hazards were stratified by the treatment indicator. Consequently, our review may have missed certain innovative proposals in this area. In addition, we have excluded methods that utilize \NEW{internal \cite[p.198]{Kalbfleisch.2002}} 
time-varying covariates which might lead to NPH over time, e.g. PKPD Models. Further, our search terms focused on terms related to NPH, which may not be a common term in other areas utilizing these methods. 
For review articles considered in this review, we manually added all investigated methods to the list of articles. Nevertheless, some of those may have been later discarded due to our in- or exclusion criteria, see Figure \ref{fig:prisma}. Consequently, some of the considered review articles may investigate methods which have not been discussed in this review.
\\
\noindent
A broad range of different methods is available for both treatment effect estimation and hypothesis testing.
However, there is no consensus on the best approaches under NPH.
Most papers reported simulation studies (158 of 211 papers). Nevertheless, the NPH scenarios and the methods under comparison differ making it difficult to aggregate and compare results across evaluations. Moreover, the NPH scenario and the competitors to newly introduced methodology might have been chosen to demonstrate superiority of the newcomer \cite{Boulesteix.2020}.
Only a few review articles comparing
different methods through simulation studies (considered to be objective) have
been identified by our review.
In particular for effect estimation methodology, independent comparison studies including neutral comparison studies covering
different NPH scenarios and a broad range of methods are not available.
Review articles of testing procedures cover a
broad range of different NPH settings and provide guidance for the choice
of the test, which, however, can be different from one comparison study to
another. 
These reviews offer some guidance on, for example, the permutation test by Ditzhaus and Friedrich \cite{Ditzhaus.2020}, and the adaptively weighted log-rank \cite{Yang.2010} for specific NPH scenarios.
Due to the hypothesis tests examined not being consistent across the comparison studies, it is difficult to make a general recommendation for the use of a specific hypothesis test. 
\\
\noindent
\NEW{The choice of an estimation method could be based on theoretical considerations.
In the absence of strong prior knowledge of the treatment/covariate effect, time-dependent treatment coefficients for the hazard rates could be flexibly modeled via a treatment spline interaction, where the corresponding basis functions are constructed on time.}
In the case of strong prior knowledge, more restrictive models might be preferred, such as a (single) change point model for a delayed treatment effect (Figure \ref{fig:delay}). 
\\
\noindent
Moreover, different summary effect measures have been proposed which offer an alternative to the hazard ratio. The constant hazard ratio estimated by a Cox PH model is commonly used for time-to-event analysis but might be misleading under NPH as the hazard ratio is time-dependent in this case. Alternatives involve, for example, the average hazard ratio and the ratio of RMSTs. These depend on the choice of the pre-specified time interval which is restricted by the maximum follow-up time. Additionally, its usefulness depends on the pattern of the treatment effect. For instance, the difference of RMST between treatment groups is not useful for delayed treatment effects \cite{Gregson.2019}. Summary effect measures can be calculated based on Kaplan-Meier curves. For more complex data, e.g. multiple continuous covariates, other methods presented in Section \ref{estimation} can be used to model the survival curves.
Depending on the choice of the estimation approach it might be difficult if not impossible to obtain specific summary effect measures, however. \NEW{Dynamic, i.e. time-varying, effect measures could be used instead and could help to communicate how survival patterns are affected by the treatment over time.
However, dynamic effect measures are less appropriate as a primary basis for binary decisions such as marketing authorizations. Nevertheless, they could be used to support a decision following a gatekeeping hypothesis test on any difference, hence disentangling the hypothesis test and estimation. As a drawback, such a decision procedure could not be clearly defined in advance. In contrast to this, single summarizing measures such as RMST difference can be used for both, hypothesis tests and estimation, and lead to an unambiguous binary decision procedure but require an upfront agreement on the most relevant measure. }
\\
\noindent 
We identified a variety of NPH approaches for both, effect estimation and testing procedures.
\NEW{Although a variety of non-proportional hazard methods are available, they are still rarely applied. Statistical practice needs to change by adopting the non-proportional hazards approaches summarized in this paper. }
Adhering to invalid assumptions\NEW{, i.e. proportional hazards,}  might lead to less reliable conclusions than choosing a non-optimal NPH approach for the data at hand \NEW{as illustrated in Section \ref{motiv}}. To fill the gap in comparisons of the methods for NPH, our further assessment will explore the advantages and disadvantages under a wide range of NPH assumptions of a selection of the identified methods, see \cite{klinglmüller2023neutral}.

\newpage
\section{ Funding sources/sponsors}

\noindent 

\noindent This work has received funding from the European Medicines Agency (Re-opening of competition EMA/2020/46/TDA/L3.02 (Lot 3))

\noindent ``This document expresses the opinion of the authors of the paper, and may not be understood or quoted as being made on behalf of or reflecting the position of the European Medicines Agency or one of its committees or working parties.''

\section{Acknowledgement}
The authors would like to thank Juan Jos\'e Abell\'an and Marcia R\"uckbeil of the European Medicines Agency as well as Andreas Brandt of the Bundesinstitut für Arzneimittel und Medizinprodukte for valuable comments and insightful discussion on the report related to the literature review performed as part of the research contract EMA/2020/46/TDA/L3.02 (Lot 3). This report formed the basis of our manuscript.

\noindent


\appendix

\setcounter{table}{0}
\renewcommand*\thetable{\Alph{section}\arabic{table}}
\section{Appendix} \label{Appendix}

\noindent 
\subsection{Literature search and data extraction} \label{detailedMethodSection}
\subsubsection{Systematic literature search and study selection}

\noindent

\noindent We performed a comprehensive literature search using the electronic databases MEDLINE and EMBASE to identify relevant publications. Pubmed was used to search the database MEDLINE. EMBASE searches in both databases MEDLINE and EMBASE. However, we excluded the database MEDLINE for our search in EMBASE. Table \ref{tab:Table3} provides details of the literature search. 
The search includes two parts. One part of the search includes topic-related search terms such as “non-proportional hazards” in various spellings (part one of the search strategy (main search in MEDLINE and EMBASE), rows 1 to 9 of \ref{tab:Table3}). In addition, in a second part of the search we used broader pre-specified topic-related terms (e.g. “crossing hazard*”, “delayed effect”, “treatment switch”).
 The part of the search with the pre-specified related search terms has been restricted to statistical journals using the Web of Science category ‘Statistics \& Probability’ (part two of the search strategy (additional search in MEDLINE), row 10 to 19 of Table 3). Reference lists of included publications and relevant reviews were checked manually to identify any additional relevant articles that were not captured by the search.

\noindent We included methodological research articles, reviews, and clinical investigations with time-to-event endpoints where methods for non-proportional hazards were applied. More specifically, the inclusion and exclusion criteria specified in Table \ref{tab:Table1} have been used.

\begin{table}[h]
    \centering
    \caption{Inclusion and exclusion criteria}
    \begin{tabular}{p{5.5cm}|p{5.5cm}}
         Inclusion criteria	& Exclusion criteria  \\
         \hline
         Methodological publications introducing methods for time-to-event analyses of clinical trials with non-proportional hazard (NPH) assumptions & 	Investigations assessing NPH assumptions without introducing new NPH methods in the analysis\\
         Applications of new NPH methods in clinical trials with time-to-event endpoints &Applications of conventional methods for NPH such as stratified Cox regressions or the use of time-dependent covariates in proportional hazards models \cite{Klein.2003}\\
         &	Methods for competing risk or recurrent event analyses\\
    \end{tabular}
    \label{tab:Table1}
\end{table}

\subsubsection{Data extraction}

\noindent

\noindent Two reviewers screened the search results for relevant articles independently. The reviewers screened titles and abstracts and excluded papers that clearly did not meet the inclusion criteria. During the abstract screening, an article was included for the full-text screening, if at least one of the two reviewers included the abstract. For all selected abstracts, we obtained the full texts and two reviewers reviewed these according to the pre-specified inclusion criteria (e.g. statistical methods, simulations, time-to-event endpoints, non-proportionality assumption). From the full text, data on basic characteristics of the investigation, on proposed methods, and, if applicable, on simulation studies conducted were collected by two independent reviewers. A third reviewer resolved all disagreements on the inclusion of the full texts and on the extracted data. Characteristics of the proposed methods include whether the methods use frequentist or Bayesian, parametric or non-parametric approaches and whether the methods adjust for covariates. We distinguished between articles including methods for treatment effect estimation, hypothesis testing and manuscripts covering both. We also extracted the availability of a software for the proposed method and differentiated whether the software was freely available, e.g. Github repository for code, R package or Code snippets in text, available, e.g. available upon request from the author, or not available or mentioned in the text. 
Additionally, the proposed methods were classified into one of the predefined categories: 
\begin{itemize}
    \item Log-rank approaches, e.g. weighted log-rank test
    \item Kaplan-Meier curve-based approaches; e.g. weighted Kaplan-Meier and Restricted Mean Survival Time (RMST)
    \item Combination approaches; e.g. Breslow test, Lee’s combo test, or MaxCombo test
    \item Regression model (parametric/semiparametric/nonparametric)
    \item Other; if the previously mentioned classes do not fit the method.
\end{itemize}

\subsubsection{Quantitative and qualitative analysis}

\noindent

\noindent Characteristics of articles and proposed methods are described by using descriptive analyses such as bar charts. The collected categorical variables are summarized by providing absolute and relative frequencies.
In order to enable an overview of the methods proposed in the identified articles, we differentiated between methods for treatment effect estimation (see Section \ref{estimation}) and hypothesis test for the comparison of survival between treatment groups (see Section \ref{test}). We additionally introduced new categories for publications focusing on treatment effect estimations via e.g. regression approaches. The categories are Kaplan-Meier for effect estimation (Section \ref{KM_est}), stratified Cox (Section \ref{StratCox}), time-varying coefficients for the hazard rates (Section \ref{time varying}), transformation models with time-covariate interaction (Section \ref{trans}), joint models (Section \ref{joint}), short- and long-term hazard ratio (Section \ref{shortHR}), frailty models (Section \ref{frailty}), parametric regression models (Section \ref{parametric}), machine learning approaches (Section \ref{ML}) and other modelling approaches (Section \ref{other_estimation}). The main characteristics of these categories are introduced in a common notation and examples for the categories are provided and described in more detail (Section \ref{estimation}).

\noindent All statistical analyses (tables and figures) are performed in R 4.1.2 (R Core Team, 2021).

\subsubsection{Data extraction guideline} \label{dataextraction}

\noindent

\noindent For the data extraction each reviewer was provided with the following instructions including the in-and exclusion criteria:

\noindent \textbf{\underbar{CONFIRMS - Literature review}}

\noindent 

\noindent AIM:

\noindent Methods and statistical software for statistical analysis and reporting of clinical trials with time-to-event endpoints with non-proportional hazards

\noindent 

\noindent INCLUSION:

\begin{itemize}
\item  Methodological publications introducing methods for time-to-event analyses of clinical trials with non-proportional hazard (NPH) assumptions 

\item  Applications of new NPH methods in clinical trials with time-to-event endpoints
\end{itemize}

\noindent EXCLUSION:

\begin{itemize}
 
\item  Investigations assessing NPH assumptions without introducing new NPH methods in the analysis 

\item  Applications of conventional methods for NPH such as stratified Cox regressions or the use of time-dependent covariates in proportional hazard models

\item  Methods for competing risk or recurrent event analyses
\end{itemize}

\noindent 

\noindent SCREENING -- 2${}^{nd}$ step - full texts AND DATA EXTRACTION:

\noindent 

\noindent Please fill in the columns L to O for all full texts 

\begin{itemize}
\item  Type of paper (methods/applied/review)

\item  For the inclusion/exclusion of the full text, please fill the columns L to N in the excel sheet:

\begin{itemize}

\item  Inclusion (yes/no)

\item  Please provide the reason for non-inclusion (if ``other'', please specify)
\end{itemize}
\end{itemize}

\noindent 

\noindent If the full text is included, please fill in the columns P to AH

\noindent 

\begin{itemize}
\item  Simulation study conducted (yes/no)

\item  Software available (not available/available/freely available)

\begin{itemize}
\item  Not available/not mentioned: No mention of code availability made in the text

\item  available: Availability mentioned, but code not freely available, e.g. ``Data available upon request from the authors''

\item  freely available: Code freely available (as stated in the paper, without background checks), e.g. Github repository for code, R package, Code snippets in the text,{\dots}.
\end{itemize}
\end{itemize}

\begin{itemize}
\item  If software freely available, software specification (text); e.g. name of R package, link to Github repository, etc.

\item  Provides program for sample size calculation (yes/no)

\item  Provides program for analysis (yes/no)
\end{itemize}

\noindent 

\noindent 

\begin{itemize}
\item  The following items capture information on the new method or methods. If multiple new methods are described, please classify each of them within this row. If the predefined categories (log-rank, Kaplan Meier,{\dots}) do not fit all of new methods well, please use the column ``Specification of other class'' and give some details on the new method or methods. 
\noindent 

\begin{itemize}
\item  Classification of the method as separate items with dropdown (yes/no)

\end{itemize}

\noindent 

\begin{itemize}
\item  Log-rank approaches, e.g. weighted log-rank test

\item  Kaplan-Meier curve-based approaches; e.g. weighted Kaplan-Meier and Restricted Mean Survival Time (RMST)

\item  Combination approaches; e.g. Breslow test, Lee's combo test, or MaxCombo test

\item  Regression model 

\begin{itemize}
\item  If regression model, parametric/semiparametric/nonparametric (dropdown)
\end{itemize}

\item  Other class; if the above-mentioned classes do not fit the method
\end{itemize}

\item  If other, specification of other class for the new NPH method or methods (text, if several new methods proposed, comment on them individually in this box)

\item  Bayesian or frequentist approach of the method (dropdown)

\item  Covariate adjustment included in the method (not discussed/ stratification for factors/ adjustment for covariates)

\item  Methods for hypothesis testing (yes/no)

\item  Methods for estimation (yes/no)

\item  Interim analysis (yes/no)

\item  Interesting case studies to inform simulation studies (yes/no); e.g. only RCTs and the data should be presented in sufficient detail to inform the simulation study, e.g. Kaplan-Meier curve with number of patients at risk

\item  Comments (text); name of the methods and optionally further details
\end{itemize}
\newpage
\subsubsection{Search term for the literature review} \label{searchterm}

\noindent

\begin{table}[hb!]
\caption{Search terms for the literature review} 
\label{tab:Table3}
\begin{tabular}{|p{0.2in}|p{2.8in}|p{1.5in}|}
\hline 
& Search terms &   \\ \hline 
1 & ``non-proportional hazard'' OR ``non-proportional hazards'' OR ``nonproportional hazard'' OR ``nonproportional hazards'' OR ``non proportional hazard'' OR ``non proportional hazards'' & Main search in MEDLINE and EMBASE without any journal restrictions \\ \hline 
2 & non-proportionality OR nonproportionality OR ``non proportionality'' &  \\ \hline 
3 & (NPH OR NPHs OR "non PH" OR "non PHs" OR non-PH OR non-PHs) AND (assumption OR assumptions) &  \\ \hline 
4 & ``proportional hazard'' OR ``proportional hazards'' OR ``hazard assumption'' OR ``hazard assumptions'' &  \\ \hline 
5 & violation OR violations OR violated OR violat* &  \\ \hline 
6 & \#4 AND \#5 &  \\ \hline 
7 & \#1 OR \#2 OR \#3 OR \#6 &  \\ \hline 
8 & ``delayed treatment effect'' OR ``delayed treatment effects'' &  \\ \hline 
9 & \#7 OR \#8 &  \\ \hline 
10 & crossing AND (hazard OR hazards) & Additional search with broader search terms in journals of the category ``Statistics \&  Probability'' \\ \hline 
11 & crossing AND (curve OR curves) &  \\ \hline 
12 & ``delayed effect'' OR ``delayed effects'' &  \\ \hline 
13 & delayed AND treatment &  \\ \hline 
14 & ``treatment switch*'' OR ``heterogeneous patient*'' OR ``heterogeneous population*'' OR ``disease progression*'' &  \\ \hline 
15 & \#10 OR \#11 OR \#12 OR \#13 OR \#14 &  \\ \hline 
16 & ``hazard assumption'' OR ``hazard assumptions'' OR time-to-event OR ``time to event'' &  \\ \hline 
17 & \#15 AND \#16 &  \\ \hline 
18 & List of journals from the Web of Science category ``Statistics \& Probability'' &  \\ \hline 
19 & \#17 AND \#18 &  \\ \hline 
20 & \#9 OR \#19                                            &   \\ \hline 
\end{tabular}
\end{table}


\appendix

\renewcommand*\thetable{\Alph{section}\arabic{table}}

\newpage
\setcounter{page}{1}
\setcounter{section}{18}
\setcounter{table}{0}
\section{Online Supplement} \label{Supplement}
\subsection{Notation}
\begin{table}[H]  
    \caption{Overview of notation used throughout the report}
    \label{tab:notation}
    \begin{tabular}{|p{1.7in}|p{2.6in}|} \hline 
$T$ & Time to event \\ \hline 
$C$ & Censoring time \\ \hline 
$x$ & Vector of covariates \\ \hline 
$Z$ & Treatment indicator, $1$ if experimental, $0$ for control \\ \hline 
$\beta $ & Vector of regression coefficients of $x$ \\ \hline 
$\gamma $ & Regression coefficient of $Z$ \\ \hline 
$N$ & Total sample size \\ \hline 
$\delta $  & Event indicator, $1$ if event, $0$ if censored \\ \hline 
${\lambda }_0(t)$  & Baseline hazard rate at time $t$ \\ \hline 
${\lambda }^{\left(Z\right)}(t)$  & Hazard rate of treatment group $Z$ at time $t$ (given $x$) \\ \hline 
${\mathrm{\Lambda }}^{\left(Z\right)}(t)$ & Cumulative hazard rate of treatment group $Z$ $\int^t_0{{\lambda }^{\left(Z\right)}(u)du}$ \\ \hline 
$S^{\left(Z\right)}(t)$  & Survival function of treatment group Z at time $t$ (given $x$) \\ \hline 
$Y_i\left(t\right)$ & At risk indicator, $Y_i\left(t\right)=1$ if $t\le {\mathrm{min} \left(T_i,\ C_i\right)\ },0$ else  \\ \hline 
$Y\left(t\right)=\sum_i{Y_i(t)}$ & Number of people at risk at time $t$ \\ \hline 
$Y^{(Z)}(t)$ & Number of people at risk in treatment group $Z$ at time $t$ \\ \hline 
$HR(t)$ & Hazard ratio at time $t$ (given $x$) \\ \hline 
$aHR(t)$ & Average $HR$ up to time $t$ \\ \hline 
$cHR(t)$ & Cumulative $HR$ up to time $t$ \\ \hline 
$RMST(t)$ & Restricted mean survival time up to $t$ \\ \hline 
$t^*$ & Specific time point \\ \hline 
$\tilde{t}$ & Maximum follow-up time \\ \hline 
$t_{\left(i\right)}$ & $i^{\text{th}}$ ordered distinct event time, $t_{\left(0\right)}$ defined as 0 \\ \hline 
\end{tabular}
    
\end{table}


\subsection{Estimation approaches for NPH treatment effects}

\subsubsection{Kaplan-Meier based estimation approaches} \label{KM_est}

\noindent 

\noindent The underlying basis of all approaches in this sub-section is the non-parametric Kaplan-Meier (KM) estimator of the survival function or the Nelson-Aalen (NA) estimator of the cumulative hazard function. The corresponding papers that fall into this category can be found in column A of Table \ref{tab:S2}. 

\noindent The non-parametric KM estimate of the survival curve is defined as $\hat{S}^{(Z)}\left(t\right)=\prod_{t_{\left(i\right)}\le t}{\bigg(1-\frac{d^{(Z)}_{(i)}}{Y^{(Z)}\left(t_{(i)}\right)}\bigg)}$, where $d^{(Z)}_{(i)}$ is the number of events of treatment group $Z$ observed at $t_{\left(i\right)}$ and the product iterates over the ordered distinct event times $t_{\left(i\right)}$. Alternatively, the NA estimate of the cumulative hazard function, i.e.  ${\widehat{\mathrm{\Lambda }}}^{\left(Z\right)}\left(t\right)=\sum_{t_{\left(i\right)}\le t}{{d^{(Z)}_{\left(i\right)}}/{Y^{(Z)}\left(t_{\left(i\right)}\right)}}$, can be utilized. The treatment stratified KM estimates do not impose any modelling assumptions on $HR\left(t\right)$. Thus, any trajectory of $HR\left(t\right)$ can be incorporated. 

\noindent 

\noindent The KM (NA) estimators result in step functions and thus the estimation of $HR\left(t\right)$ requires additional smoothing techniques, however. The $aHR{(t}^*)$ can be re-formulated as being an expression of the survival functions where the estimates might be obtained via KM \cite{Rauch.2018}. 

\noindent A doubly-weighted Nelson-Aalen estimator that accounts for dependent censoring and treament-specific covariate distributions was introduced by \cite{Schaubel.2011} via inverse probability of treatment weighting and inverse probability of censoring weighting. 
The authors suggest to process the treatment-specific cumulative hazard estimators to cumulative treatment effect measures such as the cHR, relative risk, and differences in the RMST.

\noindent 


\noindent Quantiles and $t$-year survival rates can be obtained by standard statistical software which computes KM survival curves such as the R package \texttt{survival}. The R software package \texttt{survRM2} performs two-sample comparisons with the $RMST$ and also includes a function to perform an ANCOVA-type covariate adjustment. The SAS procedure LIFETEST can be utilized to obtain non-parametric estimates of the survival function and to carry out tests of homogeneity across strata or the association of numeric covariates on survival. The R package \texttt{AHR} provides software solution for the $aHR$. 

\noindent 

\paragraph{Pseudo values}\label{pseudo} 

\noindent 

\noindent Pseudo-values are usually calculated from KM-based estimates, for example the $\widehat{RMST}(t^*)$ or $\hat{S}(t^*)$. The basic idea of the pseudo-value approach is to compute a quantity of interest at a given time $t^*$, e.g. $\widehat{RMST}(t^*)$ or $\hat{S}(t^*)$, and re-compute that measure with the $i^{th}$ observation dropped. Then, for example, $N\hat{S}\left(t^*\right)-\left(N-1\right)\ {\hat{S}}_{\left(-i\right)}(t^*)$, where ${\hat{S}}_{\left(-i\right)}$ denotes the KM estimate of the survival function with the $i^{th}$ observation dropped, is the so-called pseudo-value, and the $i^{th}$ entry of a new dependent variable. This value is an estimate of the survival function for the $i^{th}$ individual in this example. Hence, the interrelationship between the survival function and the cumulative hazard rate, $S^{\left(Z\right)}\left(t\right)={\mathrm{exp} (-{\mathrm{\Lambda }}^{\left(Z\right)}(t))\ }$, can be used to transform the pseudo value into an estimate of the log cumulative hazard rate ${\mathrm{\Lambda }}^{\left(Z\right)}(t)$ by applying log-log transformation on the pseudo values. The pseudo value procedure reduces time-to-event data, that are usually defined by two variables (time, event indicator) to a single continuous variable, and allows modelling by common regression models, i.e. linear models with the treatment indicator being a covariate, with possibly transformed pseudo-values. \cite{Andersen.2004} 

\noindent 

\noindent Application of the pseudo value procedure on $\hat{S}(t^*)$, where the pseudo value $N\hat{S}\left(t^*\right)-\left(N-1\right)\times$ ${\hat{S}}_{\left(-i\right)}(t^*)$ is analysed by the linear regression model $x^T_i\beta +Z_i\gamma $, $i=1,\dots ,\ N$,  gives estimates of $S^{\left(1\right)}\left(t^*\right)-S^{\left(0\right)}(t^*)$ through $\widehat{\gamma }$, assuming that $x$ is held constant.

\noindent 

\noindent A weighted pseudo value approach to estimate survival probabilities at time $t^*$ was introduced by \cite{Mittlbock.2022}. They used the log-log transformation of the pseudo values and gave estimates for $cHR\left(t^*\right)$. The weights were proposed to overcome missing information on group membership in stem cell transplantation patients, where early death stops the search for a suitable stem cell donor.

\noindent 

\noindent Yang et al \cite{Yang.2021} utilized the pseudo-value approach for the $RMST$: Subject to investigation was the $RMST$ given survival up to a landmark $t^*$ with (chosen) follow up time$\ t^*+\ w$, $cRMST$($t^*\ ,w)$, i.e. the expected life time within $(t^*,t^*+\ w)$ given survival up to $t^*$. Estimates of the $RMST$ were obtained via KM. Then, in a leave one out step, the $cRMST$ without the $i^{th}$ observation ${\widehat{cRMST}}_{(-i)}$($t^*,w)$ was computed. 
The individual differences $N^*\widehat{cRMST}$($t^*,w)-\left(N^*-1\right)\times$ ${\widehat{cRMST}}_{(-i)}$($t^*,w)$, with $N^*\ $being the sample size conditional on survival up to $t^*$, were then used as the dependent variable in a linear regression model $x^T_i\beta +Z_i\gamma $. 
If the multiple landmark time points are considered in a single analysis, covariate-time interactions might be utilized to obtain time-dependent covariate effects, e.g. $Z_i\gamma (t^*)$, with $\gamma(t^*) = \gamma_1 + t^*\gamma_2 +{t^*}^2 \gamma_3$, for the treatment component. 
The (time-dependent) treatment efficacy for such an approach can be evaluated via $\widehat{\gamma }(t^*)$ which estimates $cRMST^{\left(1\right)}$($t^*,w)-\ cRMST^{\left(0\right)}$($t^*,w)$.

\noindent 

\noindent Pseudo values were also considered by \cite{Potschger.2018} and \cite{Li.2020}. We did not encounter further effect measures in our literature review.

\noindent 


\noindent The R packages \texttt{pseudo} and \texttt{prodlim} as well as SAS and R functions in \cite{Klein.2008} are available for computing pseudo-values. Pseudo-values for $RMST(t)$ [but not for $S(t)$] can be calculated by PROC RMST in SAS. 

\noindent 

\paragraph{Quantile regression} \label{quantile}

\noindent 


\noindent Quantile regression is an approach to systematically compare the estimated quantiles $t^{\left(Z\right)}_{\tau }={S^{\left({Z}\right)}}^{-1}\left(1-\tau \right)$, $Z=1,2$, and $\tau \in (0,1)$.

\noindent A quantile regression approach has been established based on the generalized KM estimator \cite{Portnoy.2003} or alternatively, on martingale-based estimating equations \cite{Peng.2008}. 
In the latter case, the logarithm of the ${\tau }^{th}$ survival quantile $\ln\{t_{\tau}\}$ given covariate information is assumed to have a linear relationship with the covariates, i.e. $\ln\{t_{\tau}|x,Z\}  =$ $x^T\beta (\tau )+Z\gamma (\tau )$. 
The estimated treatment effect ${\mathrm{exp} (\widehat{\gamma }(\tau ))\ }$ gives an expression of ${{\hat{t}}^{\left(1\right)}_{\tau }}/{{\hat{t}}^{\left(0\right)}_{\tau }}$ in the case of equality in $x$. A value larger than $1$ of the above-mentioned ratio indicates a beneficial treatment effect at the respective quantile. 

\noindent See \cite{Mboup.2021} for an application to evaluate the long-term benefit of immunotherapy. Xue et al \cite{Xue.2018} developed a leave-one-out cross-validation approach for quantile regression.


\noindent An implementation of quantile regression including methods applicable to data with censored observations is available in the R package \texttt{quantreg}. The SAS procedure quantlife provides an implementation of quantile regression based on generalizations of the KM and NA estimator. 

\noindent

\subsubsection{ Stratified Cox model}\label{StratCox}

\noindent 


\noindent The stratified Cox model is a well-known model that relaxes the PH assumption for the stratification variable(s). We excluded the basic stratified Cox model, where stratification by variables other than the treatment indicator is sufficient, from our literature review. The semi-parametric stratified Cox model, in particular, does not impose modelling restrictions on the trajectory of $HR\left(t\right)$ along the stratified variables. The hazard rate is defined as ${\lambda }^{(Z)}\left(t\right)={\mathrm{exp} (x^T\beta )\ }{\lambda }^{(Z)}_0\left(t\right)$. The coefficients $\beta $ can be estimated via the partial likelihood while the stratified baseline hazard ${\widehat{\lambda }}^{(Z)}_0$ can subsequently be obtained via the non-parametric Breslow estimate for each stratum. 

\noindent Table \ref{tab:S2} column B shows papers that belong to the category `stratified Cox model'. 

\noindent 

\noindent Again, estimation of $HR\left(t\right)$ requires additional smoothing techniques due to the step function nature of the Breslow estimate. 
An estimate of $cHR\left(t^*\right)$ was suggested by \cite{Dong.2012} and \cite{Wei.2008}. 

\noindent 


\noindent The R package \texttt{survival} and the SAS procedure phreg include the stratified Cox model. 

\noindent

\subsubsection{Time varying coefficients for the hazard rates} \label{time varying}

\noindent 


\noindent Time-varying coefficients for the hazard rates $\gamma (t)$ can be utilized to incorporate, for example, a late or diluting treatment effect causing NPH. The hazard rate could be of the form ${\lambda }^{(Z)}\left(t\right)={\mathrm{exp} (Z\gamma \left(t\right)+\ x^T\beta )\ }{\lambda }_0\left(t\right)$. Note that the time-varying treatment effect $\gamma \left(t\right)$ of those models can essentially be viewed as an interaction term of (some function of) time and the treatment indicator. The specific models differ in how those interactions are constructed. 

\noindent 

\noindent The approaches to time-varying coefficients, or time-covariate interaction respectively, are plentiful and encompass change point models, fractional polynomials, spline-based approaches, other functional time-covariate interactions and the additive model. The admissible trajectory of $HR\left(t\right)={\mathrm{exp} (\gamma (t))\ }$ depends on the complexity of $\gamma \left(t\right)$. The $aHR$, or weighted Cox regression, which will be discussed in Section \ref{aHR} further below can be seen as less complex summary measure in a time-varying coefficient setting. Columns of Table \ref{tab:S2} that belong to the broader category `time-varying coefficients' of Table \ref{tab:S1} start with a `C'.

\noindent 

\noindent Common approaches to fit the models are the maximization of the partial likelihood (e.g. \cite{Zhang.2021}, \cite{Brown.2007}, or \cite{Sauerbrei.2007}), the full likelihood possibly via GLM routines (e.g. \cite{Freeman.2022}, \cite{Argyropoulos.2015}, \cite{RodriguezGirondo.2013}, or \cite{Royston.2002}) or via Bayesian approaches (see Table \ref{tab:S2} column K). We added column K to Table \ref{tab:S2} which indicates whether the corresponding paper took a Bayesian or a frequentist approach, such that the interested reader might have a look at the corresponding papers to learn more about specific fitting procedures in detail.

\noindent

\paragraph{Change point of the treatment coefficient} \label{change point}

\noindent 


\noindent A late treatment effect, for example, might be considered by $\gamma \left(t\right)=\gamma \ I(t\ge t^*)$ where $I$ denotes the indicator function, a diluting treatment effect through $\gamma \left(t\right)={\gamma }_1+\ {\gamma }_2\ I(t\ge t^*)$, with some (chosen) threshold value $t^*\ >0$, ${\gamma }_2>0$. Johannes \cite{Johannes.2007}  suggested a grid search to find the location of the change point $t^*$ associated to the highest partial likelihood value. 

\noindent A generalized version of the single change point model, introduces a piecewise constant regression coefficient ${\gamma }_k$ for each of the chosen disjoint time-intervals, i.e. $\left(t\right)=\sum^K_{k=1}{{\gamma }_k}I_k(t)$, where $I_k(t)$ is the indicator function which is unity if $t\in k^{th}$ interval and $0$ else. A tree-based method to find multiple change points was suggested by \cite{Xu.2002}; a change point is chosen such that it maximizes the score test statistic. This process is iterated over the new time partitions resulting from the former step. In order to avoid over-fitting \cite{Xu.2002} also considered stopping criteria and a pruning mechanism. 

\noindent Papers that were allocated to the category `change point for time-varying effect' can be found in Table \ref{tab:S2} column C.1.

\noindent

\paragraph{Time varying coefficients with continuous paths} \label{timeXcovariate}

\noindent 


\noindent A smooth path of $\gamma \left(t\right)$ can be obtained by choosing a continuous function (in $t$). For example, $\gamma \left(t\right)={\gamma }_0+t\ {\gamma }_1+t^2{\gamma }_2$ \cite{Stablein.1980}. Gustafson \cite{Gustafson.1998} considered an additive hazard model where $\gamma \left(t\right)=\left(1-\frac{t}{t^*}\right){\gamma }_0+\ \ \frac{t}{t^*}{\gamma }_1$ for $t<t^*$ and $\gamma \left(t\right)={\gamma }_1$ else. Sauerbrei et al \cite{Sauerbrei.2007} chose $\gamma \left(t\right)={\gamma }_0+{\mathrm{log} \left(t\right)\ }{\gamma }_1$ as default and provided an algorithm for model selection: Firstly, start with an initial model, possibly incorporating functional forms and interactions of the covariates. Secondly, re-run the model on a restricted time interval$\ (0,\ t^*$) and add relevant covariates from this step to the model of the entire time interval if necessary. Thirdly, for each covariate include time-varying coefficients, on the entire time interval, and add them to the final model if they provide a significant improvement. More flexible functions of $\gamma \left(t\right)$, obtained through fractional polynomials in time (see Section \ref{fractional}), might be chosen if the enhanced model provides a significantly better fit as indicated by a deviance test.

\noindent Papers with some functional covariate-time interaction that do not belong to one of the more specific time-varying coefficient categories above or below can be found in column C.2 of Table \ref{tab:S2}. 

\noindent

\paragraph{Fractional polynomials} \label{fractional}

\noindent 


\noindent Fractional polynomials (FP) can be used to model a time-dependent treatment effect $\gamma \left(t\right)$. A $FP1(p)$ transformation of time is defined as $t^p$ with some chosen $p$. By convention $FP1(0)$ is $\mathrm{log}\mathrm{}(t)$. The time varying treatment coefficient with $FP1(p)$ transformation in $t$ is then $\gamma \left(t\right)={\gamma }_0+t^p{\gamma }_1$. This can be extended to $FP2(p_1,\ p_2)$, this results in $\gamma \left(t\right)={\gamma }_0+t^{p_1}{\gamma }_1+t^{p_2}{\gamma }_2$. Again, by convention, $FP2(p,p)$ refers to $\gamma \left(t\right)={\gamma }_0+t^p{\gamma }_1+t^p{\mathrm{log} (t)\ }{\gamma }_2$. For applications see, for example, \cite{Berger.2003}, \cite{Austin.2022}, \cite{Freeman.2022}, or \cite{Sauerbrei.2007}. 

\noindent See Table \ref{tab:S2}, column C.3 to find papers which utilize FP to model time-varying coefficients.

\noindent

\paragraph{Splines} \label{Splines}

\noindent 


\noindent A non-parametric approach for time-varying coefficients can be incorporated through the use of splines. A popular choice are restricted cubic splines. Splines, in general, require the selection of knots that slice the time scale into disjoint intervals. In the case of restricted cubic splines, local cubic polynomials are fitted \cite{Hess.1994}. The term ``restricted'' results from handling the tails linearly, i.e. from $0$ to the first knot and from the last knot to infinity, the function is linear. We denote the corresponding time-varying treatment effect $\gamma \left(t\right)={\gamma }_0+{\gamma }_1t+\sum_d{B_d\left(t\right){\gamma }_{d+1}}$, where $B_d$ denotes the corresponding basis function and the sum iterates over all basis functions. The number of basis functions depends on the number of knots. Applications can be found in \cite{Bolard.2002} for an excess hazard model, \cite{Freeman.2022} in a network meta-analysis and \cite{Austin.2022}.

\noindent 

\noindent The spline approach can be extended to arbitrary choices of basis function. Let $\gamma \left(t\right)=\sum_d{B_d\left(t\right){\gamma }_d}$. The $B_d$ might, for example, be B-spline basis functions of arbitrary high degree and number of knots. The degree and the number of knots determine the number of basis functions or parameters, respectively, for the time-varying treatment coefficient. Overfitting can be avoided by introducing penalties on the flexibility. See, for example, \cite{RodriguezGirondo.2013} who exploited Poisson Likelihood regression, i.e. piecewise exponential hazards (see below), and penalized B-splines for survival model building. Similarly, \cite{Brown.2007} considered penalized B-splines and truncated splines. 
Argyropoulos and Unruh \cite{Argyropoulos.2015} also exploited the Poisson regression approach and include time-varying coefficients and multiple time scales via penalized cubic and thin plate splines. \cite{Giorgi.2003} considered time-varying coefficients via B-splines in an excess hazard model.

\noindent 


\noindent A spline-based time-varying coefficient in a Cox-model setting is available via the R package \texttt{dynsurv}. The R package \texttt{polspline} contains the function hare which utilizes linear splines to model the baseline hazard and covariate effects.

\noindent 

\noindent Column C.4 of Table \ref{tab:S2} shows papers that utilized spline approaches in order to model time-varying coefficients.

\noindent 

\noindent Note, that for none of the aforementioned models, the time-varying coefficient approach is restricted to the treatment effect (or binary variables) but applies to covariates more generally.

\noindent

\paragraph{Average hazard ratio and summary effect measures obtained by the \NEW{weighted} (partial) likelihood} \label{aHR}

\noindent 


\noindent The $aHR$ can be calculated as summary measure over the follow-up period $\tilde{t}$ or some  $t^*\le \tilde{t}$. The summary measure has already been mentioned in Section \ref{KM_est} based on KM estimates of the survival function. Plentiful estimation approaches of $aHR(t^*)$ based on the weighted partial likelihood are proposed in the literature. Consider the (true) model ${\lambda }^{\left(Z\right)}\left(t\right)={\mathrm{exp} \left(Z\gamma \left(t\right)\right)\ {\lambda }_0(t)\ }$. The idea is to estimate ${\lambda }^{\left(Z\right)}\left(t\right)={\mathrm{exp} \left(Z\gamma \right)\ }{\lambda }_0(t)$ instead and interpret $\gamma $ or $\mathrm{exp}\mathrm{}(\gamma )$ as an average treatment effect by introducing weights on the partial likelihood (see, for example, \cite{Boyd.2012}, \cite{Hattori.2012}, or \cite{Schemper.2009}). Results from standard Cox regression depend on the censoring distribution and a weighted Cox regression with inverse probability of censoring weights are proposed. Schemper \cite{Schemper.2009} has accordingly recommended choosing the weighting function $G\left(t\right)=S(t)/C(t)$, where $S(t)$ defines the event time \NEW{survival} function of the complete sample and $C(t)$ denotes the corresponding censoring \NEW{survival function}. For a review of differing definitions of the $aHR$, and different weighting functions \NEW{, as well as an investigation of their performance} see \cite{Schemper.2009}. 

\noindent 

\noindent Such approaches are not restricted to an estimate of $aHR(t^*)$, however. Lin and Le\'on \cite{Lin.2017} obtained adjustment factors based on weights from the log-rank test where $\mathrm{exp}(\gamma)$ is the maximum treatment effect over the course of time. 

\noindent 

\noindent An $aHR$ estimator for the Yang and Prentice model (discussed in Section \ref{shortHR}) was proposed by \cite{Yang.2011}.

\noindent 

\noindent Papers that approached the $aHR$ or similar summary effect measures in a semi-parametric fashion through the weighted partial or pseudo-likelihood can be found in column C.5 (Table \ref{tab:S2}). Note, that we classified non-KM based estimations of the $aHR$, or weighted Cox regression, as time-varying coefficients in Table \ref{tab:S1}. Further notice, that papers which based the estimation of ${aHR(t}^*)$ on KM estimates are not subsumed in column C.5 but column A of Table \ref{tab:S2}.

\noindent 


\noindent The R package \texttt{coxphw}, and the SAS macro WCM provide weighted estimation of Cox regression which might be utilized to estimate the $aHR(t^*)$ with appropriate weights. 

\noindent

\paragraph{Additive models} \label{additive}

\noindent 


\noindent Aalen's additive model also considers time-varying coefficients and NPH \cite{Aalen.2008}. The additive model exploits the martingale representation $M(t)$ of the counting process $N(t)$, i.e. $M_i\left(t\right)=N_i\left(t\right)-\ \int^t_0{Y_i\left(s\right){\lambda }^{\left(Z_i\right)}(s)ds}$. Note that $E\left(dM_i\left(t\right)\right|"history")=0$ or equivalently  ${E\left(dN_i\left(t\right)\right|"history")}$ $={Y}_i\left(t\right){\lambda }^{\left(Z_i\right)}(t)dt$, where $dt\to 0$ and $dN_i\left(t\right)=1$ denotes that the event occurred in an infinitesimal small interval following $t$, and $0$ refers to no event in the aforementioned interval. The hazard rate is assumed to be additive in the predictors, i.e. ${\lambda }^{\left(Z\right)}\left(t\right)=x^T\beta (t)+Z\gamma (t)$. This leads to a twist in the interplay of covariates on absolute and relative differences of the hazard rates as compared to the multiplicative hazard model: keeping everything else constant, the treatment effect in an absolute sense is ${\lambda }^{\left(1\right)}\left(t\right)-{\lambda }^{\left(0\right)}\left(t\right)=\ \gamma (t)$ which is independent of the level of the other covariates, whereas the relative treatment effect is $HR\left(t\right)={{\lambda }^{\left(1\right)}(t)}/{{\lambda }^{\left(0\right)}(t)}={1+\gamma \left(t\right)}/{x^T\beta }$ which does depend on the level of the remaining covariates. If, for example, ${\lambda }^{\left(Z\right)}\left(t\right)={\lambda }_0(t){\mathrm{exp}\mathrm{}(x}^T\beta (t)+Z\gamma (t))$ instead, this is the other way around.

\noindent 

\noindent Instead of optimizing the (partial) likelihood, least squares methodology is utilized. The individual squared error contribution at $t$ is equal to ${(dN}_i\left(t\right)-\ Y_i\left(t\right){\lambda }^{\left(Z_i\right)}(t){dt)}^2\ $, where ${\lambda }^{\left(Z\right)}(t)dt$ is subject to estimation. Note that the estimates of ${\lambda }^{\left(Z\right)}(t)dt$ are not restricted to the interval $[0,1]$ and thus, might not be interpretable as probabilities. 

\noindent 

\noindent The increments ${\lambda }^{\left(Z\right)}(t)dt$ will typically be estimated poorly. Estimates of ${\mathrm{\Lambda }}^{\left(Z\right)}(t)$ can be obtained through $\sum_{i\ge 1:t_{\left(i\right)}\le t}{{\widehat{\lambda }}^{\left(Z\right)}\left(t_{\left(i\right)}\right)dt}$, where the summation over the increments achieves stability in the estimates. (Chapter 4.2.1 in \cite{Aalen.2008})

\noindent 

\noindent Applications of the additive model can be found in \cite{Achilonu.2019} and \cite{Xie.2013}. Dunson and Herring \cite{Dunson.2005} placed a model selection prior on an additive-multiplicative survival model and restrict the additive part to ensure non-negative hazards. Martinussen and Pipper \cite{Martinussen.2013} developed an odds-of-concordance $\frac{P(T^{\left(0\right)}>T^{\left(1\right)})}{P(T^{\left(1\right)}>T^{\left(0\right)})}$ effect measure based on Aalen's additive model, where $T^{\left(Z\right)}$ refers to a survival time of group $Z$. 

\noindent 


\noindent The R package \texttt{timereg} contains the additive model. 

\noindent 

\noindent Papers focusing on the additive model can be found in column C.6 of table \ref{tab:S2}.

\noindent 

\subsubsection{Transformation models with time-covariate interaction} \label{trans}

\noindent 

\noindent In this section we placed the Royston-Parmar model as well as the conditional transformation model (CTM).
Both approaches have in common that time and covariate dependent model quantities are modelled via splines and spline by covariate interaction.
Also both approaches can be motivated as being generalizations of chosen parametric models.
The Royston-Parmar and the CTM are very similar for appropriately chosen reference functions.

The starting point of the Royston-Parmar model is a transformation of the survival function which is denoted by $g(S^{(Z)}(t))$.
Initially, the function is assumed to be linear in the covariates and the treatment indicator, i.e. $g(S^{(Z)}(t)) = g(S_0(t)) + x^T\beta + Z \gamma$.
The point of reference is either the Weibull distribution and PH or the log-logisitc distribution and proportional odds.
We focus on the Weibull case.
Then, $g(S^{(Z)}(t)) = \ln\{-\ln\{S^{(Z)}(t)\}\} = \ln \{ \Lambda_{0}(t)\}+x^T\beta + Z \gamma$.
Note, that the Weibull AFT arises if $\ln \{\Lambda_{0}(t)\}$ is linear in $\ln\{t\}$.
To allow for more flexibility, $\ln \{\Lambda_{0}(t)\}$ is replaced by a restricted (natural) cubic spline function of $\ln\{t\}$, $s(t;\omega) = \sum_d B_d(\ln\{t\}) \omega_d$, where $\omega$ is a parameter vector, and the accelerated failure time interpretation is lost.
In the log-logistic case, the same path is followed in analogy, except that the log-cumulative hazard function is replaced by the log-cumulative odds function, that is the log-odds of an event occurring in the interval $\left( 0,t \right)$.

\noindent NPH (or non-proportional odds) can be incorporated by an interaction of the natural cubic splines and the treatment indicator, i.e.
$s(t;\omega) + s(t,Z;\gamma),$ with  $s(t,Z;\gamma)= \sum_d Z B_d(\ln\{t\}) \gamma_{d}$, where $\gamma$ is a vector of parameters.
Further covariates might be included in $s(t,Z,X;\omega)$ in the same fashion to allow for more non-proportional effects.
For sake of comparison with the conditional transformation model (CTM) we call the functions $s(\cdot)$ transformation functions.

For more detailed information about the Royston-Parmar model see \cite{Royston.2002}. An implementation of the Royston-Parmar model is available through the R package \texttt{flexsurv} \cite{Jackson.2021}. 

\noindent The CTM formulates $S^{\left(Z\right)}\left(t\right)$ as$\ 1-G\left(\sum_d{h_d\left(t,x,Z\right)}\right),$ where $G$ is a chosen cumulative distribution function (cdf).
If $G$ is the minimum extreme value distribution, the Cox model is a special case for appropriately chosen (or estimated) transformation functions $h_d$.
In general, the transformation functions $h_d$ consist of interactions of time and covariates as well as parameters which are subject to estimation. 
The transformation function for the treatment indicator could for example be $h(t,Z) = \sum_d B_d(t)Z \gamma_d$.
M\"ost et al \cite{Most.2015} suggested interactions of penalized B-splines in time $t$ and B-splines or linear basis functions of covariates.

\noindent Consider the case where the reference cdf of the CTM is that of the standard minimum extreme value distribution $G(\cdot)=1-\exp\{-\exp\{\cdot\}\}$ and the Royston-Parmar has the Weibull model as reference as described above.
Moreover, let all covariates be included in the NPH manner as illustrated above for the Royston-Parmar and the CTM.
Then, the main difference between the Royston-Parmar and the CTM as reported here is the choice of basis functions, natural cubic splines vs penalized B-splines as proposed by \cite{Most.2015}, and the time scale on which the basis functions are computed, $\ln\{t\}$ vs $t$.

\noindent Papers that focus on the Royston-Parmar model or the CTM can be found in column F of Table \ref{tab:S2}.
Note that all papers in that column focus on the Royston-Parmar models except for the paper \cite{Most.2015}, which is about the CTM.

\subsubsection{Joint models} \label{joint}

\noindent 


\noindent A NPH model can also be obtained by jointly modeling a longitudinal and a survival outcome. Certain time and treatment-dependent components of the longitudinal model might, for example, be the covariate input of the survival model. Let, $\eta (t,Z)={\gamma }_Lg(t,Z)$ be the conditional expectation of the longitudinal outcome at $t$ (in the absence of further covariates), with $g(t,Z)$ being a chosen function of treatment and time and ${\gamma }_L$ the corresponding coefficient vector in the longitudinal model. Then, the hazard might be modelled as ${\lambda }^{\left(Z\right)}\left(t\right)={\mathrm{ex}\mathrm{p} \left(x^T\beta +Z\gamma +\eta (t,Z){\gamma }_2\right)\ }{\lambda }_0(t)$. 
In this scenario, the treatment has a direct constant impact on the $HR(t)$ through $\gamma$ as well as an indirect time-varying impact through $\eta (t,Z){\gamma }_2$. 
See \cite{Xu.2022} for a model of that kind, that also includes further covariates and random effects in the longitudinal component.

\noindent Articles focusing on joint models can be found in column D of tables \ref{tab:S2}.

\noindent Xu et al \cite{Xu.2022} also discuss the posterior distribution of the $aHR(t^*)$ in a Bayesian setting.

\noindent 


\noindent The package and macro \texttt{JM} provides a software solution for joint modeling of longitudinal and time-to-event data in R and SAS, respectively. The SAS macro is based upon NLMIXED, MIXED, GLIMMIX, and LIFEREG.

\noindent

\subsubsection{Short- and long-term HR}\label{shortHR}

\noindent 

\noindent The methods of Section \ref{time varying}  accounting for a time-varying treatment effect can essentially be considered as time-covariate interaction. Alternatively, parametric functions of the $HR(t)$ can be assumed, where the time-varying effect results from interactions of parameters with time-dependent baseline measures such as $\Lambda_0(t)$, or $S_0(t)$. 
The models in this sub-section differ from fully parametric approaches in that they have non- or semi-parametric components of the baseline measures.
We gathered those models in columns F.1 and F.2 in Table \ref{tab:S2}.

\noindent 


\noindent Such a model was introduced by \cite{Yang.2005} and further studied by \cite{Yang.2015}.
A potentially non-proportional treatment effect can be incorporated by the hazard function 
\begin{align*}
{\lambda }^{\left(Z\right)}\left(t\right)=\ \frac{{\mathrm{exp} \left(x^T\beta \right)\ }}{{\mathrm{exp} \left(-{\gamma }_1Z\right)\ }S^{\left(0\right)}\left(t\right)+{\mathrm{exp} \left(-{\gamma }_2Z\right){(1-S}^{\left(0\right)}\left(t\right)})}{\lambda }_0(t).
\end{align*}
The model is also termed short- and long-term HR model:
Here, $\exp\{\gamma_1\}$ is the short-term and $\exp\{\gamma_2\}$ is the long-term hazard ratio of the treatment variable, respectively.
This model includes strictly increasing and decreasing $HR$ scenarios as well as PH if $\gamma_1=\gamma_2$. Crossing survival curves as well as no initial treatment effect are also sub-models of the Yang and Prentice model \cite{Yang.2005}. The Yang and Prentice model readily delivers an estimate of $HR(t)$.
An estimator of $aHR(t^*)$ has also been established \cite{Yang.2019}. 

\noindent 

\noindent 
A generalization of the PH model has also been considered through 
\begin{align*}
{\lambda }^{\left(Z\right)}\left(t\right)=\ {\lambda }_0\left(t\right){\mathrm{exp} \left \{ x^T\left(\beta_1 +\beta_2 \right)+Z\ \left(\gamma_1 +\gamma_2 \right)+\left({\mathrm{exp} (x^T\beta_2 +Z\ \gamma_2 )\ }-1\right){\mathrm{log} [{\mathrm{\Lambda }}_0(t)]}\right \} },
\end{align*}
where $\beta_2 $ and $\gamma_2 $ are additional model parameters. The PH model is obtained by $\beta_2 =\gamma_2 =0$. The corresponding survival function equals $S^{(Z)}(t) = \exp\left\{ -\exp(x^T \beta_1 + Z \gamma_1) \Lambda_0(t)^{\exp(x^T \beta_2 + Z \gamma_2)} \right\}$. Crossing survival curves are also possible. The baseline hazard function can, for example, be estimated via splines. \cite{Devarajan.2013}

\noindent 
Papers that have the Yang and Prentice model in focus as well as the model introduced by \cite{Devarajan.2013} can be found in Table \ref{tab:S2} column E. 

\noindent 


\noindent R Software packages for the Yang and Prentice model are available: \texttt{YPPE} which estimates the baseline quantities via piecewise exponential distribution \cite{Demarqui.2020b}, \texttt{YPBP} which estimates the baseline distribution via Bernstein polynomials \cite{Demarqui.2020} and YPmodel.

\noindent

\subsubsection{Frailty models}\label{frailty}

\noindent 


\noindent Frailty can introduce NPH on the population level even if the PH assumption holds on the individual level given the unobservable characteristics. Frailty, or unobserved heterogeneity, might be induced by unmeasured or unmeasurable covariates. High frail individuals are prone to ``early'' events due to a high individual or conditional hazard and vice versa. We denote the population hazard as the hazard rate with the individual frailty being ``integrated out'' in this paragraph. Assume an extreme case, with lower population hazard for the treatment group early on but higher population hazard at later stages, as compared to the control group. This is not necessarily a sign of a harming long-term treatment effect. Instead, this could be an indicator of successful treatment as this might be caused by long-term survival of high-frail individuals in the treatment group. The most straightforward frailty model can be expressed as ${\lambda }^{(Z)}\left(t|U\right)=\ U{\mathrm{exp} \left(x^T\beta +Z\ \gamma \right){\lambda }_0(t)\ }$, with the frailty random variable $U\geq 0$. The corresponding population hazard rate is ${\lambda }^{\left(Z\right)}\left(t\right)=\int^{\infty }_0{{\lambda }^{\left(Z\right)}\left(t|u\right)}f_U(u|T\ge t)du$ with $f_U(u|T\geq t)$ being the frailty density of survivors. Analytical expressions for ${\lambda }^{\left(Z\right)}(t)$ are available if $f_U(u)$ is, e.g., the gamma density or another density of the Power Variance Family (see, for example, Chapter 6.2.3 in \cite{Aalen.2008}) among others. See Aalen et al \cite{Aalen.2008} or Wienke \cite{Wienke.2010} for a thorough discussion of individual frailty. A tutorial on frailty models that also discusses the population hazard ratio can be found in \cite{Balan.2020}.

\noindent We took a very broad perspective of frailty in this work and also put, for example, cure-rate models in this category which could be motivated by a binary frailty model \cite{Wienke.2010}. 

\noindent Unobserved heterogeneity has, for example, been considered through semi-parametric transformation models \cite{Choi.2012}, spatially correlated frailty models \cite{Zhang.2011}, time-varying frailty models \cite{Pennell.2006} and random delayed treatment effect two-point cure rate-  \cite{Wu.2022}, as well as responder-no-responder-models \cite{Xu.2020}. 

\noindent Papers having  a frailty perspective are marked in column G of Table \ref{tab:S2}. 


\noindent The SAS procedures phreg and nlmixed, the R packages \texttt{survival}, \texttt{coxme}, \texttt{frailtyEM}, and \texttt{frailtypack} provide frailty models. 

\noindent

\subsubsection{Fully parametric models}\label{parametric}

\noindent 

\noindent For the sake of discussion we separate parametric approaches into four classes: piecewise exponential, accelerated failure time (AFT), first hitting time (FHT) and GLMs.

\noindent

\paragraph{Piecewise exponential hazards} \label{piecewise}

\noindent 


\noindent The stratified piecewise exponential model assumes a constant hazard rate ${\lambda }^{\left(Z\right)}\left(t\right)=\ {\lambda }^{(Z)}_{(k)}$ within the specified interval $t\in [t_{k-1},t_k)$, with k$=1,\dots, K$, $t_k<t_{k+1}$, $t_0=0$ and $t_K=\infty $. Then, the $HR\left(t\right)$ is piecewise constant. Further covariates might be included in the usual PH manner: ${\lambda }^{\left(Z\right)}\left(t\right)={\mathrm{exp} (x^T{\beta }^{\left(Z\right)})\ }{\lambda }^{(Z)}_{(k)}$, where (elements of) ${\beta }^{\left(1\right)}$ might be equal to (their counterparts in) ${\beta }^{\left(0\right)}$ and $t$ is in the $k^{th}$ time interval. The stratified piecewise exponential model approaches the stratified semi-parametric Cox model if the borders of the time intervals are set by the distinct event times. 

\noindent Hagar et al \cite{Hagar.2017} discussed a Bayesian non-proportional multiresolution hazard model. The hazard is piecewise constant and treatment specific. Across partitions, the hazard parameters might be correlated what can be regarded as a smoothening attempt. A second smoothening approach is imposed through the merging of adjacent time intervals if mortality patterns are statistically similar. Constant or time-varying covariate effects might also be added to the piecewise exponential model. 

\noindent Papers that model time variant covariate effects via the piecewise exponential model can be found in Table \ref{tab:S2}, column H.1.


\noindent With respect to fitting procedures, the Poisson regression model (with an adequate offset) and its software implementations can be exploited (see, e.g., \cite{Holford.1980}, \cite{Argyropoulos.2015} or \cite{RodriguezGirondo.2013}). 
The R packages \texttt{eha} and \texttt{pch} contain the piecewise exponential model.
An R software implementation of the multi-resolution hazard model is available with the package \texttt{MRH}.

\noindent

\paragraph{AFT and generalized additive models for location scale and shape} \label{AFT}

\noindent 


\noindent The AFT model assumes a distribution $\pi$ with parameter vector ${\theta }^{\left(Z\right)}$ for the survival time $T$. The parameters in $\theta $ are (partially) dependent on group membership $Z$ and maybe of further covariates. In particular, ${\mathrm{log} \left(T\right)=x^T\beta +Z\ \gamma +\sigma \epsilon \ }$, where the distribution assumption about the error term determines $\pi $. Frequent choices of $\pi $ are the Weibull, Log-logistic, and Log-normal distribution. AFTs usually result in NPH, with the Weibull being a prominent exception. AFTs can be extended by also modelling scale and shape parameters, $\sigma$ in the above example, via link functions and covariates. In the generalized case, even the Weibull model contains NPHs \cite{Burke.2017}. An extension to interval censored data and gamma frailty can be found in \cite{Peng.2020}.

\noindent Umbrella distributions like the generalized gamma \cite{Cox.2007} and the generalized F distribution \cite{Kalbfleisch.2002}, which include ``standard'' distributions as special cases, have also been modelled in the AFT context and its generalizations. Extensions which consider all parameters as function of covariates were also discussed in \cite{Phadnis.2017}. 

\noindent An approach to dimension reduction was introduced by \cite{SpirkoBurns.2021} with a focus on genomic data, utilizing the continuum power regression (CPR) framework. The CPR-step is supposed to obtain a dimension reduction in the covariates and includes OLS, partial least squares, and principal components regression as special cases. Censoring is accounted for by adding the mean residual lifetime on censored observations in the CPR-step. The K ``remaining'' components are then the covariate input of a generalized F or a semiparametric AFT model.  

\noindent Delayed treatment effects \cite{Yoshida.2016} and cure rate models \cite{Hasegawa.2016} were also considered for the Weibull model.

\noindent Papers with a focus on the AFT model or generalizations thereof are marked in column H.2 of Table \ref{tab:S2}.


\noindent A software implementation of the generalized gamma and generalized F distribution and other (user-defined) distributions is available via the R package \texttt{flexsurv} \cite{Jackson.2021}, where also more than one parameter might depend on covariates. The R package \texttt{brms} provides Bayesian parametric survival models. The R package \texttt{spBayesSurv} offers spatial as well as non-spatial Bayesian (generalized) AFTs (and others). The R package \texttt{mpr} can be used to fit, for example, Weibull models where both parameters depend on covariates. The R package \texttt{gamlss.cens} is an add-on package to GAMLSS for the purpose of fitting censored versions of an existing GAMLSS family distribution. The SAS procedure lifereg also offers an implementation of AFT models.

\noindent

\paragraph{First hitting time} \label{FHT}

\noindent 


\noindent First hitting time (FHT) models approach the survival distribution via an underlying unobservable health process $H(t)$. An observable component of the process is the event or, more general a transition into another state. Typically, the event is defined to happen if $H\left(t\right)\le 0$ for the first time. The Wiener process is commonly chosen for $H(t)$. This leads to $T$ being inverse Gaussian distributed. The parameters of the Health process might be a function of covariates and possibly of random effects \cite{Pennell.2010}. Race and Pennell \cite{Race.2021} added random effects utilizing the Dirichlet Process to model subject-specific initial state and drift of the Wiener process. Yu et al \cite{Yu.2009} modeled the drift parameter via cubic B-splines. He et al \cite{He.2015} considered a deterministic decay path where random shocks might cause the event ``prematurely'' in a hip fracture setting. 

\noindent FHT models are marked in column H.3, Table \ref{tab:S2}.

\paragraph{GLMs \& other parametric approaches} \label{GLM}

\noindent 


\noindent If the event indicator is regarded as the target variable, a Poisson GLM with log-link and an adequate offset is equivalent to the piecewise exponential hazard likelihood. Hence, Poisson GLMs have been utilized to fit the piecewise exponential survival model or as an approximation to a general survival likelihood by letting the time intervals for the distinct hazard parameters become small \cite{Argyropoulos.2015}. We mentioned this already in the discussion about time-varying coefficients. We did not classify those papers into the GLM category as they essentially modeled the common survival likelihood via GLM routines and hence, we feel, those papers are more adequately categorized into other classes, time-varying coefficients for example.

\noindent Others, however, took a more genuine GLM approach. Among them \cite{Mandel.2013} who exploited the longitudinal nature of multiple sclerosis data-set to model transitions in the disability progression. Motivated by oncology studies with NPH, where events are observed through periodic screenings, \cite{Nguyen.2012} considered discrete time. Consequently, GLMs at the distinct time points including only at-risk individuals at the corresponding follow-up time $t_{(j)}$, i.e. $Y_i\left(t_{(j)}\right)=1$,  are suggested. 

\noindent GLMs might be incorporated into other methods, trees, for example (see Section \ref{ML}). 

\noindent Other parametric approaches including GLMs can be found in column H.4, Table \ref{tab:S2}.


\noindent GLM routines from standard statistical software can be used, for example, the SAS procedure glm.

\noindent

\subsubsection{Machine learning approaches }\label{ML}

\noindent 


\noindent Wey et al \cite{Wey.2016} suggested an average survival model which is a weighted sum of parametric, semi-parametric, and non-parametric models. The weights in the suggested model are obtained through the minimization of a loss function. Distinct treatment group survival curves are obtained by ``averaging out'' the remaining covariates or confounders respectively. Inference regarding the treatment effect is then drawn from the difference in RMST of the treatment and the control group.

\noindent Lowsky et al \cite{Lowsky.2013} discussed a K-nearest neighbor approach for estimating survival curves via weighted Kaplan-Meier. The K-nearest neighbors of a new observation are determined by its distance to the covariate vectors in the data set. Then, a weighted KM curve is estimated. The weights for each observation are reciprocal to the distance to the new covariate vector and affect the number of deaths as well as the size of the risk set at each event time. This procedure could, for example, be stratified by the treatment indicator in order to quantify the treatment effect given the remaining covariates.

\noindent The former two approaches can be found in column I.4, Table \ref{tab:S2}.

\noindent Alternative approaches include trees and survival forests. Papers concerning trees and forests are marked in column I.1, Table \ref{tab:S2}. The tree-based method by \cite{Xu.2002} to find change points for time-varying coefficients has already been discussed above. 

\noindent A Bayesian additive regression tree (BART) for survival data with NPH was considered by \cite{Sparapani.2016}. The Likelihood is specified via probit regression at the distinct event times. The probability of an event, conditional on no previous events, at a given event time is derived via the BART. The BART is an ensemble of trees where the splitting criteria at the internal nodes within each tree are set by the event time and covariates. For a given covariate-time input the function value at the terminal node is then summed up over each tree. This estimate is used as the input of the standard normal cdf to obtain the probability of an event.  

\noindent Soft BART (SBART) was considered for interval-censored data by \cite{Basak.2021}. The SBART extends the BART with a smooth regression function and better ability to remove irrelevant predictors.

\noindent Survival forests as well as an improved splitting criterion were discussed in \cite{Korepanova.2020}. A survival forest for a joint model was developed by \cite{Lin.2021}.

\noindent Neural networks for censored survival data were discussed by \cite{Trinh.2003} and \cite{Laurentiis.1994} . 
Neural networks are marked in column I.2, Table \ref{tab:S2}.
Kernel smoothing based approaches can be found in \cite{Cheng.2009} regarding the hazard ratio and \cite{Zhao.2019} with respect to differences in survival rates. Kernel smoothing based approaches are marked in column I.3, Table \ref{tab:S2}.


\noindent The R package \texttt{trtf} contains transformation trees and forests that can be utilized for time-to-event analysis. The SBART for discrete time-to-event analysis is implemented in the R package \texttt{BART}. The R package \texttt{mboost} includes a gradient boosting algorithm for right-censored data.

\noindent 

\subsubsection{Other approaches}\label{other_estimation}

\noindent 

 \noindent Some papers were difficult to categorize into one of the previous sections.

\noindent Among the papers in this section is \cite{Latminer.2018}, which discussed inverse probability of censoring weights (IPCW) approaches, the structural nested model (SNM), and the rank preserving structural failure time model (RPSFTM) in the context of treatment switching.
Similarly to the SNM, the RPSFTM assumes that the treatment decreases or increases the survival time by the factor $\gamma>0$.
More precisely, $T^{(1)}=\gamma T^{(0)}$ and $\gamma>1$ indicates a beneficial treatment effect.
With that factor, counterfactual survival times are computed, i.e. $u_i=(\text{time $i$ spent in control group}) + \frac{\text{time $i$ spent in treatment group}}{\gamma}$.
The parameter $\gamma$ itself is found via a grid search, where the optimization criterion is a test statistic that compares the estimated survival curves of the placebo and the treatment group, where for the latter the counterfactual survival times are utilized. 
The value for $\gamma$ that makes the two groups most alike in the sense of the test statistic is the point estimate. The R package \texttt{rpsftm} provides a software solution for the RPSFTM.
The IPCW approach, where the weights might be utilized to compute KM curves or weighted partial likelihood estimates, attempts to account for informative censoring.

\noindent Further proposals are a semi-parametric proportional likelihood ratio model \cite{Zhu.2014},  and concordance regression, where, brought into the two-sample setting, the likelihood is based upon $P(T^{(1)}>T^{(0)})$ \cite{Dunkler.2010}. A Bayesian non-parametric dependent Dirichlet process for modeling the time-to-event distribution was studied by \cite{Iorio.2009}. Chen and Wang \cite{Chen.2010} apply an accelerated hazard model, where the hazard is equal to $\lambda^{(Z)}(t) = \lambda_0(t \exp\{Z \gamma + x^T\beta\})$, and the baseline hazard is a smoothed non-parametric estimate. The authors suggest that the accelerated hazard model might be a good choice if hazard rates are similar after the start of follow-up but go apart due to different ageing processes. 

\noindent In column J, Table \ref{tab:S2}, we marked approaches that did not properly fit into other categories.

\newpage
\subsection{Hypothesis tests for equality of survival curves}\label{suppl.test}
\subsubsection{Log-rank tests} \label{logrank}

\noindent 

\noindent The standard log-rank test statistic \cite{Peto.1972, Mantel.1966} is the most widely used statistical test to compare the overall survival of two groups. The log-rank test is defined as  

\noindent 
\[M_w=\ \frac{\sum_{i=1}^{D}{w(t_{(i)})\big[d_{i1}-Y^{(1)}(t_{(i)})\frac{d_i}{Y(t_{(i)})}\big]}}{{\big[\sum_{i=1}^D {w(t_{(i)})^2 \frac{Y^{(1)}(t_{(i)})}{Y(t_{(i)})}(1-\frac{Y^{(1)}(t_{(i)})}{Y(t_{(i)})})\frac{Y(t_{(i)})-d_i}{Y(t_{(i)})-1}d_i\big]\ \ }}^{1/2}},\] 
with the weight function $w(t_{(i)})=1$.
The number of distinct event times of the pooled sample is denoted by $D$, $d_{i}$ is the number of events at $t_{(i)}$, and $d_{i1}$ refers to the number of events at $t_{(i)}$ in the experimental treatment group \cite{Klein.2003}.

\noindent As the shape of the survival curve influences the power of the test, multiple proposals for the weight function $w(t)$ sensitive to particular NPH patterns are available \cite{Royston.2020}.

\noindent Royston and Parmar (2020) \cite{Royston.2020} give an overview of the weight functions used for different NPH patterns, e.g. the Fleming-Harrington weight function $w(t_{(i)})=\hat{S}{\left(t^{-}_{(i)}\right)}^1{\left(1-\hat{S}\left(t^{-}_{(i)}\right)\right)}^0$, denoted $G^{1,0}$, or $w(t_{(i)})=\hat{S}{\left(t^{-}_{(i)}\right)}^0{\left(1-\hat{S}\left(t^-_{(i)}\right)\right)}^1$, denoted $G^{0,1}$, for early or late effects, respectively. Note that $t^{-}$ denotes the time just before time $t$. For testing that survival in the treatment group is stochastically less than or equal to survival in the control arm ($H_0:S_0\left(t\right)\le S_1\left(t\right)\ \forall \ t\ge 0$  ), Magirr and Burman (2019) \cite{Magirr.2019} propose \textit{modestly weighted log-rank} tests. It is a variation of the weighted log-rank test which under arbitrary weights does not control the risk to conclude that a new treatment is more efficacious than standard care when it is uniformly inferior in terms of the survival function. Treatment may be uniformly inferior in terms of the survival function, but still, there may be time points at which the treatment has a favorable hazard. The log-rank test works at the level of the hazard function, so if enough weight is put on the possibly small time interval with a favorable hazard, the treatment is declared significantly better than the control, even though this local benefit does not translate into any survival benefit. The modestly weighted test is constructed in such a way, though, that this fallacy can never happen. The weights are set to $w(t_{(i)})=\frac{1}{\mathrm{max}\mathrm{}\{\hat{S}\left(t^{-}_{(i)}\right),\hat{S}\left(t^*\right)\}}$ with ~$\hat{S}\left(t^*\right)$ denoting the KM estimate at a certain time $t^*$ based on the pooled data from both treatment arms. The choice of $t^*$ is a trade-off. A bigger value of $t^*$ results in lower weights for early events, which is useful for delayed effects, on the other hand too large $t^*$ will lead to unnecessarily high weights $w(t_{(i)})$ for late events.

\noindent 

\noindent Sample size formulas for different weighted log-rank tests are given by e.g. Yung and Liu (2020) \cite{Yung.2020} and their R package \texttt{npsurvSS}. Wei and Wu (2020) \cite{Wei.2020} and Wu and Wei (2022) \cite{Wu.2022}, Ye and Yu (2018) \cite{Ye.2018} provide R-code for their sample size formulas derived from weighted log-rank tests for cancer immunotherapy trials with delayed treatment effects.

\noindent 

\noindent Various authors also investigated the use of weighted log-rank tests in group-sequential trial designs. Group-sequential trial designs plan interim analyses at pre-specified time points. The interim analysis can lead to early efficacy or futility stopping because of either sufficiently convincing results or a further investigation not being justifiable.

\noindent The use of weighted log-rank test statistics in group-sequential designs can lead to a misspecified covariance of the test statistic due to the incorrectly estimated information fraction. The information fraction is not proportional to the number of interim events for weighted log-rank tests in general, e.g. late events on which the weight function usually places more weight in delayed treatment effect settings might be unavailable during interim analyses. Interim analyses are argued to be only sensible if a fixed time horizon for the final (primary) analysis is specified and if sufficient information up to the time horizon is available for the  interim analysis \cite{vanHouwelingen.2005}.

\noindent Brummel and Gillen (2014) \cite{Brummel.2014} focus on monitoring the weighted log-rank test statistic in group-sequential designs where information growth is nonlinear and propose using a constrained boundaries approach to maintain the planned operating characteristics of a group-sequential design. Hasegawa (2016) \cite{Hasegawa.2016} proposes a semiparametric  information fraction for group-sequential designs with delayed treatment effects. Kundu and Sarkar (2021) \cite{Kundu.2021} focus on the deviation of information fractions in weighted log-rank test from that of standard log-rank test and propose a decomposition of effects on information fractions to provide a reasonable and practically feasible range of information fractions to work with.

\noindent Li et al (2021) \cite{Li.2021} propose a group-sequential design based on the piecewise log-rank test,  Zhang and Pulkstenis (2016) \cite{Zhang.2016} provide closed-form solutions for the power and sample size calculation for group-sequential designs and Magirr and Jim\'{e}nez (2022) \cite{Magirr.2022} give practical guidance for the use of \textit{modestly weighted log-rank tests} in group-sequential trials.

\noindent 

\noindent \textit{}

\subsubsection{Kaplan-Meier based tests} \label{KM_test}

\noindent 

\noindent Kaplan-Meier-based tests are another class of tests for testing the null hypothesis of equal survival in the two treatment groups. These tests are based on KM estimates or restricted mean survival time (RMST). 

\noindent Weighted Kaplan-Meier tests are based on the weighted sum of the differences of the KM estimates of the survival curves. Uno et al (2015) \cite{Uno.2015} proposed a weighted Kaplan-Meier test with weights proportional to the observed standardized difference of the estimated survival curves at each time point. 

\noindent The unweighted Kaplan Meier test results in the difference between two RMSTs describing the mean event-free survival time up to a pre-defined time point $t^*$  \cite{Dormuth.2022}. Since equality of survival curves implies equal RMST, we can also test for the difference in RMST between treatment groups $\mathrm{\Delta }\left(t^*\right)=RMST^{\left(1\right)}\mathrm{(}{\mathrm{t}}^{\mathrm{*}}\mathrm{)}\mathrm{-}RMST^{\left(0\right)}\mathrm{(}{\mathrm{t}}^{\mathrm{*}}\mathrm{)=}\int^{t^*}_0{S^{(1)}\left(t\right)dt-\int^{t^*}_0{S^{(0)}\left(t\right)dt}}$  being zero, $H_0:\mathrm{\Delta }\left(t^*\right)=0$. The null hypothesis can be tested using the Wald statistic $M_{RMST}\left(t^*\right)=\sqrt{n}\mathrm{\ \ }\widehat{\mathrm{\Delta }}\left(t^*\right)/\widehat{\sigma }(t^*)$, with $\widehat{\sigma }\left(t^*\right)$ denoting the variance of  $\sqrt{n}\mathrm{\ }\mathrm{\{}\mathrm{\ }\widehat{\mathrm{\Delta }}\left(t^*\right)-\mathrm{\Delta }\left(t^*\right)\ \}$. Tests based on RMST do not rely on the PH assumption but are also not specifically designed to detect crossing survival curves \cite{Dormuth.2022}.

\noindent Test procedures based on RMST are proposed by e.g. Horiguchi et al (2018) \cite{Horiguchi.2018}, Lawrence et al (2019) \cite{Lawrence.2019} and Sun et al (2018) \cite{Sun.2018}.

\noindent The pre-specified time point $t^*\ $for RMST is selected data dependently in Horiguchi et al (2018). Therefore, a set of potential times $t^*=\{t^*_1,\dots,t^*_K\}$ with a fixed number K is assumed. The null hypothesis that there is no difference between 2 event time distributions against a two-sided alternative $\mathrm{\Delta }\left(t^*_k\right)\neq 0$ is tested with the test statistic $M_{RMST2}=\ {\mathop{\mathrm{max}}_{t\in t^*} |M_{RMST}\left(t^*\right)|\ }$. The distribution under the null hypothesis is obtained using a wild bootstrap procedure. The approach of Horiguchi et al (2018) \cite{Horiguchi.2018} is available in the R package \texttt{survRM2adapt}. For cure rate survival models Sun et al (2018) \cite{Sun.2018} compare different tests for cure rate survival data and showed in a simulation study that Kaplan-Meier-based tests (RMST test and weighted Kaplan Meier test) perform best among the considered test, e.g. log-rank, Wilcoxon rank test.

\noindent 

\noindent Rauch et al  \cite{Rauch.2018} compared two test statistics for the average hazard ratio $aHR\ $testing the null hypothesis $H_0:aHR\mathrm{\ }\ge 1$ to the standard log-rank test using the hazard ratio in settings with different underlying event times and censoring distributions. The two test statistics used for testing the average hazard ratio differ in their independent increments property. The comparison showed the advantage of the average hazard ratio tests in terms of power in NPH settings. 

\noindent Window mean survival time proposed by Paukner and Chappell  \cite{Paukner.2021} keeps the interpretability of RMST and unweighted log-rank tests and improves the power to detect differences in survival curves under NPH caused by late crossing or diverging curves. The difference in window mean survival time of the two treatment groups is the area between the two survival curves from $t\ =\ t^*_1$ to $t\ =\ t^*_2$, with $0\ \le t^*_1\ <\ t^*_2\ \ \le \tilde{t}$. The null hypothesis is that the difference in window mean survival time is zero. The test statistic is calculated by the ratio of the estimated difference in window mean survival time and its estimated variance. The simulation study of Paukner and Chappell \cite{Paukner.2021} showed that the test of window mean survival time has higher power compared to the weighted log-rank test if the PH assumption holds. 

\noindent 

\noindent Sample size formulas based on the RMST test are provided by e.g. Tang \cite{Tang.2022}, Royston and Parmar \cite{Royston.2013}. Yung and Liu \cite{Yung.2020} provide a R package \texttt{npsurvSS} for sample size and power calculations based on Kaplan-Meier-based tests.

\noindent 

\noindent Br\"{u}ckner and Brannath (2017) developed group-sequential designs for the hazard ratio and proved that the sequential tests based on the average hazard ratio are asymptotically multivariate normal with independent increments property.

\noindent An approach using the weighted Kaplan-Maier test for the calculation of stage-wise p-values in adaptive survival trials allowing to use discrete surrogate information for the interim analysis while controlling the type I error rate was proposed by Br\"{u}ckner et al \cite{Bruckner.2018}.  Sample size re-estimation using Kaplan-Meier-based tests is investigated by Wang \cite{Wang.2022}.

\noindent \textit{}

\subsubsection{Combination tests} \label{combination}

\noindent 

\noindent Combination tests combine tests within a class or across classes of tests. The idea underlying the combination of tests is the difficulty to predict the existence and severity of NPH caused by e.g. delayed treatment effects. Combination tests allow covering various scenarios. The maximum combination (max combo) test is an example of such combination tests and is defined as the maximum of several weighted log-rank test statistics. Using the Fleming-Harrington weighted log-rank test denoted $Z_{G^{\rho ,\ \gamma }}$ the max combo test is defined as 
\[Z_{max}={max}_{\rho ,\gamma }\{Z_{G^{{\rho }_1,\ {\gamma }_1}},Z_{G^{{\rho }_2,\ {\gamma }_2}},\ \dots ,Z_{G^{{\rho }_K,\ {\gamma }_K}}\}\ \] 
where  $Z_{G^{{\rho }_k,\ {\gamma }_k}}$ denotes one of K different weighted log-rank tests. However, the max combo test is not restricted to the Fleming-Harrington weight function in the log-rank tests. The p-value of the maximum combination test can be calculated based on the multivariate normal distribution. Ghosh et al \cite{Ghosh.2022} developed  group sequential designs using two (\textit{modestly) weighted log-rank tests }for the max combo test statistic. Ristl et al \cite{Ristl.2021} investigated different sources of non-proportionality such as e.g. delayed treatment effect, disease progression, predictive biomarker subgroups, treatment switch after progression, and their effect on the power of weighted log-rank tests and maximum combination tests. Ristl et al \cite{Ristl.2021} provide the R package \texttt{nph} to perform the statistical tests and to simulate survival data.

\noindent Sample size procedure for maximum combination tests are e.g. available in Tang (2021).

\noindent The  use of combination tests specifically the max combo test in group sequential trial designs was investigated e.g. in Li et al \cite{Li.2022}, Wang et al \cite{Wang.2021}, Prior \cite{Prior.2020}. Li et al \cite{Li.2022} investigated obtaining the group sequential boundaries and the empirical power by simulation procedures for delayed treatment effects, whereas the approach of Wang et al \cite{Wang.2021} with an R-package \texttt{GSMC} available on GitHub is simulation free. Prior \cite{Prior.2020} investigated the use of different weighting functions in the maximum combination across the time points allowing flexibility to the accrued data. 

\noindent Combination tests are not only restricted to the weighted log-rank test but can also involve other classes of tests, e.g. Royston and Parmar \cite{Royston.2016} combine the Cox test with a test of the RMST difference by obtaining the p-value of the combination test via selecting the smallest p-value of the single tests. The Cox test is based on the difference in log partial likelihoods of the Cox PH model with the binary treatment indicator as only covariate. It is closely equivalent to the standard log-rank test \cite{Royston.2016}. Le\'{o}n et al \cite{Leon.2020} combine weighted log-rank tests with the RMST test. Chi and Tsai  \cite{Chi.2001} propose the combination of weighted log-rank tests with weighted Kaplan-Meier tests. Zhang et al  \cite{Zhang.2021} propose a Cauchy combination test of multiple single change-point (CauchyCP) regression models.

\noindent

\subsubsection{Other tests} \label{other_test}

\noindent 

\noindent Besides the Kaplan-Meier-based test, the log-rank tests, and the combination approaches, we identified also articles that proposed hypothesis testing methods not fitting in either of these classes. For instance, Gorfine et al \cite{Gorfine.2020} proposed a test for $K$ groups based on sample--space partitions, which is implemented in the R package \texttt{KONPsurv}.

\noindent A modification of the Kolmogorov Smirnov test was proposed by Fleming et al \cite{Fleming.1980} who compare their proposal to the standard log-rank test and the Wilcoxon rank test for censored observations \cite{Gehan.1965} and showed higher power under NPH. 

\noindent Sooriyarachchi and Whitehead (1998) propose a binary method for testing whether the survival curves of two treatment groups are equal. This approach needs the discretization of the time. The time intervals underlying the discrete time should include equal numbers of events. The effect measuring the treatment difference is the log odds ratio of the probability surviving past time point $t^*$ in the two treatment groups. The test statistic is derived from the log-likelihood of the log odds ratio and the nuisance parameter of the probabilities. 

\noindent Permutation procedures can be used to obtain the distribution of the test statistic under the null hypothesis of equal survival functions. For combinations approaches permutation approaches were suggested by e.g. Brendel et al (2014) \cite{Brendel.2014}, Royston and Parmar (2016) \cite{Royston.2016}, Ditzhaus and Friedrich (2020) \cite{Ditzhaus.2020}, Ditzhaus et al (2021) \cite{Ditzhaus.2021} and Ditzhaus and Pauly (2019) \cite{Ditzhaus.2019}. 
The approaches \cite{Brendel.2014, Ditzhaus.2019, Ditzhaus.2020} suggest a combination of weighted log-rank statistics targeting a comprehensive range of alternatives. These tests can be applied with the R-package \texttt{mdir.logrank}. Ditzhaus et al (2021) \cite{Ditzhaus.2021}  provide the R package \texttt{GFDsurv} for their proposed approach. 

\noindent For the application of newly proposed methods providing the corresponding software is of advantage. In some settings, numerical aspects are important in the development of such software. For most existing test statistics, Riemann integration is used. However, under complex NPH pattern involving high dimensional numerical integration this approach might not be feasible. Therefore, Tang \cite{Tang.2022} proposes a sample size and power calculation method for log-rank tests and RMST tests via product integration and provides sample SAS code as online supplementary material.

\begin{landscape}

\subsection{Classification of methods proposed in selected articles}\label{S.TableS1}

\begin{longtable}{|p{0.2in}|p{0.3in}|p{0.41in}|p{0.4in}|p{0.3in}|p{0.31in}|p{0.3in}|p{0.3in}|p{0.375in}|p{0.3in}|p{0.3in}|p{0.35in}|p{0.35in}|p{0.4in}|p{0.3in}|}
\caption{Classification of methods proposed in selected articles. A description of the methods can be found in Section \ref{estimation} and Section \ref{test}. The corresponding references for the IDs are given in \ref{tab:S3}}
\label{tab:S1}
\\
\hline \textbf{} & \multicolumn{4}{|p{1.2in}|}{\textbf{Hypothesis Tests}} & \multicolumn{10}{|p{4in}|}{\textbf{Effect Estimation and Regression Models}} \\\hline 
  ID & 
  Log\-rank tests & Kaplan-Meier (KM) based tests & 
  Combi\-nation tests & Other tests & 
  KM based estimation & Strat\-ified Cox & 
  Time vary\-ing coefficients & Trans\-for\-mation models & 
  Short- and long-term HR & 
  Joint models & Frailty models & 
  Para\-metric Models & Ma\-chine Learning & 
  Other \\ 
  \hline
 1 & x &  &  &  &  &  &  &  &  &  &  &  &  &  \\ \hline 
 9 & x & x & x & x &  &  &  &  &  &  &  &  &  &  \\ \hline 
 14 &  &  &  &  &  & x & x & x &  &  &  & x &  &   \\ \hline 
 16 &  &  &  &  & x &  &  &  &  &  &  &  &  &  \\  \hline
 19 &  &  & x &  &  &  &  &  &  &  &  &  &  &  \\  \hline
 20 &  &  &  &  &  &  & x &  &  &  &  &  &  &  \\  \hline
 21 &  &  &  &  &  &  &  & x &  &  & \NEW{ } &  &  &  \\  \hline
 22 & x & x &  &  &  &  &  &  &  &  &  &  &  &  \\  \hline
 26 &  &  &  &  & x &  &  &  &  &  &  &  &  &  \\  \hline
 27 &  &  &  &  &  &  & x &  &  &  &  &  &  &  \\  \hline
 28 &  &  & x &  &  &  &  &  &  &  &  &  &  &  \\  \hline
 30 & x &  &  &  &  &  & x &  &  &  & x &  &  &  \\  \hline
 39 & x &  & x &  &  &  &  &  &  &  &  &  &  &  \\  \hline
 40 & x &  &  &  &  &  &  &  &  &  &  &  &  &  \\  \hline
 42 &  &  &  &  &  &  &  &  &  &  &  & x &  &  \\  \hline
 46 &  &  & x &  &  &  & x &  &  &  &  &  &  &   \\  \hline
 53 &  &  &  & x &  &  &  &   &   &   &  &  &  &   \\  \hline
 56 &  & x &  &  & x &  &  &  &  &  &  &  &  &  \\  \hline
 67 &  &  &  &  & x &  &  &  &  &  &  &  &  &  \\  \hline
 69 & x & x & x &  & x &  &  &  & x &  & x &  &  &  \\  \hline
 72 & x &  &  &  &  &  &  &  &  &  &  &  &  &  \\  \hline
 75 &  &  &  &  &  &  &  &  &  &  & x &  & x &  \\  \hline
 76 &  &  &  &  &  &  &  &   &   &   & x &  &  &   \\  \hline
 79 & x &  & x &  &  &  &  &  &  &  &  &  &  &  \\  \hline
 81 &  &  & x &  &  &  &  &  &  &  &  &  &  &  \\  \hline
 83 & x &  &  &  &  &  &  &  &  &  &  &  &  &  \\  \hline
 84 & x &  &  &  &  &  &  &  &  &  &  &  &  &  \\  \hline
 86 & x &  &  &  &  &  &  & x &  &  &  &  &  &   \\  \hline
 89 & x &  &  &  &  &  &  &  &  &  &  &  &  &  \\  \hline
 91 & x &  &  &  &  &  & x &  &  &  &  &  &  &  \\  \hline
92 &  &  &  &  &  &  &  &  &  &  &  & x &  &  \\  \hline
 95 & x &  &  &  &  &  &  &  &  &  &  &  &  &  \\  \hline
 96 &  &  &  & x &  &  &  &  &  &  &  &  &  &  \\  \hline
  97 &  &  &  & x & x &  &  &  &  &  &  &  &  &  \\  \hline
  103 &  &  &  &  &  &  &  &  &  & x &  &  &  &  \\  \hline
  110 &  &  &  &  &  &  & x &  &  &  & x &  &  &  \\  \hline
  113 & x &  &  &  &  &  &  &  &  &  &  &  &  &  \\  \hline
  116 & x &  &  &  &  &  &  &  &  &  &  &  &  &  \\  \hline
  118 &  & x &  &  &  &  &  &  &  &  & x & x &  &  \\  \hline
  119 & x &  & x &  &  &  &  &  &  &  &  &  &  &  \\  \hline
  122 &  &  &  &  &  &  &  &  &  &  &  & x &  &  \\  \hline
   123 &  &  &  &  &  &  &  &  &  & x &  &  & x &  \\  \hline
   126 &  &  &  &  & x &  &  &  &  &  &  &  &  &  \\  \hline
   134 &  &  &  & x &  &  &  &  &  &  &  &  &  &  \\  \hline
   136 & x &  & x &  &  &  &  &  &  &  &  &  &  &  \\  \hline
   142 &  &  &  &  &  &  &  &  &  &  &  & x &  &  \\  \hline
   145 & x &  & x &  &  &  &  &  &  &  &  &  &  &  \\  \hline
   148 & x &  &  &  &  &  &  &  &  &  &  &  &  &  \\  \hline
   149 &  &  &  & x &  &  &  &  &  &  & x &  &  &  \\  \hline
   154 & x &  &  &  &  &  &  &  &  &  &  &  &  &  \\  \hline
   157 &  &  &  &  &  &  &  &  &  &  &  & x &  &  \\  \hline
   159 & x &  &  &  &  &  &  &  &  &  &  &  &  &  \\  \hline
   162 &  &  &  & x &  &  &  &  &  &  &  &  &  &  \\  \hline
   167 & x & x &  &  &  &  &  &  &  &  &  &  &  &  \\  \hline
   169 & x &  &  &  &  &  & x &  &  &  & x &  &  &  \\  \hline
   174 & x &  &  &  &  &  &  &  &  &  &  &  &  &  \\  \hline
   176 &  &  &  &  &  &  &  &  &  &  &  & x &  &  \\  \hline
   184 &  &  &  &  &  &  &  &  &  &  &  &  & x &  \\  \hline
   188 &  &  &  &  &  &  & x &  &  &  &  &  &  &  \\  \hline
   195 &  &  &  &  &  &  &  &  &  &  &  &  & x &  \\  \hline
   201 & x &  &  &  &  &  &  &  &  &  &  &  &  &  \\  \hline
   207 &  &  &  &  &  &  & x &  &  &  &  &  &  &   \\  \hline
   213 &  & x &  &  &  &  &  &  &  &  &  &  &  &  \\  \hline
   235 &  &  &  & x &  &  &  &   &   &   &  &  &  & x \\  \hline
   242 & x &  &  &  &  &  &  &  &  &  &  &  &  &  \\  \hline
   244 & x &  &  &  &  &  &  &  &  &  &  &  &  &  \\  \hline
   245 & x &  &  &  &  &  &  &  &  &  &  &  &  &  \\  \hline
   247 &  & x &  &  & x &  &  &  &  &  &  &  &  &  \\  \hline
   251 & x &  &  &  &  &  &  &  &  &  &  &  &  &  \\  \hline
   256 & x &  &  &  &  &  &  &  &  &  &  &  &  &  \\  \hline
   265 &  & x &  &  & x &  &  &  &  &  &  &  &  &  \\  \hline
   266 &  &  &  &  &  &  & x &  &  &  &  &  &  &  \\  \hline
   267 &  & x &  &  &  &  &  &  &  &  &  &  &  &  \\  \hline
   268 &  & x & x &  &  &  & x &  & x &  &  &  &  &  \\  \hline
   269 & x &  &  &  &  &  &  &  & x &  &  &  &  &  \\  \hline
   274 &  &  &  &  &  &  &  &  &  &  &  &  &  & x \\  \hline
   284 &  &  &  &  & x &  &  &  &  &  &  &  &  &  \\  \hline
   296 & x & x &  &  &  &  &  &  &  &  &  &  &  &  \\  \hline
   313 &  &  &  &  &  &  &  &  &  &  &  & x &  &  \\  \hline
   328 &  &  &  &  & x &  &  &  &  &  &  &  &  &  \\  \hline
   335 &  &  &  &  &  &  &  &  &  &  &  & x &  &  \\  \hline
   343 &  &  &  &  &  &  &  &  &  &  &  & x &  &  \\  \hline
   347 & x &  &  &  &  &  &  &  &  &  &  &  &  &  \\  \hline
   359 & x &  &  &  &  &  & x &  &  &  & x & x &  &  \\  \hline
  360 &  &  &  &  &  &  & x &  &  &  &  &  &  &  \\  \hline
   365 &  &  &  &  & x &  &  &   &   &   &  &  &  &   \\  \hline
   367 &  &  &  &  &  &  & x &  &  &  &  & x &  &   \\  \hline
   376 &  & x &  &  & x &  &  &   &   &   &  &  &  &   \\  \hline
  378 &  &  &  &  &  &  &  &  &  &  &  &  & x &  \\  \hline
   382 &  &  & x &  &  &  &  &  &  &  &  &  &  &  \\  \hline
   383 &  &  &  &  &  &  &  &  &  &  &  &  & x &  \\  \hline
   384 & x &  &  &  &  &  &  &  &  &  &  &  &  &  \\  \hline
   392 & x &  &  &  &   &   &   &   &   &   &   &   &   &   \\  \hline
   396 &  &  &  &  &  &  &  &  &  &  &  & x &  &  \\  \hline
   397 & x &  &  &  &  &  &  &  &  &  &  & x &  &  \\  \hline
   402 &  & x &  &  &  &  &  &  &  &  &  &  &  &  \\  \hline
   404 &  &  &  &  &  &  &  & x &  &  &  &  &  &  \\  \hline
   413 & x &  &  &  &   &   &   &   &   &   &   &   &  &   \\  \hline
   417 &  &  &  &  &  &  & x &  &  &  &  &  &  &  \\  \hline
   422 &  &  &  &  &  &  & x &  &  &  &  &  &  &  \\  \hline
   424 &  &  &  &  &  &  &  & x &  &  &  &  &  &  \\  \hline
   425 &  &  &  &  &  &  &  &  & x &  &  &  &  &  \\  \hline
   427 & x &  & x &  &  &  &  &  &  &  &  &  &  &  \\  \hline
   433 &  &  &  &  &  &  &  &  &  &  &  &  &  & x \\  \hline
   435 &  &  &  &  &  &  &  &  &  &  &  & x &  &  \\  \hline
   444 & x &  & x &  &  &  &  &  &  &  &  &  &  &  \\  \hline
   445 &  &  &  &  & x &  &  &  &  &  &  &  &  &  \\  \hline
   446 &  & x &  &  &  &  &  &   &   &   &  & x &  &   \\  \hline
   452 &  &  &  &  &  &  &  &  &  &  &  & x &  &  \\  \hline
   453 &  &  &  &  &  &  &  &  & x &  &  &  &  &  \\  \hline
   456 &  &  &  &  &  &  & x &  &  &  &  &  &  &  \\  \hline
   464 &  &  &  &  &  &  &  &  &  &  &  &  &  & x \\  \hline
   465 & x &  &  &  &  &  &  &  &  &  &  &  &  &  \\  \hline
   467 & x &  &  &  &  &  &  &  &  &  &  &  &  &  \\  \hline
   469 &  &  &  &  &  &  &  &  & x &  &  &  &  &  \\  \hline
   470 &  & x &  &  &  &  &  &  &  &  &  & x &  &  \\  \hline
   482 &  &  &  &  &  &  & x &  &  &  &  &  &  &  \\  \hline
   501 &  &  &  &  & x &  &  &  &  &  &  &  & x &  \\  \hline
   510 &  &  &  &  &  &  &  &   &   &   &  & x &  &   \\  \hline
   511 & x &  &  &  &  &  &  &  &  &  &  &  &  &  \\  \hline
   513 &  &  &  &  &  &  & x &  &  &  &  &  &  &  \\  \hline
   514 &  &  &  &  &  &  &  &  &  &  &  & x &  &  \\  \hline
   518 &  &  &  &  &  &  &  &  &  & x &  &  &  &  \\  \hline
   520 & x &  &  &  &   &   &   &   &   &   &   &   &  &   \\  \hline
   524 &  &  &  &  &  &  & x &  &  &  &  &  &  &   \\  \hline
   525 &  &  &  &  &  &  &  & x &  &  &  &  &  &  \\  \hline
   528 &  &  &  &  &  &  & x &  &  &  &  &  &  &  \\  \hline
   529 &  &  &  &  &  &  &  &  &  &  & x &  &  &  \\  \hline
   531 &  & x &  &  & x &  &  &  &  &  &  &  &  &  \\  \hline
   535 &  &  &  &  &  &  &  &  &  &  &  & x &  &  \\  \hline
   539 &  &  &  &  &  &  & x &  &  &  &  &  &  &  \\  \hline
   542 &  &  &  &  &  &  & x &  &  &  &  &  &  &  \\  \hline
   550 &  &  &  &  &  &  &  & x &  &  &  &  &  &  \\  \hline
   553 &  &  &  &  &  & x &  &  &  &  &  &  &  &  \\  \hline
   568 &  &  &  &  & x &  &  & x &  &  &  &  &  &  \\  \hline
   571 &  &  &  &  &  &  &  &   &   &   &  & x &  &   \\  \hline
   573 &  &  &  &  &  &  &  &  &  &  & x & x &  &  \\  \hline
   575 &  &  &  &  &  &  &  &  & x &  &  &  &  &  \\  \hline
   586 &  &  &  &  &  &  &  &  &  &  &  & x &  &  \\  \hline
   588 &  &  &  &  &  &  & x &  & x &  &  &  &  &  \\  \hline
   594 &  &  &  &  & x &  &  &   &   &   &  &  &  &  \\  \hline
   596 &  &  &  &  &  &  &  &  &  &  &  & x &  &  \\  \hline
   607 &  &  &  &  &  &  &  &   &   &   &  &  &  & x \\ \hline 
   609 &  &  &  &  &  &  &  &  &  &  &  &  &  & x \\  \hline
   611 &  &  &  &  &  &  &  &  &  &  & x & x &  &  \\  \hline
   613 &  &  &  &  &  &  &  &  &  &  &  &  & x &  \\  \hline
   614 &  &  &  &  &  &  &  &  &  &  & x & x &  &  \\  \hline
   618 &  &  &  &  &  &  &  &  &  &  &  & x &  &  \\  \hline
   622 &  &  &  &  &  &  & x &  &  &  &  &  &  &  \\  \hline
   626 &  &  &  &  &  &  &  &   &   &   &  &  &  & x \\  \hline
   630 & x &  &  &  &  &  &  &  &  &  &  &  &  &  \\  \hline
   645 &  &  &  &  &  & x &  &  &  &  &  &  &  &  \\  \hline
   655 &  &  &  &  &  &  & x &  &  &  &  &  &  &  \\  \hline
   656 &  &  &  &  &  &  & x &  &  &  &  &  &  &  \\  \hline
   658 &  &  &  &  &  &  &  &  &  &  & x &  &  &  \\  \hline
   664 &  &  &  &  &  &  &  &  &  &  & x &  &  &  \\  \hline
   665 &  &  &  &  &  &  & x &  &  &  &  &  &  &  \\  \hline
   666 &  &  &  &  &  &  &  &  &  &  &  & x &  &  \\  \hline
   667 &  &  &  &  &  &  & x &  &  &  & x &  &  &   \\  \hline
   676 & x &  &  &  &  &  &  &  &  &  &  &  &  &  \\  \hline
   677 &  &  &  &  &  &  & x &  &  &  &  &  &  &  \\  \hline
   681 & x &  &  &  &  &  &  &  &  &  &  &  &  &  \\  \hline
   682 & x &  &  &  &   &   &   &   &   &   &   &   &  &   \\  \hline
   689 & x & x &  &  &  &  & x &  &  &  &  &  &  &  \\  \hline
   694 &  &  &  &  &  &  & x &  &  &  &  &  &  &  \\  \hline
   699 &  &  &  &  &  &  &  &  &  &  &  & x &  &  \\  \hline
   700 &  &  &  &  &  &  &  &   &   &   &  &  &  & x \\  \hline
   701 & x &  &  &  &  &  &  &  &  &  &  &  &  &  \\  \hline
   711 &  &  &  &  &  &  &  &  &  &  &  & x &  &  \\  \hline
   716 &  &  &  &  &  &  &  &  &  &  & x & x &  &  \\  \hline
   717 &  &  &  &  &  &  &  &  &  &  &  &  & x &  \\  \hline
   719 &  &  &  &  &  &  & x &  &  &  &  &  &  &  \\  \hline
   720 &  &  &  &  &  &  & x &  &  &  &  &  &  &  \\  \hline
   722 &  &  &  &  &  &  & x &  &  &  &  &  &  &  \\  \hline
   723 &  &  &  &  &  &  & x &  &  &  &  &  &  &  \\  \hline
   732 &  &  &  &  &  &  &  & x &  &  &  &  &  &  \\  \hline
    739 &  &  &  &  &  &  & x &  &  &  &  &  &  &   \\  \hline
    740 &  &  &  &  &  &  & x &  &  &  &  &  & x &  \\  \hline
    745 &  &  &  &  &  &  & x &  &  &  &  &  &  &  \\  \hline
    751 &  &  &  &  &  &  & x &  &  &  &  &  &  &  \\  \hline
    755 &  &  &  &  &  &  & x &  &  &  &  &  &  &   \\  \hline
    773 & x &  &  &  &  &  &  &  &  &  &  & x &  &  \\  \hline
    776 & x &  &  &  &  &  &  &  &  &  &  &  &  &  \\  \hline
    777 &  &  &  &  &  &  & x &  &  &  &  &  &  &  \\  \hline
    778 & x &  &  &  &  &  &  &  &  &  &  &  &  &   \\  \hline
  781 &  & x &  &  &  &  &  &  &  &  &  &  &  &  \\ \hline
  791 &  &  &  &  &  &  &  &  &  &  &  & x &  &  \\ \hline
  796 &  & x &  &  &  &  &  &   &   &   &  & x &  &   \\  \hline
    797 & x &  &  &  &  &  &  &  &  &  &  &  &  &  \\  \hline
    804 &  &  &  & x &  &  &  &  &  &  &  &  &  &  \\  \hline
    813 &  &  &  &  &  &  &  &  &  &  &  &  & x &  \\  \hline
    821 &  &  &  &  &  &  &  &  &  &  &  & x &  &  \\  \hline
    834 &  &  &  &  &  &  & x &  &  &  &  &  &  &  \\  \hline
    M3 &  & x &  &  &  &  &  &  &  &  &  &  &  &  \\  \hline
    M5 & x &  & x &  &  &  &  &  &  &  &  &  &  &  \\  \hline
    M6 &  &  &  & x &  &  &  &  &  &  &  &  &  &  \\  \hline
    M8 &  &  &  & x &  &  &  &  &  &  &  &  &  &  \\  \hline
    M9 &  & x &  &  &  &  &  &  &  &  &  &  &  &  \\  \hline
    M10 &  &  &  & x &  &  &  &  &  &  &  &  &  &  \\  \hline
    M11 &  &  & x &  &  &  &  &  &  &  &  &  &  &  \\  \hline
    M13 &  &  &  &  &  &  &  &  & x &  &  &  &  &  \\  \hline
    E2 &  & x &  &  &  &  &  &  &  &  &  &  &  &  \\  \hline
    E3 &  &  &  &  &  &  & x &  &  &  &  &  &  &  \\  \hline
    E11 & x & x & x &  &  &  & x &  &  &  &  &  &  &  \\  \hline
    E15 & x & x & x &  &  &  &  &  &  &  &  &  &  &  \\  \hline
    E24 &  & x &  &  &  &  &  &  &  &  &  &  &  &  \\  \hline
    E31 & x &  &  &  &  &  & x &  &  &  &  &  &  &  \\  \hline
    E33 & x &  & x &  &  &  &  &  &  &  &  &  &  &  \\  \hline
    E37 & x &  &  &  &  &  &  &  &  &  &  &  &  &  \\  \hline
    E61 &  &  &  &  &  &  &  &  &  &  &  & x &  &  \\  \hline
    E65 & x &  &  &  &  &  &  &  &  &  &  &  &  &  \\ 
   \hline

\end{longtable}

\subsection{Classification of estimation methods}

    
\begin{longtable}{|p{0.305in}|p{0.15in}|p{0.15in}|p{0.15in}|p{0.15in}|p{0.15in}|p{0.15in}|p{0.15in}|p{0.15in}|p{0.15in}|p{0.15in}|p{0.15in}|p{0.15in}|p{0.15in}|p{0.15in}|p{0.15in}|p{0.15in}|p{0.15in}|p{0.15in}|p{0.15in}|p{0.15in}|p{0.15in}|p{0.15in}|} 
\caption{Classification of estimation methods (according to Section \ref{estimation}).\\ 
A: KM based estimation approaches (\ref{KM_est}); B: stratified Cox (\ref{StratCox}); C.1: change point for time varying effect (\ref{change point}); C.2: other time covariate interaction for time-varying coefficients (\ref{timeXcovariate}); C.3 fractional polynomials (\ref{fractional}); C.4 splines (\ref{Splines}); C.5: aHR and summary effect measures obtained through weighted partial likelihood (\ref{aHR}); C.6: additive model (\ref{additive}); D: joint models (\ref{joint}); E: Short- and long-term HR (\ref{shortHR}); F: transformation models with time-covariate interaction (\ref{trans}); G: frailty models (\ref{frailty}); H.1: piecewise exponential hazard (\ref{piecewise}); H.2: AFT and GAMLSS (\ref{AFT}); H.3: FHT (\ref{FHT}); H.4: GLMs and other parametric approaches (\ref{GLM}); 
I.1 trees and forests (\ref{ML}); 
I.2 neural networks (\ref{ML}); 
I.3 kernel smoothing (\ref{ML}); 
I.4 other machine learning (\ref{ML}); 
J: other modelling approaches (\ref{other_estimation}); 
K: Bayesian fitting approach} 
\label{tab:S2} 
\\
\hline 
ID & A & B & C.1 & C.2 & C.3 & C.4 & C.5 & C.6 & D & E & F & G & H.1 & H.2 & H.3 & H.4 & I.1 & I.2 & I.3 & I.4 & J & K \\ \hline 
\textbf{14} &  & x &  &  & x &  &  &  &  &  & x &  & x & x &  &  &  & ~ & ~ & ~ & ~ &  \\ \hline 
\textbf{16} & x &  &  &  &  &  &  &  &  &  &  &  &  &  &  &  &  &  &  &  &  &  \\ \hline 
\textbf{20} &  &  & x &  &  & &  &  &  &  &  &  &  &  &  &  &  &  &  &  &  & x \\ \hline 
\textbf{21} &  &  &  &  &  &  &  &  &  &  & x & \NEW{ } &  &  &  &  &  &  &  &  &  &  \\ \hline 
\textbf{26} & x &  &  &  &  &  &  &  &  &  &  &  &  &  &  &  &  &  &  &  &  &  \\ \hline 
\textbf{27} &  &  &  & x & x & x &  &  &  &  &  &  &  &  &  &  &  &  &  &  &  &  \\ \hline 
\textbf{30} &  &  & x &  &  &  &  &  &  &  &  & x &  &  &  &  &  &  &  &  &  &  \\ \hline 
\textbf{42} &  &  &  &  &  &  &  &  &  &  &  &  &  &  &  & x &  &  &  &  &  &  \\ \hline 
\textbf{46} &  &  & x & x &  &  &  &  &  &  &  &  &  &  &  &  &  & ~ & ~ & ~ & ~ &  \\ \hline 
\textbf{56} & x &  &  &  &  &  &  &  &  &  &  &  &  &  &  &  &  &  &  &  &  &  \\ \hline 
\textbf{67} & x &  &  &  &  &  &  &  &  &  &  &  &  &  &  &  &  &  &  &  &  &  \\ \hline 
\textbf{69} & x &  &  &  &  &  &  &  &  & x &  & x &  &  &  &  &  &  &  &  &  &  \\ \hline 
\textbf{75} &  &  &  &  &  &  &  &  &  &  &  & x &  &  &  &  & x &  &  &  &  & x \\ \hline 
\textbf{76} &  &  & ~ & ~ & ~ &  & ~ & ~ & ~ & ~ & ~ & x &  &  &  &  &  & ~ & ~ & ~ & ~ &  \\ \hline 
\textbf{86} &  &  &  &  &  &  &  &  &  &  & x &  &  &  &  &  &  & ~ & ~ & ~ & ~ &  \\ \hline 
\textbf{91} &  &  &  &  &  &  & x &  &  &  &  &  &  &  &  &  &  &  &  &  &  &  \\ \hline 
\textbf{92} &  &  &  &  &  &  &  &  &  &  &  &  &  &  & x &  &  &  &  &  &  & x \\ \hline 
\textbf{97} & x &  &  &  &  &  &  &  &  &  &  &  &  &  &  &  &  &  &  &  &  &  \\ \hline 
\textbf{103} &  &  &  &  &  &  &  &  & x &  &  &  & x &  &  &  &  &  &  &  &  & x \\ \hline 
\textbf{110} &  &  & x &  &  &  &  &  &  &  &  & x &  &  &  &  &  &  &  &  &  &  \\ \hline 
\textbf{118} &  &  &  &  &  &  &  &  &  &  &  & x &  &  &  & x &  &  &  &  &  &  \\ \hline 
\textbf{122} &  &  &  &  &  &  &  &  &  &  &  &  & x &  &  & x &  &  &  &  &  &  \\ \hline 
\textbf{123} &  &  &  &  &  &  &  &  & x &  &  &  &  &  &  &  & x &  &  &  &  &  \\ \hline 
\textbf{126} & x &  &  &  &  &  &  &  &  &  &  &  &  &  &  &  &  &  &  &  &  &  \\ \hline 
\textbf{142} &  &  &  &  &  &  &  &  &  &  &  &  &  & x &  &  &  &  &  &  &  &  \\ \hline 
\textbf{149} &  &  &  &  &  &  &  &  &  &  &  & x &  &  &  &  &  &  &  &  &  &  \\ \hline 
\textbf{157} &  &  &  &  &  &  &  &  &  &  &  &  &  & x &  &  &  &  &  &  &  &  \\ \hline 
\textbf{169} &  &  & x &  &  &  &  &  &  &  &  & x &  &  &  &  &  &  &  &  &  &  \\ \hline 
\textbf{176} &  &  &  &  &  &  &  &  &  &  &  &  &  &  &  & x &  &  &  &  &  &  \\ \hline 
\textbf{184} &  &  &  &  &  &  &  &  &  &  &  &  &  &  &  &  &  &  & x &  &  &  \\ \hline 
\textbf{188} &  &  &  &  &  &  &  & x &  &  &  &  &  &  &  &  &  &  &  &  &  &  \\ \hline 
\textbf{195} &  &  &  &  &  &  &  &  &  &  &  &  &  &  &  &  & x &  &  &  &  &  \\ \hline 
\textbf{207} &  &  &  &  & x &  &  &  &  &  &  &  & ~ & ~ & ~ &  &  & ~ & ~ & ~ & ~ &  \\ \hline 
\textbf{235} &  &  & ~ & ~ & ~ &  & ~ & ~ & ~ & ~ & ~ &  &  &  &  &  &  &  &  &  & x &  \\ \hline 
\textbf{247} & x &  &  &  &  &  &  &  &  &  &  &  &  &  &  &  &  &  &  &  &  &  \\ \hline 
\textbf{265} & x &  &  &  &  &  &  &  &  &  &  &  &  &  &  &  &  &  &  &  &  &  \\ \hline 
\textbf{266} &  &  &  &  &  &  & x &  &  &  &  &  &  &  &  &  &  &  &  &  &  &  \\ \hline 
\textbf{268} &  &  &  &  &  &  & x &  &  & x &  &  &  &  &  &  &  &  &  &  &  &  \\ \hline 
\textbf{269} &  &  &  &  &  &  &  &  &  & x &  &  &  &  &  &  &  &  &  &  &  &  \\ \hline 
\textbf{274} &  &  &  &  &  &  &  &  &  &  &  &  &  &  &  &  &  &  &  &  & x &  \\ \hline 
\textbf{284} & x &  &  &  &  &  &  &  &  &  &  &  &  &  &  &  &  &  &  &  &  &  \\ \hline 
\textbf{313} &  &  &  &  &  &  &  &  &  &  &  &  &  & x &  &  &  &  &  &  &  &  \\ \hline 
\textbf{328} & x &  &  &  &  &  &  &  &  &  &  &  &  &  &  &  &  &  &  &  &  &  \\ \hline 
\textbf{335} &  &  &  &  &  &  &  &  &  &  &  &  & x &  &  &  &  &  &  &  &  & x \\ \hline 
\textbf{343} &  &  &  &  &  &  &  &  &  &  &  &  &  & x &  &  &  &  &  &  &  &  \\ \hline 
\textbf{359} &  &  & x &  &  &  &  &  &  &  &  & x &  & x &  &  &  &  &  &  &  &  \\ \hline 
\textbf{360} &  &  &  &  &  & x &  &  &  &  &  &  &  &  &  &  &  &  &  &  &  & x \\ \hline 
\textbf{365} & x &  & ~ & ~ & ~ &  & ~ & ~ & ~ & ~ & ~ &  &  &  &  &  &  & ~ & ~ & ~ & ~ &  \\ \hline 
\textbf{367} &  &  &  & x &  &  &  &  &  &  &  &  &  & x &  &  &  & ~ & ~ & ~ & ~ &  \\ \hline 
\textbf{376} & x &  & ~ & ~ & ~ &  & ~ & ~ & ~ & ~ & ~ &  &  &  &  &  &  & ~ & ~ & ~ & ~ &  \\ \hline 
\textbf{378} &  &  &  &  &  &  &  &  &  &  &  &  &  &  &  &  &  &  &  & x &  &  \\ \hline 
\textbf{383} &  &  &  &  &  &  &  &  &  &  &  &  &  &  &  &  & x &  &  &  &  & x \\ \hline  
\textbf{396} &  &  &  &  &  &  &  &  &  &  &  &  &  & x &  &  &  &  &  &  &  &  \\ \hline 
\textbf{397} &  &  &  &  &  &  &  &  &  &  &  &  &  & x &  &  &  &  &  &  &  &  \\ \hline 
\textbf{404} &  &  &  &  &  &  &  &  &  &  & x &  &  &  &  &  &  &  &  &  &  &  \\ \hline 
\textbf{417} &  &  &  &  &  & x &  &  &  &  &  &  &  &  &  &  &  &  &  &  &  &  \\ \hline 
\textbf{422} &  &  &  &  &  & x &  &  &  &  &  &  &  &  &  &  &  &  &  &  &  &  \\ \hline 
\textbf{424} &  &  &  &  &  &  &  &  &  &  & x &  &  &  &  &  &  &  &  &  &  &  \\ \hline 
\textbf{425} &  &  &  &  &  &  &  &  &  & x &  &  &  &  &  &  &  &  &  &  &  &  \\ \hline 
\textbf{433} &  &  &  &  &  &  &  &  &  &  &  &  &  &  &  &  &  &  &  &  & x &  \\ \hline 
\textbf{435} &  &  &  &  &  &  &  &  &  &  &  &  &  &  & x &  &  &  &  &  &  &  \\ \hline 
\textbf{445} & x &  &  &  &  &  &  &  &  &  &  &  &  &  &  &  &  &  &  &  &  &  \\ \hline 
\textbf{446} &  &  & ~ & ~ & ~ &  & ~ & ~ & ~ & ~ & ~ &  & x &  &  &  &  & ~ & ~ & ~ & ~ &  \\ \hline 
\textbf{452} &  &  &  &  &  &  &  &  &  &  &  &  & x &  &  &  &  &  &  &  &  &  \\ \hline 
\textbf{453} &  &  &  &  &  &  &  &  &  & x &  &  &  &  &  &  &  &  &  &  &  &  \\ \hline 
\textbf{456} &  &  &  &  &  & x &  &  &  &  &  &  &  &  &  &  &  &  &  &  &  &  \\ \hline 
\textbf{464} &  &  &  &  &  &  &  &  &  &  &  &  &  &  &  &  &  &  &  &  & x &  \\ \hline 
\textbf{469} &  &  &  &  &  &  &  &  &  & x &  &  &  &  &  &  &  &  &  &  &  &  \\ \hline 
\textbf{470} &  &  &  &  &  &  &  &  &  &  &  &  & x &  &  &  &  &  &  &  &  &  \\ \hline 
\textbf{482} &  &  &  &  &  & x &  &  &  &  &  &  &  &  &  &  &  &  &  &  &  &  \\ \hline 
\textbf{501} & x &  &  &  &  &  &  &  &  &  &  &  &  &  &  &  &  &  &  & x &  &  \\ \hline 
\textbf{510} &  &  & ~ & ~ & ~ &  & ~ & ~ & ~ & ~ & ~ &  &  &  &  & x &  & ~ & ~ & ~ & ~ &  \\ \hline 
\textbf{513} &  &  &  &  &  &  &  & x &  &  &  &  &  &  &  &  &  &  &  &  &  &  \\ \hline 
\textbf{514} &  &  &  &  &  &  &  &  &  &  &  &  &  &  &  & x &  &  &  &  &  &  \\ \hline 
\textbf{518} &  &  &  &  &  &  &  &  & x &  &  &  &  &  &  &  &  &  &  &  &  & x \\ \hline 
\textbf{524} &  &  &  &  &  &  &  & x &  &  &  &  &  &  &  &  &  & ~ & ~ & ~ & ~ &  \\ \hline 
\textbf{525} &  &  &  &  &  &  &  &  &  &  & x &  &  &  &  &  &  &  &  &  &  &  \\ \hline 
\textbf{528} &  &  &  & x &  &  &  &  &  &  &  &  &  &  &  &  &  &  &  &  &  &  \\ \hline 
\textbf{529} &  &  &  &  &  &  &  &  &  &  &  & x &  &  &  &  &  &  &  &  &  &  \\ \hline 
\textbf{531} & x &  &  &  &  &  &  &  &  &  &  &  &  &  &  &  &  &  &  &  &  &  \\ \hline 
\textbf{535} &  &  &  &  &  &  &  &  &  &  &  &  &  &  &  & x &  &  &  &  &  &  \\ \hline 
\textbf{539} &  &  &  &  &  &  & x &  &  &  &  &  &  &  &  &  &  &  &  &  &  &  \\ \hline 
\textbf{542} &  &  &  &  &  &  & x &  &  &  &  &  &  &  &  &  &  &  &  &  &  &  \\ \hline 
\textbf{550} &  &  &  &  &  & x &  &  &  &  &  &  &  &  &  &  &  &  &  &  &  &  \\ \hline 
\textbf{553} &  & x &  &  &  &  &  &  &  &  &  &  &  &  &  &  &  &  &  &  &  &  \\ \hline 
\textbf{568} & x &  &  &  &  &  &  &  &  &  & x &  &  &  &  &  &  &  &  &  &  &  \\ \hline 
\textbf{571} &  &  & ~ & ~ & ~ &  & ~ & ~ & ~ & ~ & ~ &  &  & x &  &  &  & ~ & ~ & ~ & ~ &  \\ \hline 
\textbf{573} &  &  &  &  &  &  &  &  &  &  &  & x &  & x &  &  &  &  &  &  &  & x \\ \hline 
\textbf{575} &  &  &  &  &  &  &  &  &  & x &  &  &  &  &  &  &  &  &  &  &  &  \\ \hline 
\textbf{586} &  &  &  &  &  &  &  &  &  &  &  &  & x &  &  &  &  &  &  &  &  &  \\ \hline 
\textbf{588} &  &  &  &  &  &  & x &  &  & x &  &  &  &  &  &  &  &  &  &  &  &  \\ \hline 
\textbf{594} & x &  & ~ & ~ & ~ &  & ~ & ~ & ~ & ~ & ~ &  & ~ & ~ & ~ &  &  &  &  &  &  &  \\ \hline 
\textbf{596} &  &  &  &  &  &  &  &  &  &  &  &  &  & x &  &  &  &  &  &  &  &  \\ \hline 
\textbf{607} &  &  & ~ & ~ & ~ &  & ~ & ~ & ~ & ~ & ~ &  & ~ & ~ & ~ &  &  &  &  &  & x &  \\ \hline 
\textbf{609} &  &  &  &  &  &  &  &  &  &  &  &  &  &  &  &  &  &  &  &  & x &  \\ \hline 
\textbf{611} &  &  &  &  &  &  &  &  &  &  &  & x &  & x &  &  &  &  &  &  &  &  \\ \hline 
\textbf{613} &  &  &  &  &  &  &  &  &  &  &  &  &  &  &  &  &  &  & x &  &  &  \\ \hline 
\textbf{614} &  &  &  &  &  &  &  &  &  &  &  & x &  &  & x &  &  &  &  &  &  & x \\ \hline 
\textbf{618} &  &  &  &  &  &  &  &  &  &  &  &  &  &  & x &  &  &  &  &  &  &  \\ \hline 
\textbf{622} &  &  &  &  &  &  & x &  &  &  &  &  &  &  &  &  &  &  &  &  &  &  \\ \hline 
\textbf{626} &  &  & ~ & ~ & ~ &  & ~ & ~ & ~ & ~ & ~ &  &  &  &  &  &  &  &  &  & x & x \\ \hline 
\textbf{645} &  & x &  &  &  &  &  &  &  &  &  &  &  &  &  &  &  &  &  &  &  &  \\ \hline 
\textbf{655} &  &  &  &  & x &  &  &  &  &  &  &  &  &  &  &  &  &  &  &  &  &  \\ \hline 
\textbf{656} &  &  &  &  &  & x &  &  &  &  &  &  &  &  &  &  &  &  &  &  &  &  \\ \hline 
\textbf{658} &  &  &  &  &  &  &  &  &  &  &  & x &  &  &  &  &  &  &  &  &  & x \\ \hline 
\textbf{664} &  &  &  &  &  &  &  &  &  &  &  & x &  &  &  &  &  &  &  &  &  & x \\ \hline 
\textbf{665} &  &  & x &  &  &  &  &  &  &  &  &  &  &  &  &  &  &  &  &  &  &  \\ \hline 
\textbf{666} &  &  &  &  &  &  &  &  &  &  &  &  &  & x &  &  &  &  &  &  &  &  \\ \hline 
\textbf{667} &  &  &  & x & x & x &  &  &  &  &  & x &  &  &  &  &  & ~ & ~ & ~ & ~ &  \\ \hline 
\textbf{677} &  &  &  & x &  &  &  &  &  &  &  &  &  &  &  &  &  &  &  &  &  &  \\ \hline 
\textbf{689} &  &  &  &  &  &  & x &  &  &  &  &  &  &  &  &  &  &  &  &  &  &  \\ \hline 
\textbf{694} &  &  &  &  &  &  &  & x &  &  &  &  &  &  &  &  &  &  &  &  &  & x \\ \hline 
\textbf{699} &  &  &  &  &  &  &  &  &  &  &  &  &  &  &  & x &  &  &  &  &  &  \\ \hline 
\textbf{700} &  &  & ~ & ~ & ~ &  & ~ & ~ & ~ & ~ & ~ &  &  &  &  &  &  &  &  &  & x &  \\ \hline 
\textbf{711} &  &  &  &  &  &  &  &  &  &  &  &  &  & x &  &  &  &  &  &  &  &  \\ \hline 
\textbf{716} &  &  &  &  &  &  &  &  &  &  &  & x &  & x &  &  &  &  &  &  &  &  \\ \hline 
\textbf{717} &  &  &  &  &  &  &  &  &  &  &  &  &  &  &  &  &  & x &  &  &  & x \\ \hline 
\textbf{719} &  &  &  &  &  & x &  &  &  &  &  &  &  &  &  &  &  &  &  &  &  &  \\ \hline 
\textbf{720} &  &  &  &  &  &  & x &  &  &  &  &  &  &  &  &  &  &  &  &  &  &  \\ \hline 
\textbf{722} &  &  &  &  &  & x &  &  &  &  &  &  &  &  &  &  &  &  &  &  &  &  \\ \hline 
\textbf{723} &  &  &  &  & x &  &  &  &  &  &  &  &  &  &  &  &  &  &  &  &  &  \\ \hline 
\textbf{732} &  &  &  &  &  &  &  &  &  &  & x &  &  &  &  &  &  &  &  &  &  &  \\ \hline 
\textbf{739} &  &  &  &  &  & x &  &  &  &  &  &  &  &  &  &  &  & ~ & ~ & ~ & ~ &  \\ \hline 
\textbf{740} &  &  & x &  &  &  &  &  &  &  &  &  &  &  &  &  & x &  &  &  &  &  \\ \hline 
\textbf{745} &  &  & x &  &  &  &  &  &  &  &  &  &  &  &  &  &  &  &  &  &  &  \\ \hline 
\textbf{751} &  &  & x &  &  &  &  &  &  &  &  &  &  &  &  &  &  &  &  &  &  &  \\ \hline 
\textbf{755} &  &  & x &  &  &  &  &  &  &  &  &  &  &  &  &  &  & ~ & ~ & ~ & ~ & x \\ \hline 
\textbf{773} &  &  &  &  &  &  &  &  &  &  &  &  &  & x &  &  &  &  &  &  &  &  \\ \hline 
\textbf{777} &  &  &  & x &  &  &  &  &  &  &  &  &  &  &  &  &  &  &  &  &  & x \\ \hline 
\textbf{791} &  &  &  &  &  &  &  &  &  &  &  &  &  &  &  & x &  &  &  &  &  &  \\ \hline 
\textbf{796} &  &  & ~ & ~ & ~ &  & ~ & ~ & ~ & ~ & ~ &  & x &  &  &  &  & ~ & ~ & ~ & ~ &  \\ \hline 
\textbf{813} &  &  &  &  &  &  &  &  &  &  &  &  &  &  &  &  &  & x &  &  &  &  \\ \hline 
\textbf{821} &  &  &  &  &  &  &  &  &  &  &  &  &  & x &  &  &  &  &  &  &  &  \\ \hline 
\textbf{834} &  &  &  & x &  &  &  &  &  &  &  &  &  &  &  &  &  &  &  &  &  &  \\ \hline 
\textbf{M13} &  &  &  &  &  & x &  &  &  & x &  &  &  &  &  &  &  &  &  &  &  &  \\ \hline 
\textbf{E3} &  &  &  &  &  &  & x &  &  &  &  &  &  &  &  &  &  &  &  &  &  &  \\ \hline 
\textbf{E11} &  &  &  &  &  &  & x &  &  &  &  &  &  &  &  &  &  &  &  &  &  &  \\ \hline 
\textbf{E31} &  &  &  &  &  &  & x &  &  &  &  &  &  &  &  &  &  &  &  &  &  &  \\ \hline 
\textbf{E61} &  &  &  &  &  &  &  &  &  &  &  &  &  &  &  & x &  &  &  &  &  &  \\ \hline 
\end{longtable}
    

\newpage
\subsection{Classification of articles including hypothesis tests}
    
    \begin{longtable}{|p{0.3in}|p{0.4in}|p{0.4in}|p{0.4in}|p{0.3in}|p{0.4in}|p{0.4in}|p{1.0in}|p{0.65in}|p{0.8in}|}
    \caption{Classification of articles including hypothesis tests including the definition of the null hypothesis.}
    \label{tab:S4} 
    \\
    \hline 
\textbf{} & \multicolumn{4}{|p{1.8in}|}{\textbf{Hypothesis Tests}} & \textbf{} & \textbf{} & \textbf{} & \textbf{} & \textbf{} \\ \hline 
ID & Log\-rank tests\textbf{} & Kaplan-Meier based tests\textbf{} & Combi\-nation tests\textbf{} & Other tests\textbf{} & $S_1(t)=S_2(t)$ & $S_1\left(t\right)\le S_2(t$) & Other average based null hypothesis & Other/Not explicitly mentioned & Approach for sample size calculation \\ \hline 
1 & x &  &  &  &  & x &  &  & x \\ \hline 
9 & x & x & x & x & x &  & x (RMST) &  &  \\ \hline 
19 &  &  & x &  &  & x &  &  &  \\ \hline 
22 & x & x &  &  & x  &  & x(milestone survival) &  & x \\ \hline 
28 &  &  & x &  & x  &  &  &  & x \\ \hline 
30 & x &  &  &  & x &  &  &  & x \\ \hline 
39 & x &  & x &  & x &  &  &  &  \\ \hline 
40 & x &  &  &  & x &  &  &  &  \\ \hline 
46 &  &  & x &  &  &  & x(regression coefficient) &  &  \\ \hline 
53 &  &  &  & x &  &  & x(relative time) &  & x \\ \hline 
56 &  & x &  &  &  &  & x(window mean survival time) &  &  \\ \hline 
69 & x & x & x &  & x &  & x(RMST) &  & x \\ \hline 
72 & x &  &  &  & x &  &  &  & x \\ \hline 
79 & x &  & x &  &  &  & x(HR) &  & x \\ \hline 
81 &  &  & x &  &  &  & x(HR) &  & x \\ \hline 
83 & x &  &  &  &  &  & x(HR) &  & x \\ \hline 
84 & x &  &  &  &  &  & x(HR) &  &  \\ \hline 
86 & x &  &  &  & x &  &  &  &  \\ \hline 
89 & x &  &  &  & x &  &  &  & x \\ \hline 
91 & x &  &  &  &  &  & x(HR) &  &  \\ \hline 
95 & x &  &  &  &  &  & x(HR) &  & x \\ \hline 
96 &  &  &  & x &  & x &  &  & x \\ \hline 
97 &  &  &  & x &  &  & x(Survival median) &  &  \\ \hline 
113 & x &  &  &  &  &  & x(HR non inferiority) &  & x \\ \hline 
116 & x &  &  &  & x &  &  &  & x \\ \hline 
118 &  & x &  &  &  &  & x(RMST) &  &  \\ \hline 
119 & x &  & x &  &  & x &  &  &  \\ \hline 
134 &  &  &  & x &  &  &  & x &  \\ \hline 
136 & x &  & x &  & x &  &  &  &  \\ \hline 
145 & x &  & x &  & x &  &  &  &  \\ \hline 
148 & x &  &  &  & x &  &  &  &  \\ \hline 
149 &  &  &  & x &  &  &  & x &  \\ \hline 
154 & x &  &  &  & x &  &  &  & x \\ \hline 
159 & x &  &  &  & x &  &  &  & x \\ \hline 
162 &  &  &  & x & x &  &  &  &  \\ \hline 
167 & x & x &  &  & x &  &  &  & x \\ \hline 
169 & x &  &  &  & x &  &  &  & x \\ \hline 
174 & x &  &  &  & x &  &  &  & x \\ \hline 
201 & x &  &  &  &  & x &  &  &  \\ \hline 
213 & x & x &  &  &  &  & x(RMST) &  &  \\ \hline 
235 &  &  &  & x & x &  &  &  &  \\ \hline 
242 & x &  &  &  & x &  &  &  &  \\ \hline 
244 & x &  &  &  & x &  &  &  &  \\ \hline 
245 & x &  &  &  &  &  & x(HR) &  &  \\ \hline 
247 &  & x &  &  & x &  &  &  &  \\ \hline 
251 & x &  &  &  & x &  &  &  & x \\ \hline 
256 & x &  &  &  & x &  &  &  & x \\ \hline 
265 &  & x &  &  &  &  & x(aHR non inferiority) &  &  \\ \hline 
267 &  & x &  &  &  &  & x(RMST) &  &  \\ \hline 
268 &  & x & x &  & x &  &  &  &  \\ \hline 
269 & x &  &  &  & x &  &  &  &  \\ \hline 
296 & x & x &  &  &  &  & x(HR) &  & x \\ \hline 
347 & x &  &  &  & x &  &  &  &  \\ \hline 
359 & x &  &  &  & x &  &  &  &  \\ \hline 
376 &  & x &  &  &  &  & x(aHR) &  &  \\ \hline 
382 &  &  & x &  & x &  &  &  &  \\ \hline 
384 & x &  &  &  & x &  &  &  & x \\ \hline 
392 & x &  &  &  & x &  &  &  &  \\ \hline 
397 & x &  &  &  & x &  &  &  &  \\ \hline 
402 &  & x &  &  & x &  &  &  &  \\ \hline 
413 & x &  &  &  &  &  & x(parameter) &  &  \\ \hline 
427 & x &  & x &  & x &  &  &  &  \\ \hline 
444 & x &  & x &  &  &  & x(HR) &  &  \\ \hline 
446 &  & x &  &  &  &  & x(RMST) &  &  \\ \hline 
465 & x &  &  &  & x &  &  &  & x \\ \hline 
467 & x &  &  &  & x &  &  &  &  \\ \hline 
470 &  & x &  &  &  &  & x(RMST) &  & x \\ \hline 
511 & x &  &  &  & x &  &  &  &  \\ \hline 
520 & x &  &  &  & x &  &  &  &  \\ \hline 
531 &  & x &  &  & x &  &  &  &  \\ \hline 
630 & x &  &  &  & x &  &  &  & x \\ \hline 
676 & x &  &  &  &  &  & x(HR non inferiority) &  &  \\ \hline 
681 & x &  &  &  &  & x &  &  & x \\ \hline 
682 & x &  &  &  & x &  &  &  &  \\ \hline 
689 & x & x &  &  & x &  &  &  &  \\ \hline 
701 & x &  &  &  & x &  &  &  & x \\ \hline 
773 & x &  &  &  & x &  &  & x &  \\ \hline 
776 & x &  &  &  & x &  &  &  & x \\ \hline 
778 & x &  &  &  &  &  & x(regression coefficient and HR) &  &  \\ \hline 
781 &  & x &  &  & x &  &  &  &  \\ \hline 
796 &  & x &  &  &  &  & x(RMST) &  &  \\ \hline 
797 & x &  &  &  & x &  &  &  &  \\ \hline 
804 &  &  &  & x & x &  &  &  &  \\ \hline 
M3 &  & x &  &  & x &  &  &  &  \\ \hline 
M5 & x &  & x &  & x &  &  &  &  \\ \hline 
M6 &  &  &  & x & x &  &  &  &  \\ \hline 
M8 &  &  &  & x & x &  &  &  &  \\ \hline 
M9 &  & x &  &  & x &  &  &  &  \\ \hline 
M10 &  &  &  & x & x &  &  &  &  \\ \hline 
M11 &  &  & x &  & x &  &  &  &  \\ \hline 
E2 &  & x &  &  & x &  &  &  & x \\ \hline 
E11 & x & x & x &  & x &  & x(RMST) &  &  \\ \hline 
E15 & x & x & x &  & x &  & x(RMST) &  &  \\ \hline 
E24 &  & x &  &  &  &  & x(RMST) &  &  \\ \hline 
E31 & x &  &  &  &  &  & x(HR) &  &  \\ \hline 
E33 & x &  & x &  & x &  &  &  &  \\ \hline 
E37 & x &  &  &  &  &  & x(HR) &  & x \\ \hline 
E65 & x &  &  &  & x &  &  &  &  \\ \hline 
\end{longtable}

\newpage
\subsection{Bibliographic information for all articles included in our systematic review}\label{completebib}

       
    \begin{longtable}{|p{0.3in}|p{1.8in}|p{2.4in}|p{0.68in}|p{0.89in}|p{0.3in}|}
     \caption{Bibliographic information for all articles included in our systematic review. IDs correspond to the original set of 907 articles screened during the review. IDs without any letters are the articles identified by the PUBMED search. The letter ``M'' indicates the results of the manual search and ``E'' the identified articles of the EMBASE search. }
    \label{tab:S3}
    \\
    \hline 
ID & DOI & Title & First Author & Source & Year \\ \hline 
1 & 10.1177/17407745211072848          & Design and analysis of group-sequential clinical trials based on a modestly weighted log-rank test in anticipation of a delayed separation of survival curves: A practical guidance & Magirr D & Clin Trials & 2022 \\ \hline 
9 & 10.1186/s12874-022-01520-0           & Which test for crossing survival curves? A user's guideline & Dormuth I & BMC Med Res Methodol & 2022 \\ \hline 
14 & 10.1177/09622802211070253          & Challenges of modelling approaches for network meta-analysis of time-to-event outcomes in the presence of non-proportional hazards to aid decision making: Application to a melanoma network & Freeman SC & Stat Methods Med Res & 2022 \\ \hline 
16 & 10.1200/PO.20.00164         & Insights for Quantifying the Long-Term Benefit of Immunotherapy Using Quantile Regression & Mboup B & JCO Precis Oncol & 2021 \\ \hline 
19 & 10.1002/bimj.202000169      & Robust group sequential designs for trials with survival endpoints and delayed response & Ghosh P & Biom J & 2022 \\ \hline 
20 & 10.1016/j    .ejca.2021.11.002    & Bayesian interpretation of immunotherapy trials with dynamic treatment effects & Casta\~{n}on E & Eur J Cancer & 2022 \\ \hline 
21 & 10.1001/\-jamanetworkopen.2021.39573     & Assessment of Treatment Effects and Long-term Benefits in Immune Checkpoint Inhibitor Trials Using the Flexible Parametric Cure Model: A Systematic Review & Filleron T & JAMA Netw Open & 2021 \\ \hline 
22 & 10.1002/sim.9256          & Complex survival trial design by the product integration method & Tang Y & Stat Med & 2022 \\ \hline 
26 & 10.1177/09622802211041756         & Weighted pseudo-values for partly unobserved group membership in paediatric stem cell transplantation studies & Mittlb\"{o}ck M & Stat Methods Med Res & 2022 \\ \hline 
27 & 10.1002/sim.9259        & Using fractional polynomials and restricted cubic splines to model non-proportional hazards or time-varying covariate effects in the Cox regression model & Austin PC & Stat Med & 2022 \\ \hline 
28 & 10.1002/sim.9251       & M\&M: A maximum duration design with the Maxcombo test for a group sequential trial of an immunotherapy with a random delayed treatment effect & Li B & Stat Med & 2022 \\ \hline 
30 & 10.1002/sim.9258       & Cancer immunotherapy trial design with random delayed treatment effect and cure rate & Wu J & Stat Med & 2022 \\ \hline 
39 & 10.1111/biom.13575      & CASANOVA: Permutation inference in factorial survival designs & Ditzhaus M & Biometrics & 2021 \\ \hline 
40 & 10.1080/10543406.2021      .\-1979575 & Multi-arm multi-stage clinical trials for time-to-event outcomes & Dixit V & J Biopharm Stat & 2021 \\ \hline 
42 & 10.1002/bimj.202000292       & Nonproportional hazards model with a frailty term for modeling subgroups with evidence of long-term survivors: Application to a lung cancer dataset & Gazon AB & Biom J & 2022 \\ \hline 
46 & 10.1177/09622802211037076       & CauchyCP: A powerful test under non-proportional hazards using Cauchy combination of change-point Cox regressions & Zhang H & Stat Methods Med Res & 2021 \\ \hline 
53 & 10.1002/bimj.202000043      & Sample size calculation for two-arm trials with time-to-event endpoint for nonproportional hazards using the concept of Relative Time when inference is built on comparing Weibull distributions & Phadnis MA & Biom J & 2021 \\ \hline 
56 & 10.1002/sim.9138      & Window mean survival time & Paukner M & Stat Med & 2021 \\ \hline 
67 & 10.1016/j     .cmpb.2021.106155 & Dynamic prediction and analysis based on restricted mean survival time in survival analysis with nonproportional hazards & Yang Z & Comput Methods Programs Biomed & 2021 \\ \hline 
69 & 10.1016/j    .critrevonc.2021.\-103350 & Critical review of oncology clinical trial design under non-proportional hazards & Anantha\-krishnan R & Crit Rev Oncol Hematol & 2021 \\ \hline 
72 & 10.1186/s12874-021-01286-x        & Using the geometric average hazard ratio in sample size calculation for time-to-event data with composite endpoints & Cort\'{e}s Mart\'{i}nez J & BMC Med Res Methodol & 2021 \\ \hline 
75 & 10.1111/biom.13478     & Semiparametric analysis of clustered interval-censored survival data using soft Bayesian additive regression trees (SBART) & Basak P & Biometrics & 2021 \\ \hline 
76 & 10.1001/jamaoncol.2021.0289     & Development and Evaluation of a Method to Correct Misinterpretation of Clinical Trial Results With Long-term Survival & Hsu CY & JAMA Oncol & 2021 \\ \hline 
79 & 10.1177/0962280220988570        & A unified approach to power and sample size determination for log-rank tests under proportional and nonproportional hazards & Tang Y & Stat Methods Med Res & 2021 \\ \hline 
81 & 10.1002/pst.2116     & A simulation-free group sequential design with max-combo tests in the presence of non-proportional hazards & Wang L & Pharm Stat & 2021 \\ \hline 
83 & 10.1080/10543406.2020       .\-1815035 & Statistical and practical considerations in designing of immuno-oncology trials & Mukho\-padhyay P & J Biopharm Stat & 2020 \\ \hline 
84 & 10.1002/sim.8905        & On information fraction for Fleming-Harrington type weighted log-rank tests in a group-sequential clinical trial design & Kundu MG & Stat Med & 2021 \\ \hline 
86 & 10.1016/j  .esmoop.2020   .\-100043 & Cediranib in addition to chemotherapy for women with relapsed platinum-sensitive ovarian cancer (ICON6): overall survival results of a phase III randomised trial & Ledermann JA & ESMO Open & 2021 \\ \hline 
89 & 10.1007/s10985-021-09517-5       & Testing equivalence of survival before but not after end of follow-up & Furberg JK & Lifetime Data Anal & 2021 \\ \hline 
91 & 10.1002/pst.2092    & A weighted log-rank test and associated effect estimator for cancer trials with delayed treatment effect & Yu C & Pharm Stat & 2021 \\ \hline 
92 & 10.1007/s10985-020-09514-0       & Semi-parametric survival analysis via Dirichlet process mixtures of the First Hitting Time model & Race JA & Lifetime Data Anal & 2021 \\ \hline 
95 & 10.1177/0962280220980780       & A group sequential design and sample size estimation for an immunotherapy trial with a delayed treatment effect & Li B & Stat Methods Med Res & 2021 \\ \hline 
96 & 10.1002/pst.2091   & Non-proportional hazards in immuno-oncology: Is an old perspective needed? & Magirr D & Pharm Stat & 2021 \\ \hline 
97 & 10.1177/0962280220980784     & Inferring median survival differences in general factorial designs via permutation tests & Ditzhaus M & Stat Methods Med Res & 2021 \\ \hline 
103 & 10.1093/biostatistics/kxaa044    & Bayesian design of clinical trials using joint models for longitudinal and time-to-event data & Xu J & Biostatistics & 2020 \\ \hline 
110 & 10.1002/sim.8694       & Design for immuno-oncology clinical trials enrolling both responders and nonresponders & Xu Z & Stat Med & 2020 \\ \hline 
113 & 10.1002/pst.2069     & Sample size calculation for log-rank test and prediction of number of events over time & Lu K & Pharm Stat & 2021 \\ \hline 
116 & 10.1002/pst.2060    & Cancer immunotherapy trial design with long-term survivors & Ding X & Pharm Stat & 2021 \\ \hline 
118 & 10.1186/s12874-020-01098-5       & Dynamic RMST curves for survival analysis in clinical trials & Liao JJZ & BMC Med Res Methodol & 2020 \\ \hline 
119 & 10.1002/pst.2062    & Delayed treatment effects, treatment switching and heterogeneous patient populations: How to design and analyze RCTs in oncology & Ristl R & Pharm Stat & 2021 \\ \hline 
122 & 10.1016/j  .jval.2020.03.010   & Nonproportional Hazards in Network Meta-Analysis: Efficient Strategies for Model Building and Analysis & Wiksten A & Value Health & 2020 \\ \hline 
123 & 10.1177/0962280220941532       & Functional survival forests for multivariate longitudinal outcomes: Dynamic prediction of Alzheimer's disease progression & Lin J & Stat Methods Med Res & 2021 \\ \hline 
126 & 10.1136/bmjopen-2019-033965      & Moving beyond the Cox proportional hazards model in survival data analysis: a cervical cancer study & Li L & BMJ Open & 2020 \\ \hline 
134 & 10.1111/biom.13315       & Optimality of testing procedures for survival data in the nonproportional hazards setting & Arf\`{e} A & Biometrics & 2021 \\ \hline 
136 & 10.1177/0962280220931560        & Group sequential monitoring based on the maximum of weighted log-rank statistics with the Fleming-Harrington class of weights in oncology clinical trials & Prior TJ & Stat Methods Med Res & 2020 \\ \hline 
142 & 10.1002/sim.8508        & A multiparameter regression model for interval-censored survival data & Peng D & Stat Med & 2020 \\ \hline 
145 & 10.1186/s13063-020-4153-2        & A simulation study comparing the power of nine tests of the treatment effect in randomized controlled trials with a time-to-event outcome & Royston P & Trials & 2020 \\ \hline 
148 & 10.1177/0962280220907355        & K-sample omnibus non-proportional hazards tests based on right-censored data & Gorfine M & Stat Methods Med Res & 2020 \\ \hline 
149 & 10.1186/s13045-020-0847-x        & Treating non-responders: pitfalls and implications for cancer immunotherapy trial design & Xu Z & J Hematol Oncol & 2020 \\ \hline 
154 & 10.1002/pst.2003       & Designing cancer immunotherapy trials with delayed treatment effect using maximin efficiency robust statistics & Ding X & Pharm Stat & 2020 \\ \hline 
157 & 10.1109/TCBB.2020.2965934       & "Supervised Dimension Reduction for Large-Scale ""Omics"" Data With Censored Survival Outcomes Under Possible Non-Proportional Hazards" & Spirko-Burns L & IEEE/ACM Trans Comput Biol Bioinform & 2021 \\ \hline 
159 & 10.1002/pst.1976       & Study design of single-arm phase II immunotherapy trials with long-term survivors and random delayed treatment effect & Chu C & Pharm Stat & 2020 \\ \hline 
162 & 10.1002/pst.2000       & A resampling-based test for two crossing survival curves & Liu T & Pharm Stat & 2020 \\ \hline 
167 & 10.1111/biom.13196       & Sample size and power for the weighted log-rank test and Kaplan-Meier based tests with allowance for nonproportional hazards & Yung G & Biometrics & 2020 \\ \hline 
169 & 10.1002/sim.8440       & Cancer immunotherapy trial design with cure rate and delayed treatment effect & Wei J & Stat Med & 2020 \\ \hline 
174 & 10.1002/pst.1982       & Cancer immunotherapy trial design with delayed treatment effect & Wu J & Pharm Stat & 2020 \\ \hline 
176 & 10.1177/0962280219883905      & Long-term frailty modeling using a non-proportional hazards model: Application with a melanoma dataset & Calsavara VF & Stat Methods Med Res & 2020 \\ \hline 
184 & 10.1002/sim.8363      & Quantifying treatment effects using the personalized chance of longer survival & Zhao YQ & Stat Med & 2019 \\ \hline 
188 & 10.3389/fpubh.2019.00201      & Modeling Long-Term Graft Survival With Time-Varying Covariate Effects: An Application to a Single Kidney Transplant Centre in Johannesburg, South Africa & Achilonu OJ & Front Public Health & 2019 \\ \hline 
195 & 10.1177/0962280219862586      & Survival forests under test: Impact of the proportional hazards assumption on prognostic and predictive forests for amyotrophic lateral sclerosis survival & Korepanova N & Stat Methods Med Res & 2020 \\ \hline 
201 & 10.1002/sim.8186      & Modestly weighted log-rank tests & Magirr D & Stat Med & 2019 \\ \hline 
207 & 10.1016/j.jamcollsurg.2019     .\-04.019 & A Tale of Two Kidneys: Differences in Graft Survival for Kidneys Allocated to Simultaneous Liver Kidney Transplant Compared with Contralateral Kidney from the Same Donor & Cannon RM & J Am Coll Surg & 2019 \\ \hline 
213 & 10.1186/s13063-019-3251-5      & Combined test versus log-rank/Cox test in 50 randomised trials & Royston P & Trials & 2019 \\ \hline 
235 & 10.1111/biom.13007     & Nonparametric group sequential methods for evaluating survival benefit from multiple short-term follow-up windows & Xia M & Biometrics & 2019 \\ \hline 
242 & 10.1002/sim.7975     & Design and monitoring of survival trials in complex scenarios & Luo X & Stat Med & 2019 \\ \hline 
244 & 10.1002/sim.7958      & Interim monitoring using the adaptively weighted log-rank test in clinical trials for survival outcomes & Yang S & Stat Med & 2019 \\ \hline 
245 & 10.1002/sim.7937      & Designing cancer immunotherapy trials with random treatment time-lag effect & Xu Z & Stat Med & 2018 \\ \hline 
247 & 10.1002/sim.7936     & Nonparametric adaptive enrichment designs using categorical surrogate data & Br\"{u}ckner M & Stat Med & 2018 \\ \hline 
251 & 10.1002/pst.1878    & Weighted log-rank test for time-to-event data in immunotherapy trials with random delayed treatment effect and cure rate & Liu S & Pharm Stat & 2018 \\ \hline 
256 & 10.1111/biom.12916   & A robust approach to sample size calculation in cancer immunotherapy trials with delayed treatment effect & Ye T & Biometrics & 2018 \\ \hline 
265 & 10.3414/ME17-01-0058   & The Average Hazard Ratio - A Good Effect Measure for Time-to-event Endpoints when the Proportional Hazard Assumption is Violated? & Rauch G & Methods Inf Med & 2018 \\ \hline 
266 & 10.1016/j .conctc.2017.09.004 & Estimation of treatment effects in weighted log-rank tests & Lin RS & Contemp Clin Trials Commun & 2017 \\ \hline 
267 & 10.1002/sim.7661   & A flexible and coherent test/estimation procedure based on restricted mean survival times for censored time-to-event data in randomized clinical trials & Horiguchi M & Stat Med & 2018 \\ \hline 
268 & 10.1002/sim.7676    & Improving testing and description of treatment effect in clinical trials with survival outcomes & Yang S & Stat Med & 2019 \\ \hline 
269 & 10.1177/1740774518769865    & A new modeling and inference approach for the Systolic Blood Pressure Intervention Trial outcomes & Yang S & Clin Trials & 2018 \\ \hline 
274 & 10.1007/s10985-018-9428-5    & Improved precision in the analysis of randomized trials with survival outcomes, without assuming proportional hazards & D\'{i}az I & Lifetime Data Anal & 2019 \\ \hline 
284 & 10.1186/s12874-017-0430-5   & Assessing the effect of a partly unobserved, exogenous, binary time-dependent covariate on survival probabilities using generalised pseudo-values & P\"{o}tschger U & BMC Med Res Methodol & 2018 \\ \hline 
296 & 10.1002/pst.1840  & Design considerations in clinical trials with cure rate survival data: A case study in oncology & Sun S & Pharm Stat & 2018 \\ \hline 
313 & 10.1002/sim.7421   & A clinical trial design using the concept of proportional time using the generalized gamma ratio distribution & Phadnis MA & Stat Med & 2017 \\ \hline 
328 & 10.1136/bmj.j2250  & Life expectancy difference and life expectancy ratio: two measures of treatment effects in randomised trials with non-proportional hazards & Dehbi HM & BMJ & 2017 \\ \hline 
335 & 10.1177/0962280216688034   & Flexible modeling of the hazard rate and treatment effects in long-term survival studies & Hagar Y & Stat Methods Med Res & 2017 \\ \hline 
343 & 10.1111/biom.12625  & Multi-parameter regression survival modeling: An alternative to proportional hazards & Burke K & Biometrics & 2017 \\ \hline 
347 & 10.1002/sim.7157    & Designing therapeutic cancer vaccine trials with delayed treatment effect & Xu Z & Stat Med & 2017 \\ \hline 
359 & 10.1002/pst.1760   & Group sequential monitoring based on the weighted log-rank test statistic with the Fleming-Harrington class of weights in cancer vaccine studies & Hasegawa T & Pharm Stat & 2016 \\ \hline 
360 & 10.1186/s40064-016-2210-9      & Bayesian test for hazard ratio in survival analysis & Kim G & Springerplus & 2016 \\ \hline 
365 & 10.1177/0962280216648724      & A censored quantile regression approach for the analysis of time to event data & Xue X & Stat Methods Med Res & 2018 \\ \hline 
367 & 10.1177/0962280216642264      & Assessing methods for dealing with treatment switching in clinical trials: A follow-up simulation study & Latimer NR & Stat Methods Med Res & 2018 \\ \hline 
376 & 10.1007/s10985-016-9360-5       & Sequential tests for non-proportional hazards data & Br\"{u}ckner M & Lifetime Data Anal & 2017 \\ \hline 
378 & 10.1002/sim.6929       & Estimating restricted mean treatment effects with stacked survival models & Wey A & Stat Med & 2016 \\ \hline 
382 & 10.1186/s12874-016-0110-x       & Augmenting the logrank test in the design of clinical trials in which non-proportional hazards of the treatment effect may be anticipated & Royston P & BMC Med Res Methodol & 2016 \\ \hline 
383 & 10.1002/sim.6893       & Nonparametric survival analysis using Bayesian Additive Regression Trees (BART) & Sparapani RA & Stat Med & 2016 \\ \hline 
384 & 10.1002/sim.6889       & Design and analysis of clinical trials in the presence of delayed treatment effect & Sit T & Stat Med & 2016 \\ \hline 
392 & 10.1002/sim.6801        & Gaining power and precision by using model-based weights in the analysis of late stage cancer trials with substantial treatment switching & Bowden J & Stat Med & 2016 \\ \hline 
396 & 10.1177/0962280215605107 & An ensemble survival model for estimating relative residual longevity following stroke: Application to mortality data in the chronic dialysis population & Phadnis MA & Stat Methods Med Res & 2017 \\ \hline 
397 & 10.1080/10543406.2015.\-1094809  & Interim analysis based on the weighted log-rank test for delayed treatment effects under staggered patient entry & Yoshida M & J Biopharm Stat & 2016 \\ \hline 
402 & 10.1002/sim.6591        & A versatile test for equality of two survival functions based on weighted differences of Kaplan-Meier curves & Uno H & Stat Med & 2015 \\ \hline 
404 & 10.1016/S1470-2045(15)00086-8 & Standard chemotherapy with or without bevacizumab for women with newly diagnosed ovarian cancer (ICON7): overall survival results of a phase 3 randomised trial & Oza AM & Lancet Oncol & 2015 \\ \hline 
413 & 10.1016/j       .cct.2015.05.018  & Design of clinical trials with failure-time endpoints and interim analyses: An update after fifteen years & He P & Contemp Clin Trials & 2015 \\ \hline 
417 & 10.1371/journal.\-pone.0123784  & Analysis of time to event outcomes in randomized controlled trials by generalized additive models & Argyro\-poulos C & PLoS One & 2015 \\ \hline 
422 & 10.1111/tri.12568    & Post-transplant survival is improved for hepatitis C recipients who are RNA negative at time of liver transplantation & Fortune BE & Transpl Int & 2015 \\ \hline 
424 & 10.1515/ijb-2014-0006    & Conditional transformation models for survivor function estimation & M\"{o}st L & Int J Biostat & 2015 \\ \hline 
425 & 10.1002/sim.6453  & Assessing potentially time-dependent treatment effect from clinical trials and observational studies for survival data, with applications to the Women's Health Initiative combined hormone therapy trial & Yang S & Stat Med & 2015 \\ \hline 
427 & 10.1371/journal.pone .\-0116774  & Statistical inference methods for two crossing survival curves: a comparison of methods & Li H & PLoS One & 2015 \\ \hline 
433 & 10.1177/0962280214557578        & Adjusting for treatment switching in randomised controlled trials - A simulation study and a simplified two-stage method & Latimer NR & Stat Methods Med Res & 2017 \\ \hline 
435 & 10.1002/sim.6356        & A model for time to fracture with a shock stream superimposed on progressive degradation: the Study of Osteoporotic Fractures & He X & Stat Med & 2015 \\ \hline 
444 & 10.1186/1745-6215-15-314       & An approach to trial design and analysis in the era of non-proportional hazards of the treatment effect & Royston P & Trials & 2014 \\ \hline 
445 & 10.1007/s10985-014-9301-0      & Quantifying the average of the time-varying hazard ratio via a class of transformations & Chen Q & Lifetime Data Anal & 2015 \\ \hline 
446 & 10.1200/JCO.2014.55.2208    & Moving beyond the hazard ratio in quantifying the between-group difference in survival analysis & Uno H & J Clin Oncol & 2014 \\ \hline 
452 & 10.1016/S1470-2045(14)70189-5 & Ipilimumab versus placebo after radiotherapy in patients with metastatic castration-resistant prostate cancer that had progressed after docetaxel chemotherapy (CA184-043): a multicentre, randomised, double-blind, phase 3 trial & Kwon ED & Lancet Oncol & 2014 \\ \hline 
453 & 10.1016/j .spl.2013.03.007  & On penalized likelihood estimation for a non-proportional hazards regression model & Devarajan K & Stat Probab Lett & 2013 \\ \hline 
456 & 10.1002/sim.6178   & Impact of the model-building strategy on inference about nonlinear and time-dependent covariate effects in survival analysis & Wynant W & Stat Med & 2014 \\ \hline 
464 & 10.1002/sim.6105    & Likelihood approaches for proportional likelihood ratio model with right-censored data & Zhu H & Stat Med & 2014 \\ \hline 
465 & 10.1002/pst.1609   & Sample size determination for the weighted log-rank test with the Fleming-Harrington class of weights in cancer vaccine studies & Hasegawa T & Pharm Stat & 2014 \\ \hline 
467 & 10.1080/07474946.2014 .\-856635   & Flexibly Monitoring Group Sequential Survival Trials When Testing is Based Upon a Weighted Log-Rank Statistic & Brummel SS & Seq Anal & 2014 \\ \hline 
469 & 10.1111/biom.12097   & Efficient semiparametric estimation of short-term and long-term hazard ratios with right-censored data & Diao G & Biometrics & 2013 \\ \hline 
470 & 10.1186/1471-2288-13-152    & Restricted mean survival time: an alternative to the hazard ratio for the design and analysis of randomized trials with a time-to-event outcome & Royston P & BMC Med Res Methodol & 2013 \\ \hline 
482 & 10.1002/sim.5961    & Model building in nonproportional hazard regression & Rodr\'{i}guez-Girondo M & Stat Med & 2013 \\ \hline 
501 & 10.1002/sim.5673    & A K-nearest neighbors survival probability prediction method & Lowsky DJ & Stat Med & 2013 \\ \hline 
510 & 10.1080/10543406.2011 .\-616973   & A threshold hazard model for estimating serious infection risk following anti-tumor necrosis factor therapy in rheumatoid arthritis patients & Fu B & J Biopharm Stat & 2013 \\ \hline 
511 & 10.1080/10543406.2011.\-616975   & Power comparisons for group sequential tests with nonparametric statistics in case of nonproportional hazards & Demirhan H & J Biopharm Stat & 2013 \\ \hline 
513 & 10.1155/2013/796270    & Additive hazard regression models: an application to the natural history of human papillomavirus & Xie X & Comput Math Methods Med & 2013 \\ \hline 
514 & 10.1111/biom.12002   & Estimating time to disease progression comparing transition models and survival methods--an analysis of multiple sclerosis data & Mandel M & Biometrics & 2013 \\ \hline 
518 & 10.1177/0962280212473348   & Estimating efficacy in the presence of non-ignorable non-trial interventions in the Helsinki Psychotherapy Study & H\"{a}rk\"{a}nen T & Stat Methods Med Res & 2016 \\ \hline 
520 & 10.1186/2051-1426-1-18   & Statistical issues and challenges in immuno-oncology & Chen TT & J Immunother Cancer & 2013 \\ \hline 
524 & 10.1007/s10985-012-9234-4   & Estimation of odds of concordance based on the Aalen additive model & Martinussen T & Lifetime Data Anal & 2013 \\ \hline 
525 & 10.1371/journal.pone .\-0047804  & A flexible alternative to the Cox proportional hazards model for assessing the prognostic accuracy of hospice patient survival & Miladinovic B & PLoS One & 2012 \\ \hline 
528 & 10.1186/1471-2369-13-130   & All-cause and cause-specific mortality associated with diabetes in prevalent hemodialysis patients & Sattar A & BMC Nephrol & 2012 \\ \hline 
529 & 10.1111/j.1541-0420.2012.01784.x   & A general class of semiparametric transformation frailty models for nonproportional hazards survival data & Choi S & Biometrics & 2012 \\ \hline 
531 & 10.1177/1740774512455464   & Utilizing the integrated difference of two survival functions to quantify the treatment contrast for designing, monitoring, and analyzing a comparative clinical study & Zhao L & Clin Trials & 2012 \\ \hline 
535 & 10.1007/s10985-012-9224-6   & Robust inference in discrete hazard models for randomized clinical trials & Nguyen VQ & Lifetime Data Anal & 2012 \\ \hline 
539 & 10.1002/sim.5440   & Estimation of treatment effect under non-proportional hazards and conditionally independent censoring & Boyd AP & Stat Med & 2012 \\ \hline 
542 & 10.1007/s10985-012-9222-8   & Estimation of treatment effects based on possibly misspecified Cox regression & Hattori S & Lifetime Data Anal & 2012 \\ \hline 
550 & 10.1056/NEJMoa1103799   & A phase 3 trial of bevacizumab in ovarian cancer & Perren TJ & N Engl J Med & 2011 \\ \hline 
553 & 10.1111/j.1541-0420.2011.01696.x   & Empirical likelihood for cumulative hazard ratio estimation with covariate adjustment & Dong B & Biometrics & 2012 \\ \hline 
568 & 10.1002/sim.4274   & The use of restricted mean survival time to estimate the treatment effect in randomized clinical trials when the proportional hazards assumption is in doubt & Royston P & Stat Med & 2011 \\ \hline 
571 & 10.1007/s00405-011-1601-3   & A critical appraisal of different survival techniques in oral cancer patients & K\"{o}hler HF & Eur Arch Otorhinolaryngol & 2012 \\ \hline 
573 & 10.1080/02664760903521476    & Bayesian Parametric Accelerated Failure Time Spatial Model and its Application to Prostate Cancer & Zhang J & J Appl Stat & 2011 \\ \hline 
575 & 10.1007/s10985-011-9195-z    & Efficiency improvement in a class of survival models through model-free covariate incorporation & Garcia TP & Lifetime Data Anal & 2011 \\ \hline 
586 & 10.1001/archinternmed.2010  .\-352  & Similar outcomes with hemodialysis and peritoneal dialysis in patients with end-stage renal disease & Mehrotra R & Arch Intern Med & 2011 \\ \hline 
588 & 10.1093/biostatistics/kxq061    & Estimation of the 2-sample hazard ratio function using a semiparametric model & Yang S & Biostatistics & 2011 \\ \hline 
594 & 10.1111/j.1541-0420.2010.01449.x      & Double inverse-weighted estimation of cumulative treatment effects under nonproportional hazards and dependent censoring & Schaubel DE & Biometrics & 2011 \\ \hline 
596 & 10.1111/j.1365-2753.2009.01190.x     & SAS macros for point and interval estimation of area under the receiver operating characteristic curve for non-proportional and proportional hazards Weibull models & Mannan H & J Eval Clin Pract & 2010 \\ \hline 
607 & 10.1093/bioinformatics/   \-btq035 & Gene selection in microarray survival studies under possibly non-proportional hazards & Dunkler D & Bioinformatics & 2010 \\ \hline 
609 &  & Estimating a scale-change effect for time-varying phenotypes in genome-wide association studies & Chen YQ & J Appl Statist Sci & 2010 \\ \hline 
611 & 10.1002/bimj.200800244    & An accelerated failure time mixture cure model with masked event & Zhang JJ & Biom J & 2009 \\ \hline 
613 & 10.1007/s10985-009-9132-6     & Confidence intervals for the first crossing point of two hazard functions & Cheng MY & Lifetime Data Anal & 2009 \\ \hline 
614 & 10.1093/biostatistics/kxp041    & Bayesian random-effects threshold regression with application to survival data with nonproportional hazards & Pennell ML & Biostatistics & 2010 \\ \hline 
618 & 10.1002/sim.3686     & A semi-parametric threshold regression analysis of sexually transmitted infections in adolescent women & Yu Z & Stat Med & 2009 \\ \hline 
622 & 10.1002/sim.3623     & The estimation of average hazard ratios by weighted Cox regression & Schemper M & Stat Med & 2009 \\ \hline 
626 & 10.1111/j.1541-0420.2008.01166.x     & Bayesian nonparametric nonproportional hazards survival modeling & De Iorio M & Biometrics & 2009 \\ \hline 
630 & 10.1002/sim.3501     & Power and sample size calculation for log-rank test with a time lag in treatment effect & Zhang D & Stat Med & 2009 \\ \hline 
645 & 10.1111/j.1541-0420.2007.00947.x     & Estimating cumulative treatment effects in the presence of nonproportional hazards & Wei G & Biometrics & 2008 \\ \hline 
655 & 10.1002/bimj.200610328    & A new proposal for multivariable modelling of time-varying effects in survival data based on fractional polynomial time-transformation & Sauerbrei W & Biom J & 2007 \\ \hline 
656 & 10.1002/bimj.200510325    & A partial likelihood approach to smooth estimation of dynamic covariate effects using penalised splines & Brown D & Biom J & 2007 \\ \hline 
658 & 10.1007/s10985-007-9035-3     & Bayesian analysis of generalized odds-rate hazards models for survival data & Banerjee T & Lifetime Data Anal & 2007 \\ \hline 
664 & 10.1111/j.1541-0420.2006.00571.x      & Bayesian semiparametric dynamic frailty models for multiple event time data & Pennell ML & Biometrics & 2006 \\ \hline 
665 & 10.1534/genetics.106.059808      & Mapping temporally varying quantitative trait loci in time-to-failure experiments & Johannes F & Genetics & 2007 \\ \hline 
666 & 10.1002/pst   .213 & Comparing proportional hazards and accelerated failure time models: an application in influenza & Patel K & Pharm Stat & 2006 \\ \hline 
667 & 10.1002/sim.2729     & Approaches in modelling long-term survival: an application to breast cancer & Perperoglou A & Stat Med & 2007 \\ \hline 
676 & 10.1080/10543400600614791     & Testing for crossover of two hazard functions using Gail and Simon's method & Chen YH & J Biopharm Stat & 2006 \\ \hline 
677 & 10.1007/s10549-006-9231-y     & Effects of estrogen receptor expression and histopathology on annual hazard rates of death from breast cancer & Anderson WF & Breast Cancer Res Treat & 2006 \\ \hline 
681 & 10.1002/sim.2517     & Evaluation of sample size and power for multi-arm survival trials allowing for non-uniform accrual, non-proportional hazards, loss to follow-up and cross-over & Barthel FM & Stat Med & 2006 \\ \hline 
682 & 10.1002/gepi.20132   & Family-based association test for time-to-onset data with time-dependent differences between the hazard functions & Jiang H & Genet Epidemiol & 2006 \\ \hline 
689 & 10.1002/sim.2248     & Interim analysis on survival data: its potential bias and how to repair it & van Houwelingen HC & Stat Med & 2005 \\ \hline 
694 & 10.1007/s10985-004-0384-x     & Bayesian model selection and averaging in additive and proportional hazards models & Dunson DB & Lifetime Data Anal & 2005 \\ \hline 
699 & 10.1016/j    .compbiolchem.\-2005.02.001 & Survival analysis of microarray expression data by transformation models & Xu J & Comput Biol Chem & 2005 \\ \hline 
700 & 10.1111/j.0006-341X.2005.030814.x     & Survival model predictive accuracy and ROC curves & Heagerty PJ & Biometrics & 2005 \\ \hline 
701 & 10.1111/j.0006-341X.2005.031206.x     & A sample size formula for the supremum log-rank statistic & Eng KH & Biometrics & 2005 \\ \hline 
711 & 10.1023/b:    lida.0000019255.\-21735.9b & Incorporating follow-up time in M-estimation for survival data & Heller G & Lifetime Data Anal & 2004 \\ \hline 
716 & 10.1002/sim.1618     & Parametric randomization-based methods for correcting for treatment changes in the assessment of the causal effect of treatment & Walker AS & Stat Med & 2004 \\ \hline 
717 &  & The effect of sample size and MLP architecture on Bayesian learning for cancer prognosis--a case study & Trinh QA & Stud Health Technol Inform & 2003 \\ \hline 
719 & 10.1002/sim.1484     & A relative survival regression model using B-spline functions to model non-proportional hazards & Giorgi R & Stat Med & 2003 \\ \hline 
720 & 10.1093/biostatistics/1.4.423   & Estimating average regression effect under non-proportional hazards & Xu R & Biostatistics & 2000 \\ \hline 
722 &  & Assessing time-by-covariate interactions in relative survival models using restrictive cubic spline functions & Bolard P & J Cancer Epidemiol Prev & 2002 \\ \hline 
723 & 10.1002/sim.1411    & Dynamic Cox modelling based on fractional polynomials: time-variations in gastric cancer prognosis & Berger U & Stat Med & 2003 \\ \hline 
732 & 10.1002/sim.1203    & Flexible parametric proportional-hazards and proportional-odds models for censored survival data, with application to prognostic modelling and estimation of treatment effects & Royston P & Stat Med & 2002 \\ \hline 
739 & 10.1111/j.0006-341x.2002.00443.x    & Generalized additive models with interval-censored data and time-varying covariates: application to human immunodeficiency virus infection in hemophiliacs & Bacchetti P & Biometrics & 2002 \\ \hline 
740 & 10.1111/j.0006-341x.2002.00305.x   & Survival analysis with time-varying regression effects using a tree-based approach & Xu R & Biometrics & 2002 \\ \hline 
745 & 10.1002/sim.1061   & Estimation of the survival function for Gray's piecewise-constant time-varying coefficients model & Valenta Z & Stat Med & 2002 \\ \hline 
751 & 10.1016/s0895-4356(01)00363-8   & Modelling time-dependent hazard ratios in relative survival: application to colon cancer & Bolard P & J Clin Epidemiol & 2001 \\ \hline 
755 & 10.1111/j.0006-341x.1999.00585.x   & Bayesian analysis and model selection for interval-censored survival data & Sinha D & Biometrics & 1999 \\ \hline 
773 & 10.1002/(  sici)1097-0258(19991015)18:19 $\mathrm{<}$2617\-::aid-sim187$\mathrm{>}$3.0.co;2-e" & Randomization-based methods for correcting for treatment changes: examples from the Concorde trial & White IR & Stat Med & 1999 \\ \hline 
776 & 10.1002/( sici)1097-0258(19981115)17:21$\mathrm{<}$2525\-::aid-sim936$\mathrm{>}$3.0.co;2-e" & Sample size determination in complex clinical trials comparing more than two groups for survival endpoints & Ahnn S & Stat Med & 1998 \\ \hline 
777 & 10.1023/a:1009673932333    & Flexible Bayesian modelling for survival data & Gustafson P & Lifetime Data Anal & 1998 \\ \hline 
778 & 10.1023/a:1009689630516    & Frailty models and rank tests & Oakes D & Lifetime Data Anal & 1998 \\ \hline 
781 &  & A method for sequential analysis of survival data with nonproportional hazards & Soori\-yarachchi MR & Biometrics & 1998 \\ \hline 
791 & 10.1002/(sici)1097-0258(19970830)16:16$\mathrm{<}$1831\-::aid-sim617$\mathrm{>}$3.0.co;2-m & On a non-proportional hazards regression model for repeated medical random counts & Mackenzie G & Stat Med & 1997 \\ \hline 
796 & 10.1016/s0197-2456(96)00089-x   & Use of Irwin's restricted mean as an index for comparing survival in different treatment groups--interpretation and power considerations & Karrison TG & Control Clin Trials & 1997 \\ \hline 
797 & 10.1002/(sici)1097-0258(19970330)16:6$\mathrm{<}$695\-::aid-sim436$\mathrm{>}$3.0.co;2-c & A partially grouped log-rank test & Sposto R & Stat Med & 1997 \\ \hline 
804 &  & Some permutation tests for survival data & Sun Y & Biometrics & 1996 \\ \hline 
813 & 10.1016/0304-3835(94)90095-7   & A technique for using neural network analysis to perform survival analysis of censored data & De Laurentiis M & Cancer Lett & 1994 \\ \hline 
821 &  & A nonproportional hazards Weibull accelerated failure time regression model & Anderson KM & Biometrics & 1991 \\ \hline 
834 &  & Survival analysis of drug combinations using a hazards model with time-dependent covariates & Stablein DM & Biometrics & 1980 \\ \hline 
M3 & 10.1111/j.0006-341X.2001.00837.x ~  & Maximum of the Weighted Kaplan-Meier Tests with Application to Cancer Prevention and Screening Trials & Shen Y & Biometrics & 2001 \\ \hline 
M5 & 10.1093/biomet/64.1.156  & On distribution-free tests for equality of survival distributions & Tarone RE & Biometrika & 1977 \\ \hline 
M6 & 10.1002/bimj.200310053  & A new testing approach for comparing the overall homogeneity of survival curves & Lin X & Biometrical & 2004 \\ \hline 
M8 & 10.2307/2556114  & Modified Kolmogorov-Smirnov test procedures with application to arbitrarily right-censored data & Fleming TR & Biometrics & 1980 \\ \hline 
M9 & 10.1002/pst.376  & A new method for the comparison of survival distributions & Lin X & Pharm Stat 9(1): 67--76 & 2010 \\ \hline 
M10 &  & A two-stage procedure for comparing hazard rate functions & Qiu P & J R Statist Soc B 70(1): 191--208 & 2008 \\ \hline 
M11 & 10.1081/SAC-100107779   & Some versatile tests based on the simultaneous use of weighted log-rank and weighted kaplan-meier statistics & Chi Y & Commun Stat 30(4):743--759 & 2001 \\ \hline 
M13 & 10.1093/biomet/92.1.1   & Semiparametric analysis of short-term and long-term hazard ratios with two-sample survival data & Yang S & Biometrika 2005; 92: 1--17 & 2005 \\ \hline 
E2 & 10.1007/s12561-0 21-09316-4 & Sample Size Re-estimation with the Com-Nougue Method to Evaluate Treatment Effect & Wang J. & Statistics in Biosciences (2022) 14:1 (90-103). Date of Publication: 1 Apr 2022 & 2022 \\ \hline 
E3 & 10.1080/19466315.2020.\-1801496  & An Assessment of the Treatment Effect in Treatment of Physician Choice Trials & Mpofu P., Karuri S.W. & Statistics in Biopharmaceutical Research (2022) 14:1 (103-113). Date of Publication: 2022 & 2022 \\ \hline 
E11 & 10.1007/s12561-020-09276-1  & On Weighted Log-Rank Combination Tests and Companion Cox Model Estimators & Le\'{o}n L.F., Lin R., Anderson K.M. & Statistics in Biosciences (2020) 12:2 (225-245). Date of Publication: 1 Jul 2020 & 2020 \\ \hline 
E15 & 10.1080/19466315.2019 .\-1697738  & Alternative Analysis Methods for Time to Event Endpoints Under Nonproportional Hazards: A Comparative Analysis & Lin R.S., Lin J., Roychoudhury S., Anderson K.M., Hu T., Huang B., Leon L.F., Liao J.J.Z., Liu R., Luo X., Mukhopadhyay P., Qin R., Tatsuoka K., Wang X., Wang Y., Zhu J., Chen T.-T., Iacona R. & Statistics in Biopharmaceutical Research (2020) 12:2 (187-198). Date of Publication: 2 Apr 2020 & 2020 \\ \hline 
E24 & 10.1080/19466315.2018 .\-1527249  & Difference in Restricted Mean Survival Time: Small Sample Distribution and Asymptotic Relative Efficiency & Lawrence J., Qiu J., Bai S., Hung H.M.J. & Statistics in Biopharmaceutical Research (2019) 11:1 (61-66). Date of Publication: 2 Jan 2019 & 2019 \\ \hline 
E31 & 10.1016/j.conctc.2017.09.004   & Estimation of treatment effects in weighted log-rank tests & Lin R.S., Le\'{o}n L.F. & Contemporary Clinical Trials Communications (2017) 8 (147-155). Date of Publication: 1 Dec 2017 & 2017 \\ \hline 
E33 & 10.1080/19466315.2016 .\-1257436  & Testing Treatment Effect in Randomized Clinical Trials With Possible Nonproportional Hazards & Callegaro A., Spiessens B. & Statistics in Biopharmaceutical Research (2017) 9:2 (204-211). Date of Publication: 3 Apr 2017 & 2017 \\ \hline 
E37 & 10.1080/19466315.2016.\-1207560  & Sample Size and Power of Survival Trials in Group Sequential Design With Delayed Treatment Effect & Zhang J., Pulkstenis E. & Statistics in Biopharmaceutical Research (2016) 8:3 (268-275). Date of Publication: 2 Jul 2016 & 2016 \\ \hline 
E61 & 10.2202/1557-4679.1125  & Fitting smooth-in-time prognostic risk functions via logistic regression & Hanley J.A., Miettinen O.S. & International Journal of Biostatistics (2009) 5:1 Article Number: 3. Date of Publication: 2009 & 2009 \\ \hline 
E65 & 10.1177/009286150\-704100412  & Consequences of delayed treatment effects on analysis of time-to-event endpoints & Fine G.D. & Drug Information Journal (2007) 41:4 (535-539). Date of Publication: 2007 & 2007 \\ \hline 
\end{longtable}

\end{landscape}

\nocite{}

\end{document}